\input harvmac.tex 
\input amssym.def
\input amssym
\baselineskip 14pt
\magnification\magstep1
\parskip 6pt
\newdimen\itemindent \itemindent=32pt
\def\textindent#1{\parindent=\itemindent\let\par=\resetpar%
\indent\llap{#1\enspace}\ignorespaces}

\let\oldpar=\par
\def\resetpar{\oldpar\parindent=20pt\let\par=\oldpar}

\font\ninerm=cmr9 \font\ninesy=cmsy9
\font\eightrm=cmr8 \font\sixrm=cmr6
\font\eighti=cmmi8 \font\sixi=cmmi6
\font\eightsy=cmsy8 \font\sixsy=cmsy6
\font\eightbf=cmbx8 \font\sixbf=cmbx6
\font\eightit=cmti8
\def\eightpoint{\def\rm{\fam0\eightrm}
 \textfont0=\eightrm \scriptfont0=\sixrm \scriptscriptfont0=\fiverm
 \textfont1=\eighti  \scriptfont1=\sixi  \scriptscriptfont1=\fivei
 \textfont2=\eightsy \scriptfont2=\sixsy \scriptscriptfont2=\fivesy
 \textfont3=\tenex   \scriptfont3=\tenex \scriptscriptfont3=\tenex
 \textfont\itfam=\eightit  \def\it{\fam\itfam\eightit}%
 \textfont\bffam=\eightbf  \scriptfont\bffam=\sixbf
 \scriptscriptfont\bffam=\fivebf  \def\bf{\fam\bffam\eightbf}%
 \normalbaselineskip=9pt
 \setbox\strutbox=\hbox{\vrule height7pt depth2pt width0pt}%
 \let\big=\eightbig  \normalbaselines\rm}
\catcode`@=11 %
\def\eightbig#1{{\hbox{$\textfont0=\ninerm\textfont2=\ninesy
 \left#1\vbox to6.5pt{}\right.\n@@space$}}}
\def\vfootnote#1{\insert\footins\bgroup\eightpoint
 \interlinepenalty=\interfootnotelinepenalty
 \splittopskip=\ht\strutbox %
 \splitmaxdepth=\dp\strutbox %
 \leftskip=0pt \rightskip=0pt \spaceskip=0pt \xspaceskip=0pt
 \textindent{#1}\footstrut\futurelet\next\fo@t}
\catcode`@=12 %

\font \bigbf=cmbx10 scaled \magstep1

\def \de{\delta}
\def \De{\Delta}
\def \si{\sigma}

\def \ga{\gamma}

\def \tr{{\rm tr }}

\def \J{{\rm J}}

\def \r{\big \rangle}

\def \vep{\varepsilon}
\def \half{{\textstyle {1 \over 2}}}

\def \ts{\textstyle}

\def \d{{\rm d}}
\def \v{{\rm v}}
\def \x{{\rm x}}
\def \y{{\rm y}}

\def \u{{\rm u}}
\def \v{{\rm v}}

\def \A{{\cal A}}

\def \D{{\cal D}}

\def \H{{\cal H}}
\def \I{{\cal I}}
\def \J{{\cal J}}
\def \K{{\cal K}}

\def \M{{\cal M}}
\def \N{{\cal N}}

\def \Q{{\cal Q}}
\def \R{{\cal R}}
\def \S{{\cal S}}
\def \U{{\cal U}}
\def \V{{\cal V}}

\def \bQ{{\bar{\cal Q}}}
\def \bS{{\bar {\cal S}}}
\def \bsi{\bar \sigma}

\def \e{{\rm e}}
\def \x{{\rm x}}
\def \y{{\rm y}}

\def \z{{\rm z}}

\def \bga{{\bar \gamma}}

\def \bga{{\bar \gamma}}

\def \bsi{\bar \sigma}

\def \al{{\alpha}}
\def \tr{{\rm Tr}}
\def \be{{\beta}}

\def \rr{{\rm r}}

\def \blambda{{\underline \lambda}}
\def \brho{{\underline \rho}}
\def \bsi{{\underline \sigma}}
\def \ba{{\bar r}}
\def \bb{{\bar s}}
\def \hc{{\hat k}}
\def \hd{{\hat l}}
\def \ha{{\hat m}}
\def \hb{{\hat n}}

\def\fo#1{\mathrel{\mathop{\longrightarrow}\limits_{\scriptstyle{#1}}}}

\lref\grant{L. Grant, P.A. Grassi, S. Kim and S. Minwalla,
 {\it Comments on $1/16$ BPS Quantum States and Classical Configurations},
  JHEP 0805 (2008) 049,
  arXiv:0803.4183 [hep-th].}

\lref\harary{{\it Graphical Enumeration}, F. Harary and E. M. Palmer,
Academic Press New York and London, 1973.}

\lref\domokos{M. Domokos, {\it Multisymmetric Syzygies}, arXiv:math/0602303.}

\lref\counting{F.A. Dolan,
  {\it Counting BPS operators in $\N=4$ SYM},
  Nucl. Phys.  B 790 (2008) 432,
  arXiv:0704.1038 [hep-th].}
  
\lref\indexfh{  
  F.A. Dolan and H. Osborn,
  {\it Applications of the Superconformal Index for Protected Operators and
  $q$-Hypergeometric Identities to $\N=1$ Dual Theories},
  arXiv:0801.4947 [hep-th].}

\lref\wyb{B.G. Wybourne, N. Flocke, J. Karwowski,
{\it Characters of Two-row Representations of the Symmetric Group},
Int. J. Quant. Chem. 62:3  (1997) 261.}

\lref\dobsez{V.K. Dobrev and E. Sezgin, {\it Spectrum and Character Formulae
of $SO(3,2)$ Unitary Representations}, Lecture Notes in Physics, Vol. 379,
eds. J.D Hennig, W. L\"ucke and J. Tolar, Springer-Verlag, Berlin, 1990.}

\lref\mep{F.A. Dolan, {\it Character Formulae and Partition Functions in
Higher Dimensional Conformal Field Theory},
 J. Math. Phys. 47 (2006) 062303, hep-th/0508031.}

\lref\bofengt{B. Feng, A. Hanany, Y.H. He,
 {\it Counting Gauge Invariants: The Plethystic Program}, hep-th/0701063.}

\lref\bofeng{S. Benvenuti, B. Feng, A. Hanany, Y.H. He,
 {\it Counting $BPS$ Operators in Gauge Theories: Quivers, Syzygies and
 Plethystics}, hep-th/0608050.}
 
 \lref\poly{
A.M. Polyakov,
 {\it Gauge Fields and Space-Time},
 Int. J. Mod. Phys.  A 17S1 (2002) 119,
 hep-th/0110196.}

\lref\mald{
J.~Kinney, J.M.~Maldacena, S.~Minwalla and S.~Raju,
{\it An Index for $4$ Dimensional Super Conformal Theories},
hep-th/0510251.}

\lref\char{M. Bianchi, F.A. Dolan, P.J. Heslop and H. Osborn,
{\it $\N=4$ Superconformal Characters and Partition Functions},
Nucl. Phys. B 767 (2007) 163, hep-th/0609179.}

\lref\ska{ B.-S. Skagerstam, {\it On the Large $N_c$ Limit of the $SU(N_c)$
Colour Quark-Gluon Partition Function}, Z. Phys. C 24 (1984) 97.}

\lref\sund{B. Sundborg,
 {\it The Hagedorn Transition, Deconfinement and $N = 4$ SYM Theory},
 Nucl.  Phys.  B 573 (2000) 349, hep-th/9908001.}

\lref\forc{D. Forcella, {\it BPS Partition Functions for Quiver Gauge Theories:
Counting Fermionic Operators}, arXiv:0705.2989 [hep-th].}

\lref\luc{J. Lucietti and M. Rangamani,
 {\it Asymptotic Counting of BPS Operators in Superconformal Field Theories},
  arXiv:0802.3015 [hep-th].}

\lref\minw{J. Bhattacharya and S. Minwalla,
 {\it Superconformal Indices for ${\cal N}=6$ Chern Simons Theories},
  arXiv:0806.3251 [hep-th].}
  
\lref\warner{M. Gunaydin and N.P. Warner, {\it Unitary Supermultiplets
of $Osp(8|4,\Bbb{R})$ and the Spectrum of the $S^7$ Compactification
of $11$-Dimensional Supergravity}, Nucl. Phys. B 272 (1986) 99.}

\lref\unitm{S.~Minwalla,
  {\it Restrictions Imposed by Superconformal Invariance on Quantum Field
  Theories},
  Adv.\ Theor.\ Math.\ Phys.\  2 (1998) 781,
 hep-th/9712074.}
  
\lref\lamb{J. Bagger and N. Lambert,
  {\it Comments On Multiple M$2$-branes},
  JHEP 0802 (2008) 105,
  arXiv:0712.3738 [hep-th].}
  
 \lref\abjm{O. Aharony, O. Bergman, D.L. Jafferis and J. Maldacena,
 {\it $\N=6$ Superconformal Chern-Simons-Matter Theories, M$2$-Branes and Their
  Gravity Duals},
  arXiv:0806.1218 [hep-th].}

\lref\yak{Y. Nakayama,
  {\it Index for Non-relativistic Superconformal Field Theories},
  arXiv: 0807.3344 [hep-th].}
  
\lref\bak{J.A. Minahan and K. Zarembo,
  {\it The Bethe Ansatz for Superconformal Chern-Simons},
  JHEP 0809 (2008) 040,
  arXiv:0806.3951 [hep-th]\semi
  D. Bak and S.J. Rey,
 {\it Integrable Spin Chain in Superconformal Chern-Simons Theory},
  arXiv:0807.2063 [hep-th]\semi
  N.~Gromov and P.~Vieira,
 {\it The All Loop AdS$_4$/CFT$_3$ Bethe Ansatz},
  arXiv:0807.0777 [hep-th]\semi
  G. Grignani, T. Harmark, M. Orselli and G. W. Semenoff,
 {\it Finite Size Giant Magnons in the String Dual of $\N=6$ Superconformal
  Chern-Simons Theory},
  arXiv:0807.0205 [hep-th]\semi
  D. Gaiotto, S. Giombi and X. Yin,
  {\it Spin Chains in $\N=6$ Superconformal Chern-Simons-Matter Theory},
  arXiv:0806.4589 [hep-th]\semi
  C.~Krishnan,
  {\it AdS$_4$/CFT$_3$ at One Loop},
  JHEP 0809 (2008) 092,
  arXiv:0807.4561 [hep-th]\semi 
  T. McLoughlin, R. Roiban and A.A. Tseytlin,
 {\it Quantum Spinning Strings in AdS$_4 \times$CP$^3$: Testing the Bethe Ansatz
 Proposal},
  arXiv:0809.4038 [hep-th].}
  
\lref\hana{A. Hanany, N. Mekareeya and A. Zaffaroni,
  {\it Partition Functions for Membrane Theories},
  JHEP 0809 (2008) 090,
  arXiv:0806.4212 [hep-th].}

\lref\NO{M. Nirschl and H. Osborn, {\it Superconformal Ward Identities and
their Solution}, Nucl. Phys. B711 (2005) 409, hep-th/0407060\semi
F.A. Dolan, M. Nirschl and H. Osborn, {\it Conjectures for Large $N$ 
$\N=4$ Superconformal  Chiral Four Point Functions}.
Nucl. Phys. B749 (2006) 109, hep-th/0601148.
}

\lref\benna{M. Benna, I. Klebanov, T. Klose and M. Smedback,
  {\it Superconformal Chern-Simons Theories and AdS$_4$/CFT$_3$ Correspondence},
  JHEP 0809 (2008) 072
  arXiv:0806.1519 [hep-th]\semi
  O. Aharony, O. Bergman and D.L. Jafferis,
 {\it Fractional M$2$-Branes},
  arXiv:0807.4924 [hep-th]\semi
  A. Giveon and D. Kutasov,
  {\it Seiberg Duality in Chern-Simons Theory},
  arXiv: 0808.0360 [hep-th]\semi
  V. Niarchos,
 {\it Seiberg Duality in Chern-Simons Theories with Fundamental and Adjoint
  Matter},
  JHEP 0811 (2008) 001, arXiv:0808.2771 [hep-th].}

\lref\flato{M. Flato and C. Fronsdal, {\it One Massless Particle Equals Two Dirac Singletons},
Lett. Math. Phys. 2 (1978) 421.}

\lref\fadhoy{F.A. Dolan and H. Osborn,
  {\it On Short and Semi-short Representations for Four Dimensional Superconformal
  Symmetry},
  Annals Phys.  {307} (2003) 41,
  hep-th/0209056.}

\lref\hama{G.~Grignani, T.~Harmark and M.~Orselli,
  {\it The $SU(2)\times  SU(2)$ Sector in the String Dual of $\N=6$ Superconformal
  Chern-Simons Theory},
  arXiv:0806.4959 [hep-th].}
  
\lref\minwero{
  J. Bhattacharya, S. Bhattacharyya, S. Minwalla and S. Raju,
 {\it Indices for Superconformal Field Theories in $3,5$ and $6$ Dimensions},
  JHEP 0802 (2008) 064,
  arXiv:0801.1435 [hep-th].}

\lref\pope{B.E.W. Nilsson and C.N. Pope, 
{\it Hopf Fibration of Eleven-Dimensional Supergravity}, 
Class. Quantum Grav. 1(1984) 499-515.}

\lref\sokatchevr{
S. Ferrara and E. Sokatchev,
  {\it Representations of Superconformal Algebras in the AdS$_{(7/4)}$/CFT$_{(6/3)}$
  Correspondence},
  J. Math. Phys.  42 (2001) 3015,
  hep-th/0010117.}

\lref\guny{
M. G\"unaydin, B. Nilsson, G. Sierra and P. Townsend, 
{\it Singletons and Superstrings}, Phys. Lett. B176 (1986) 45.}

\lref\mald{
J.~Kinney, J.M.~Maldacena, S.~Minwalla and S.~Raju,
{\it An Index for $4$ Dimensional Superconformal Theories},
hep-th/0510251.}

{\nopagenumbers
\rightline{ITFA-08-46}
\rightline{arXiv:0811.2740 [hep-th]}
\vskip 1.5truecm
\centerline{\bigbf On Superconformal Characters and Partition Functions in Three Dimensions }
\vskip  6pt
\vskip 2.0 true cm
\centerline {F.A. Dolan}
\vskip 12pt
\centerline {Institute for Theoretical Physics, University of Amsterdam,}
\centerline {Valckenierstraat 65, 1018 XE Amsterdam, The Netherlands}
\vskip 1.5 true cm
{\eightpoint
\parindent 1.5cm{
{\narrower\smallskip\parindent 0pt

Possible short and semi-short positive energy, unitary representations
of the $Osp(2N|4)$ superconformal group in three dimensions are discussed.
Corresponding character formulae are obtained, consistent
with character formulae for the $SO(3,2)$ conformal group, 
revealing long multiplet decomposition
at unitarity bounds in a simple way.   Limits, corresponding to reduction
to various $Osp(2N|4)$ subalgebras, are taken in the characters
that isolate contributions from fewer states, at a given unitarity
threshold, leading to considerably simpler formulae.  Via these
limits, applied to partition functions, 
closed formulae for the generating functions for numbers of 
BPS operators in the free field limit of
superconformal $U(n)\times U(n)$ $\N=6$ Chern Simons theory 
and its supergravity dual are obtained in the large $n$ limit.  
Partial counting of semi-short
operators is performed and consistency
between operator counting for the free field and 
supergravity limits with long multiplet
decomposition rules is explicitly demonstrated.  Partition functions
counting certain protected scalar primary semi-short
operators, and their superconformal descendants, are proposed and 
computed for large $n$.  Certain  chiral ring partition functions
are discussed from a combinatorial perspective.

Keywords:
Superconformal Characters, Gauge Invariant Operator Counting,
Superconformal Chern Simons Theory
\narrower}}
\vfill
\line{\hskip0.2cm E-mail:
{{\tt
F.A.H.Dolan@uva.nl}}\hfill}
}

\eject}

\pageno=1

\newsec{Introduction}

With the resurgence of interest recently in three dimensional superconformal field
theory, due largely to the discovery by Bagger and Lambert \lamb\ of a new superconformal $\N=8$
Chern Simons theory and, more recently, by Aharony {\it et al} \abjm\
 of a superconformal $\N=6$ Chern-Simons theory, with $U(n)\times U(n)$ gauge symmetry,
much attention has been devoted to uncovering dualities \refs{\benna,\abjm}, investigating integrability
\refs{\bak, \hama}
and spectra {\refs{\minw,\hana,\yak}} {\it etc.}  One issue that has hitherto perhaps
not been explored in very much
detail is the representation theory, underlying these theories, for positive
energy, unitary representations of $Osp(\N|4)$.  
This paper is an attempt to address in some detail this issue
and focuses on the case of even $\N$.{\foot{Note that there is some overlap 
in the discussion here with some related issues explored in \minwero\
particularly with regard to long multiplet decomposition formulae.
Odd $\N$ would require some modification of the analysis here
particularly with regard to possible short/semi-short multiplets.  While
straightforward, an extension to odd $\N$ is avoided here to ensure
notation is not overly cumbersome.}}

For $\N=4$ superconformal symmetry in four dimensions
such a 
detailed study of the representation theory \fadhoy\ 
proved fruitful in many
ways, {\it e.g.} investigating the operator product expansion, 
in terms of conformal partial wave expansions of four point functions, see \NO,
for example, 
in partially determining  the operator spectrum of
$\N=4$ super Yang Mills.
In this context, superconformal characters provide for an alternative
and arguably more straightforward and powerful way of organising the
sometimes detailed multiplet structure that arises.
They easily lead to the rules for long multiplet decomposition
into short/semi-short multiplets at unitarity bounds, formulae for decompositions
in terms of subalgebra representations
and indicate how partition functions may be computed for decoupled
sectors.  

The outline of this paper is as follows.  In section 2 the superconformal algebra for
$Osp(2N|4)$, {\it i.e.} for $\N=2N$,
 is given in detail.  Mainly for the purposes of discussing shortening conditions, 
constructing Verma module characters and finding
subalgebras, it proves easiest to use the orthonormal
basis for the $SO(2N)$ $R$-symmetry algebra.  Section 3 gives a brief account of
the possible shortening conditions for semi-short
representations, denoted by $(N,A,n)$, $n=1,\dots, N{-1}$, and $(N,A,\pm)$,
and short or BPS representations, denoted by $(N,B,n)$ and $(N,B,\pm)$,
as well as for the conserved current multiplet, denoted $(N,{\rm cons.})$.  
Here, highest weight states in the
$(N,A,n)$, respectively $(N,A,\pm)$, $(N,B,n)$, $(N,B,\pm)$, 
representations are annihilated by $n/4N$,
respectively $1/4$, $n/2N$, $1/2$ of the supercharges.  (Hence,
the $(N,B,\pm)$ representations are referred to here as half BPS.
For $N=4$, it is well known that there are two types of half BPS representation, 
see \guny, and \sokatchevr\ for a discussion.)  

In section 4 the characters for positive energy,
unitary representations of the $Osp(2N|4)$ superconformal
group are obtained through use of Verma
module characters and the Weyl symmetriser for the maximal compact subgroup,
$U(1)\otimes SU(2)\otimes SO(2N)$. BPS and conserved conserved current multiplet 
characters are decomposed in
terms of  $SO(3,2)\otimes SO(2N)$ characters.  
The decomposition rules for long multiplets at the
unitarity bounds are found.  In particular, they imply that in any $\N=2N$ superconformal
field theory in three dimensions,  operators in all
$(N,B,n)$, $n>1$, as well as in $(N,B,\pm)$ and certain $(N,B,1)$,
BPS multiplets 
must have protected conformal dimensions.

In section 5, various limits are taken in the
characters that lead to non-vanishing formulae only for 
subsets of the characters corresponding to different short/semi-short multiplets.  
These are shown, from appendix A, to match with various
$Osp(2N|4)$ subalgebra characters and, in terms of partition
functions, isolate different sectors of operators.{\foot{I thank Troels Harmark for 
pointing out to me that different sectors, along with the $SU(2)\times SU(2)$ one, 
were also briefly considered in \hama, for $\N=6$ 
superconformal Chern Simons theory.}}
For the subset being $\{(N,B,m),\,\, m\geq n$,
$(N,B,+), (N,B,-)\}$ the subalgebra is $U(1)\otimes SO(2N{-2m})$
while for the subsets $\{(N,B,\pm)\}$ there are two $U(1)$ subalgebras.
Similarly, for the subset being $\{(N,A,m),\,\, m\geq n$,
$(N,A,+)$, $(N,A,-)$, $(N,B,l)$, $1\leq l\leq N$,  $(N,B,+)$, $(N,B,-)$, $(N,{\rm cons.})\}$
where
$(N,B,l)$, $l<m$,
 multiplets have particular $SO(2N)$ $R$-symmetry eigenvalues, the subalgebra is 
$U(1)\otimes Osp(2N{-2m}|2)$.  Finally, for the subsets 
$\{(N,A,\pm)$, $(N,B,l)$, $1\leq l\leq N$,  $(N,B,+)$,
$(N,B,-)$, $(N,{\rm cons.})\}$,
where
$(N,B,l)$, $l<m$, $(N,B,\mp)$
 multiplets have particular $SO(2N)$ $R$-symmetry eigenvalues,
  there are
two $U(1)\otimes SU(1,1)$ subalgebras.
The limits are shown to be consistent with
long multiplet decomposition rules and also with the index of \minw.

In section 6, the general free field multi-particle partition function
for $\N=6$ superconformal Chern Simons theory is 
written in detail in terms of superconformal characters and, 
for large $n$, computed using  
symmetric polynomial techniques.  Also, the corresponding partition function
is given in detail for the supergravity limit. 

In section 7, operator counting
for short and certain semishort multiplets is investigated using the limits taken in
section 5, and other symmetric polynomial techniques,
described in appendix B, and shown to be consistent with
expectations from long multiplet decomposition rules.

In the conclusion, partition functions counting certain 
protected semi-short operators are discussed. Some related comments
are made about $\N=4 $ super Yang Mills, explained in more detail
in appendix C, where a certain class of chiral ring partition functions
are computed using the Polya enumeration theorem.

\newsec{The Superconformal Algebra in Three Dimensions}

In $d$ dimensions, the standard non-zero commutators of the conformal group
$SO(d,2)$ are
given by, for $\eta_{ab}={\rm diag}.(-1,1,\dots,1)$, $a,b=0,1,\dots, d-1$,
\eqn\confalg{\eqalign{
[M_{ab},P_c]&{}=i(\eta_{ac}P_b-\eta_{b c}P_a)\, ,\qquad 
[M_{ab},K_c]=i(\eta_{ac}K_b-\eta_{b c}K_a)\, , \cr
[M_{ab},M_{cd}]&{}=
i(\eta_{a c}\,M_{bd}-\eta_{bc}\,M_{ad}-\eta_{ad}\,M_{bc}+\eta_{bd}\,M_{ac})\, ,
\cr
[D,P_a]&{}=P_a\, ,\quad [D,K_a]=-K_a\, ,\quad [K_a,P_b]=-2iM_{ab}+2 \eta_{a b}D\, ,}
}
where the generators of translations are $P_a$, those of special conformal transformations 
are $K_a$, those of $SO(d-1,1)$ are $M_{ab}=-M_{ba}$
while that of scale transformations is $D$.  

For later application it is convenient to write the 
$Sp(4,\Bbb{R})/\Bbb{Z}_2\simeq SO(3,2)$ algebra for
three dimensions in the spinor basis so that for,{\foot{The conventions used here
for the gamma matrices are that $\gamma^0=1$,
$\gamma^1=\si_1$, $\gamma^2=\si_3$, in terms of Pauli matrices $\si_1,\si_3$,
 so that $\ga^a$ are real, symmetric
matrices.  We take $({\bar \ga}^a)^{\al \be}=\vep^{\al\ga}\, 
(\ga^a)_{\ga\de}\,\vep^{\de\be}$ 
where $\vep^{\al\be}=-\vep^{\be\al}$, $\vep^{12}=1$,
$\vep_{\al\be}\vep^{\ga\de}=-\de_{\al}{}^\ga\,\de_{\be}{}^\de+\de_{\al}{}^\de\,
\de_{\be}{}^\ga$.  
We thus have $\ga^a{\bar \ga}^b+\ga^b{\bar \ga^a}=\eta^{a b}1$,
$\bga^a{\ga}^b+\bga^b{\ga^a}=\eta^{a b}1$, and the completeness relation
$(\ga^{a})_{\al \be}(\bga_a)^{\ga\de}=
\de_{\al}{}^\ga\,\de_{\be}{}^\de+\de_{\al}{}^\de\,\de_{\be}{}^\ga$.}}
\eqn\spinorgen{
P_{\al\be}=(\ga^a)_{\al\be}\,P_a\, ,\quad K^{\al\be}=({\bar \ga}^a)^{\al \be}\,K_a\, ,
\quad M_\al{}^{\be}={\ts{i\over 2}}(\ga^a \, {\bar \ga}^b)_\al{}^\be\, M_{a b}\, ,
}
the algebra \confalg\ becomes,
\eqn\confpart{\eqalign{
[M_\al{}^{\be},P_{\ga\de}]&{}=
 \de_{\ga}{}^{\be} \, P_{\al \de} 
+  \de_{\de}{}^{\be} \, P_{\al \ga}- \de_\al{}^{\be} \, P_{\ga\de} \, ,\cr 
[M_\al{}^{\be},K^{\ga\de}]&{}=
 - \de_{\al}{}^{\ga} \, K^{\be \de} 
- \de_{\al}{}^{\de} \, K^{\be \ga} + \de_\al{}^{\be} \, K^{\ga\de}\, ,\cr
[M_{\al}{}^{\be},M_{\ga}{}^{\de}]{}&=-\de_{\al}{}^{\de}\, M_{\ga}{}^{\be}
+ \de_{\ga}{}^{\be}\, M_{\al}{}^{\de}\, ,\quad 
[D,P_{\al\be}]=P_{\al\be}\, ,\quad 
[D,K^{\al\be}]=- K^{\al\be}\, ,\cr
[K^{\al \be},P_{\ga \de}]&{}=4 \, \de_{(\ga}{}^{(\al}\, M_{\de)}{}^{\be)}+4  \,
\de_{(\ga}{}^{\al}\, \de_{\de)}{}^{\be}\, D\, .}
}

For supercharges and their superconformal extensions the non-zero
(anti-)commutators are given by,
\eqn\form{\eqalign{
\{Q_{r\al},Q_{s\be}\}&{}=2\de_{rs}\, P_{\al\be}\, ,
\qquad \qquad\qquad\qquad\,
\{S_{r}{}^{\al},S_{s}{}^{\be}\}=2\de_{rs}\,K^{\al\be}\, ,\cr
[K^{\al \be},Q_{r\ga}]&{}=i (\de_{\ga}{}^{\al}\,S_{r}{}^{\be}
+ \de_{\ga}{}^{\be}\,S_{r}{}^{\al})\, ,
\qquad \quad\,\,
[P_{\al\be},S_{r}{}^{\ga}]=-i(\de_\al{}^{\ga}\,Q_{r\be}
+\de_{\be}{}^{\ga}\,Q_{r\al})\, ,\cr
[M_\al{}^{\be},Q_{r\ga}]&{}= \de_\ga{}^{\be}\,Q_{r\al}
-\half \de_\al{}^\be\, Q_{r\ga}\, ,
\qquad \,\,\,\,\,\,\,
[M_{\al}{}^{\be},S_{r}{}^{\ga}]=- \de_\al{}^\ga\, S_{r}{}^{\be}
+\half \de_{\al}{}^{\be} S_{r}{}^{\ga}\, ,\cr
[D,Q_{r\al}]&{}=\half Q_{r\al}\, , 
\qquad \qquad\qquad\qquad\qquad\,\,\,\,
[D,S_r{}^{\al}]=-\half S_r{}^{\al}\, ,\cr
[R_{rs},Q_{t\al}]{}&=i (\de_{rt}\,Q_{s\al}-\de_{s t}\, Q_{r\al})\, ,
\qquad\qquad 
[R_{rs},S_{t}{}^{\al}]=i (\de_{rt}\,S_{s}{}^{\al}-\de_{s t}\, S_{r}{}^{\al})\, ,}
 }
along with
\eqn\CT{
\{Q_{r\al},S_s{}^{\be}\}=2 i (M_\al{}^{\be}\de_{rs}
-i \de_\al{}^\be\, R_{r s}+\de_\al{}^\be \, \de_{rs}\, D)
\, ,
}
where the $SO(n)$ $R$-symmetry generators 
$R_{rs}=-R_{rs}=R_{rs}{}^{\dagger}$ satisfy
\eqn\rsym{
[R_{rs},R_{tu}]=i(\de_{rt}\,R_{su}-\de_{st}\,R_{ru}-\de_{ru}\,R_{st}+\de_{su}\,R_{rt})\, .
}

In order to discuss highest weight states and shortening conditions
\rsym\ is now rewritten in terms of generators in the orthonormal basis of $SO(2N)$,
which has rank $N$.
In this basis, the Cartan subalgebra $H_n$,
and raising/lowering operators $E^{+\pm}_{mn}$/$E^{-\pm}_{mn}$, $m< n$,  
are given by, for $m,n=1,\dots,N$,{\foot{$E^{+\pm}_{mn}$/$E^{-\pm}_{mn}$
correspond to the
positive/negative roots $\e_m\pm\e_n$/$-\e_m\pm\e_n$, $m<n$,
where $\e_n,\e_m$ are usual $\Bbb{R}^N$ orthonormal vectors.
A linearly independent basis for raising/lowering operators is
 $\{E^{+-}_{n\,n+1}, E^{++}_{N-1\,N}\}$/$\{E^{-+}_{n\,n+1}, E^{--}_{N-1\,N}\}$, 
 $n=1,\dots,N{-1}$, corresponding to the positive/negative simple roots.}}
\eqn\crl{\eqalign{
H_n&{}=R_{2n-1\,2n}\, , \quad E^{+\pm}_{mn}=R_{2m-1\,2n-1}+i\,R_{2m\,2n-1}
\pm i\, R_{2m-1\,2n}\mp R_{2m\,2n}\, ,\cr
\qquad E^{-\pm}_{mn}&{}=E^{+\mp}_{mn}{}^\dagger\, ,\qquad 
E^{-\pm}_{mn}=-E^{\pm -}_{nm}\, ,\qquad E^{+\pm}_{mn}=-E^{\pm+}_{nm}\, ,}
}
so that the non-zero commutators from \rsym\ are, for $l=1,\dots, N$,
\eqn\orthoalg{\eqalign{
[H_l,E^{+\pm}_{{mn}}]&{}=(\de_{lm}\pm\de_{ln})
E^{+\pm}_{mn}\, ,\qquad [H_l,E^{-\pm}_{{mn}}]=(-\de_{lm}\pm\de_{ln})
E^{-\pm}_{mn}\, ,\cr
[E^{+\pm}_{mn},E^{-\mp}_{mn}]&{}=4(H_m\pm H_n)\, ,\cr
[E^{+\pm}_{lm},E^{-\pm}_{ln}]&{}=2i\,
E^{\pm\pm}_{mn}\, ,
\qquad 
[E^{+\pm}_{lm},E^{-\mp}_{ln}]=2i\,
E^{\pm\mp}_{mn}\, ,
}
}
where $m\neq n$ in the last line.

 In this basis a convenient choice for the supercharges and their superconformal
 extensions is given by
 \eqn\supercharges{\eqalign{
 \Q_{n\al}&{}={1\over \sqrt{2}}\Big(Q_{2n-1\,\al}+i\,Q_{2n\,\al}\Big)\, ,\qquad
 \bQ_{n\al}={1\over \sqrt{2}}\Big(Q_{2n{-1}\,\al}-i\,Q_{2n\,\al}\Big)
\, ,\cr 
 \S_{n}{}^{\al}&{}={1\over \sqrt{2}}\Big(S_{2n-1}{}^{\al}+i\,S_{2n}{}^{\al}\Big)\, ,
 \qquad\,\,\,\,
 \bS_{n}{}^{\al}={1\over \sqrt{2}}
 \Big(S_{2n-1}{}^{\al}-i\,S_{2n}{}^{\al}\Big)
\, .}
 }
Trivially, from \form,
 \eqn\formp{\eqalign{
[K^{\al \be},\Q_{n\ga}]&{}=i (\de_{\ga}{}^{\al}\,\S_{n}{}^{\be}
+\de_{\ga}{}^{\be}\,\S_{n}{}^{\al})\, ,
\qquad \quad\,\,
[P_{\al\be},\S_{n}{}^{\ga}]=-i(\de_\al{}^{\ga}\,\Q_{n\be}+
\de_{\be}{}^{\ga}\,\Q_{n\al})\, ,\cr
[M_\al{}^{\be},\Q_{n\ga}]&{}=\de_\ga{}^{\be}\,\Q_{n\al}
-\half \de_\al{}^\be\, \Q_{n\ga}\, ,
\qquad \quad
[M_{\al}{}^{\be},\S_{n}{}^{\ga}]=- \de_\al{}^\ga\, \S_{n}{}^{\be}
+\half \de_{\al}{}^{\be} \S_{n}{}^{\ga}\, ,\cr
[D,\Q_{n\al}]&{}=\half \Q_{n\al}\, , 
\qquad \qquad\qquad\qquad\qquad\,\,\,\,
[D,\S_n{}^{\al}]=-\half \S_n{}^{\al}\, ,}
 }
along with identical equations for $\Q_{n\al},\S_{n}{}^{\al}\to\bQ_{n\al},\bS_{n}{}^{\al}$.
The non-zero anti-commutators among the supercharges and
their superconformal extensions are,
\eqn\supernew{\eqalign{
\{\Q_{m\al},\bQ_{n\be}\}&{}=2\de_{mn}P_{\al\be}\, ,\qquad
\{\S_{m}{}^{\al},\bS_{n}{}^{\be}\}=2\de_{mn}K^{\al\be}\, ,\cr
\{\Q_{m\al},\bS_{m}{}^{\be}\}&{}=2 i\big(M_\al{}^\be
+\de_\al{}^\be D-\de_\al{}^\be H_m\big)\, ,\cr
\{\bQ_{m\al},\S_m{}^{\be}\}&{}=2i\big(M_\al{}^\be
+\de_\al{}^\be D+\de_\al{}^\be H_m\big)\, ,\cr
\{\Q_{m\al},\S_n{}^{\be}\}&{}=\de_{\al}{}^\be E^{++}_{mn}\, ,\qquad\,
\{\bQ_{m\al},\bS_n{}^{\be}\}=\de_{\al}{}^\be E^{--}_{mn}\, ,
\cr
\{\Q_{m\al},\bS_n{}^{\be}\}&{}=\de_{\al}{}^\be E^{+-}_{mn}\, ,\qquad\,
\{\bQ_{m\al},\S_n{}^{\be}\}=\de_{\al}{}^\be E^{-+}_{mn}\, ,}
}
where $m\neq n$ in the last two lines.
Finally, under the action of the generators \crl\ the non-zero commutators are 
\eqn\chargesrsym{\eqalign{
[H_m,\Q_{n\al}]&{}=\de_{mn}\,\Q_{n\al}\, ,
\qquad \qquad\qquad\qquad\quad
[H_m,\S_{n}{}^{\al}]=\de_{mn}\,\S_{n}{}^{\al}\, ,\cr
[H_m,\bQ_{n\al}]&{}=-\de_{mn}\,\bQ_{n\al}\, ,
\qquad \qquad\qquad\qquad\,\,\,
[H_m,\bS_{n}{}^{\al}]=-\de_{mn}\,\bS_{n}{}^{\al}\, ,\cr
[E^{-+}_{lm},\Q_{n\al}]&{}=-[E^{+-}_{ml},\Q_{n\al}]=2i\,\de_{ln}\Q_{m\al}\, ,
\quad
[E^{-+}_{lm},\S_{n}{}^{\al}]=-[E^{+-}_{ml},\S_{n}{}^{\al}]=2i\,\de_{ln}\S_{m}{}^{\al}\, ,\cr
[E^{--}_{lm},\Q_{n\al}]&{}=2i(\de_{ln}\bQ_{m\al}-\de_{mn}\bQ_{l\al})\, ,
\qquad\quad
[E^{--}_{lm},\S_{n}{}^{\al}]=2i(\de_{ln}\bS_{m}{}^{\al}-\de_{mn}\bS_{l}{}^{\al})\, ,\cr
[E^{+-}_{lm},\bQ_{n\al}]&{}=-[E^{-+}_{ml},\bQ_{n\al}]=2i\,\de_{ln}\bQ_{m\al}\, ,
\quad
[E^{+-}_{lm},\bS_{n}{}^{\al}]=-[E^{-+}_{ml},\bS_{n}{}^{\al}]=
2i\,\de_{ln}\bS_{m}{}^{\al}\, ,\cr
[E^{++}_{lm},\bQ_{n\al}]&{}=2i(\de_{ln}\Q_{m\al}-\de_{mn}\Q_{l\al})\, ,
\qquad\quad
[E^{++}_{lm},\bS_{n}{}^{\al}]=2i(\de_{ln}\S_{m}{}^{\al}-\de_{mn}\S_{l}{}^{\al})\, .}
}

Suppressing spinor indices, the action of the linearly independent
set of lowering operators on $\Q_n$ may be expressed  by the following diagram.
\eqn\harry{
\Q_{1}\fo{E^{-+}_{12}}\Q_{2}\fo{E^{-+}_{23}}\Q_3\longrightarrow\cdots
\longrightarrow\!\!\!\!\!\!\!\!\matrix{
&\!\!\!\!\!\! \!\!\!\!\!\!\!\!\!  
\Q_N  \!\!\!\!\!\!\!\!\!\!\!\!\!\!\! &       \cr
\quad{}^{E^{-+}_{N{-1}\,N}}\!\!\! \nearrow   & &  \searrow\!\!\!{}^{E^{--}_{N{-1}\,N}}\cr
\!\!\!\! \!\!\!\!\!\! \!\!\Q_{N{-1}}  \!\!\!\!\!\!\!\!\!\!\!\!\!&  &\!\!\!\! \bQ_{N{-1}} \!\!\!\!\!\!\!\!\!\!\!\!\cr
\quad {}_{E^{--}_{N{-1}\,N}}\!\!\!\!\searrow   &  &
\nearrow\!\!\!{}_{E^{-+}_{N{-1}\,N}}\cr
&\!\!\!\!\!\! \!\!\!\!\!\!\!\!\!  
\bQ_{N}  \!\!\!\!\!\!\!\!\!\!\!\! \!\!\!&     \cr
}\longrightarrow
\cdots\longrightarrow\bQ_3\fo{E^{-+}_{23}}\bQ_{2}\fo{E^{-+}_{12}}\bQ_{1}
 }
Of course, an identical diagram applies to $\S_n, \,\bS_n$. (This diagram is helpful later for discussing the shortening conditions and the
computation of superconformal characters.)

The spin generators in terms of $SU(2)$ generators may be expressed as
follows,
\eqn\spingen{
[M_{\al}{}^{\be}]=\pmatrix{J_3 & J_+\cr J_- & -J_3}\, ,\qquad 
[J_+,J_-]=2 J_3\, ,
\qquad [J_3,J_{\pm}]=\pm J_\pm\, ,
}
where $(\Q_{n1},\Q_{n2})$, $(\bQ_{n1},\bQ_{n2})$, $(\S_n{}^2,-\S_n{}^1)$, 
$(\bS_n{}^2,-\bS_n{}^1)$ transform as usual spin $\half$ doublets, each with $J_3$
eigenvalues $(\half,-\half)$.

\newsec{Shortening Conditions for Unitary Multiplets}

For physical applications, we require states with positive real conformal dimensions, 
non-negative half integer spin eigenvalues and in finite dimensional
irreducible representations of the $R$-symmetry group $SO(2N)$.
Hence it is sufficient
to consider superconformal representations defined by highest weight states
$|\De,j,\rr\rangle^{\rm h.w.}$ with,
\eqn\eigenvalues{\eqalign{
&(K^{\al\be},\,\,\S_n{}^{\al},
\,\,\bS_n{}^{\al},\,\,J_+,\,\,E^{+\pm}_{mn})|\De,j,\rr\rangle^{\rm h.w.}=0\, ,
\qquad 1\leq m<n\leq N\, ,\cr 
&(D,\,J_3,\,H_m)|\De,j,\rr\rangle^{\rm h.w.}=(\Delta,\,j,\,r_m)
|\Delta,j,\rr\rangle^{\rm h.w.}\, ,\cr
&\rr=(r_1,\dots,r_N)\,,}
}
where $\De$ is the conformal dimension, $j\in \half \Bbb{N}$ is the spin, and
with $SO(2N)$ Dynkin labels expressed in terms of
$r_j\in \half \Bbb{Z}$ required to satisfy,
\eqn\rsymeigen{
[r_1-r_2,\dots,r_{N{-2}}-r_{N{-1}},r_{N{-1}}+r_N,r_{N{-1}}-r_N]\in \Bbb{N}^N\, ,
}
so that, in particular, $r_1\geq r_2\geq\dots\geq r_{N-1}\geq |r_N|$.

For a representation space with basis $V_{(\De;j;\rr)}$, the states are given by
\eqn\restrictedverma{
V_{(\De;j;\rr)}=\left\{\prod_{\al,\be,\ga,\de=1,2\atop \ga\leq \de,\, 1\leq n,m,r,s\leq N} 
(\Q_{n\al})^{\kappa_{n\al}} (\bQ_{m\be})^{{\bar \kappa}_{m\be}}
(P_{\ga\de})^{k_{\ga\de}}(J_-)^K(E^{-\pm}_{rs})^{K_{rs}}|\De,j,\rr\rangle^{\rm h.w.}
\right\}\, ,
}
for 
$\kappa_{n\al},{\bar \kappa}_{m\be}\in\{0, 1\}$ and $k_{\ga\de},K,K_{rs}\in\Bbb{N}$.

Unitarity requires that \unitm,{\foot{In the basis \confpart, \form, \CT, \rsym, all the generators of $Osp(2N|4)$
are hermitian, apart from $D$ and $M_\al{}^{\be}$ which are anti-hermitian.
In order to impose the physically necessary unitarity conditions
arising from a scalar product defined by the two point function
for the conformal fields, it is sufficient to perform a similarity   transformation, 
see \fadhoy\ for example, where, in particular, $D$ and $M_\al{}^{\be}$
become hermitian so that $M_\al{}^{\be}$ generates $SO(3)$, rather than $SO(2,1)$,
and $D$ has real eigenvalues which are required to be positive except for the
trivial representation.
Of course, such a similarity transformation does not affect the shortening conditions derived here.}}
\eqn\unitarity{
\De\geq \cases{r_1+j+1\, , & for $j>0$ \cr r_1 \, , & for $j=0$}\,.
}

The superconformal multiplets may be truncated by various shortening
conditions which are considered now. 
For the BPS shortening condition,
\eqn\shortcond{
\Q_{n\al}|\De,j,\rr\rangle^{\rm h.w.}=0\, , \qquad \al=1,2\, ,
} 
then this leads to the following equations, using  \supernew,
\eqn\shortc{
(E^{+\pm}_{nm},\,\,J_{\pm},\,\,D\pm J_3-H_n)|\De,j,\rr\rangle^{\rm h.w.}=0\, ,
\qquad 1\leq m\leq N\, ,\qquad m\neq n\, , 
}
so that, using \orthoalg, \spingen\ and \eigenvalues,
\eqn\solshort{
\De=r_n\, ,\qquad r_1=r_2=\dots=r_n\, ,\qquad j=0\, .
}
Clearly, \shortcond\ with \shortc\ implies that $\Q_{m\al}|\De,j,\rr\rangle^{\rm h.w.}=0$,
$m<n$.

Imposing the constraint, 
\eqn\shortcondo{
\bQ_{N\al}|\De,0,\rr\rangle^{\rm h.w.}=0\, , \qquad \al=1,2\, ,
} 
leads to, using \supernew, 
\eqn\shortd{
(E^{\pm-}_{mN},\,\,J_{\pm},\,\,D\pm J_3+H_N)|\De,j,\rr\rangle^{\rm h.w.}=0\, ,
\qquad 1\leq m < N\, ,
}
so that, using \orthoalg, \spingen\ and \eigenvalues,
\eqn\solshorte{
\De=-r_N\, ,\qquad r_1=r_2=\dots=-r_N\, , \qquad  j=0\, .
} 

We may also consider the semi-short multiplet condition,
\eqn\semishortcond{
\Big(\Q_{n2}-{1\over 2j} \Q_{n1}J_-\Big)|\De,j,\rr\rangle^{\rm h.w.}=0\, ,
} 
whereby applying $\bS_l{}^2$ and using \supernew\
we obtain,
\eqn\semishortc{
(E^{+\pm}_{nm},\,\,D-J_3-H_n-{\ts{1\over 2j}}\,J_+J_- )|\De,j,\rr\rangle^{\rm h.w.}=0\, ,
\qquad 1\leq m\leq N\, ,\qquad m\neq n\,,
}
so that, using \orthoalg, \spingen\ and \eigenvalues,
\eqn\solsemi{
\De=r_n+j+1\, ,\qquad r_1=r_2=\dots=r_n\, .
}

We may also consider,
\eqn\semishortcondo{
\Big(\bQ_{N2}-{1\over 2j} \bQ_{N1}J_-\Big)|\De,j,\rr\rangle^{\rm h.w.}=0\, ,
} 
whereby applying $\S_l{}^2$ and using \supernew\
then
\eqn\semishortco{
(E^{\pm-}_{mN},\,\,D-J_3+H_N-{\ts{1\over 2j}}J_+J_- )|\De,j,\rr\rangle^{\rm h.w.}=0\, ,
\qquad 1\leq m < N\, ,
}
so that, using \orthoalg, \spingen\ and \eigenvalues,
\eqn\solsemio{
\De=-r_N+j+1\, ,\qquad r_1=r_2=\dots=-r_N\,.
}

Imposing both,
\eqn\semishortconde{
\Big(\Q_{N2}-{1\over 2j} \Q_{N1}J_-\Big)|\De,j,\rr\rangle^{\rm h.w.}=0\, ,\qquad
\Big(\bQ_{N2}-{1\over 2j} \bQ_{N1}J_-\Big)|\De,j,\rr\rangle^{\rm h.w.}=0\, ,
} 
leads to a conservation condition on the highest weight state, using \supernew,
\eqn\conservationcond{
\Big((2j-1)\big(2j\,P_{22}-2\,P_{12}J_-\big)+P_{11}J_-{}^2\Big)
|\De,j,\rr\rangle^{\rm h.w.}=0\, ,
}
and also to, using \solsemi, for $n=N$, and \solsemio,
\eqn\conserv{
\De=j+1\, , \qquad \rr=0\, .
}

It is easy to see that imposing  BPS or semi-shortening conditions,
other than the above,
using other $\bQ_{n\al}$, $n\neq N$, leads to violations of \unitarity,
implying that corresponding multiplets are non-unitary.  These are
not considered here.

For the truncated supermultiplets $\M$, the
Verma modules 
 $V_{(\De;j;\rr)}\to \V^{\M}_{(\De;j;\rr)}$ are generated by a subset of the
 generators
in \restrictedverma\ so that it is sufficient to set some
  $\kappa_{n\al},{\bar \kappa}_{m\be}, k_{\ga\de}$  
to zero.

Using the information encapsulated in \harry, 
then the condition \shortcond, for BPS multiplets, entails
omitting $\Q_{j\al},\,j=1,\dots, n$ from \restrictedverma, so that $\kappa_{j\al}=0$. 
Similarly, for \shortcondo, then $\kappa_{j\al},\bar \kappa_{N\al}=0,\,j=1,\dots, N{-1}$, while  
for the semi-shortening conditions, \semishortcond, $\kappa_{j2}=0,\, j=1,\dots,n$, and
\semishortcondo, $\kappa_{j2},\bar \kappa_{N2}=0,\,j=1,\dots,N{-1}$.
Corresponding to \semishortconde\ with \conservationcond\ then
$\kappa_{j2},\bar \kappa_{N2}=0,\,j=1,\dots,N$, $k_{22}=0$ for the multiplet
of conserved currents.
This information along with notation is summarised in the table below.

\medskip
\vbox{
\hskip0cm   Table 1
\nobreak

\hskip0cm
\vbox{\tabskip=0pt \offinterlineskip
\hrule
\halign{&\vrule# &\strut \ \hfil#\  \cr
height2pt&\omit&&\omit&&\omit&&\omit&& \omit &\cr
&\   Type  \ \hfil   && \  $ \De $ \ \hfil && \ $ \rr $  \  \hfil && \ Omitted \ \hfil  
&& \ Denoted \ \hfil &\cr
height2pt&\omit&&\omit&&\omit&&\omit&& \omit &\cr
\noalign{\hrule}
height2pt&\omit&&\omit&&\omit&&\omit&& \omit &\cr
& \ Long  \ \hfil &&  $\geq r_1+j+1$  \hfil     &&    $r_1\geq\dots\geq |r_N|$      \hfil            
&&  None \hfil  &&  $(N,A,0)$    \hfil    &\cr
height3pt&\omit&&\omit&&\omit&&\omit&& \omit &\cr
& \ Semi-Short  \ \hfil &&  $r_1+j+1$  \hfil     &&    $r_1=\dots =r_n$    \hfil 
&&  $\{\Q_{i2}\}_{i=1}^n$ \hfil   &&  $(N,A,n)$    \hfil    &\cr
&   &&   &&  &&   &&  for $n<N$    \hfil    &\cr
height3pt&\omit&&\omit&&\omit&&\omit&& \omit &\cr
& \ Semi-Short  \ \hfil &&  $r_1+j+1$  \hfil     &&    $r_1=\dots =r_N$     \hfil 
&&  $\{\Q_{i2}\}_{i=1}^N$\hfil   &&  $(N,A,+)$    \hfil    &\cr
height3pt&\omit&&\omit&&\omit&&\omit&& \omit &\cr
& \ Semi-Short  \ \hfil &&  $r_1+j+1$  \hfil     &&    $r_1=\dots =-r_N$     \hfil 
&&  $\{\Q_{i2},\bQ_{N2}\}_{i=1}^{N-1}$ \hfil   &&  $(N,A,-)$    \hfil    &\cr
height3pt&\omit&&\omit&&\omit&&\omit&& \omit &\cr
& \ BPS  \ \hfil &&  $r_1$  \hfil     &&    $r_1=\dots =r_n$     \hfil 
&&  $\{\Q_{i\al}\}_{i=1}^n$ \hfil   &&  $(N,B,n)$    \hfil    &\cr
&   &&   &&  &&   &&  for $n<N$    \hfil    &\cr
height3pt&\omit&&\omit&&\omit&&\omit&& \omit &\cr
& \ $\half$ BPS  \ \hfil &&  $r_1$  \hfil     &&    $r_1=\dots =r_N$     \hfil 
&&  $\{\Q_{i\al}\}_{i=1}^N$\hfil   &&  $(N,B,+)$    \hfil    &\cr
height3pt&\omit&&\omit&&\omit&&\omit&& \omit &\cr
& \ $\half$ BPS  \ \hfil &&  $r_1$  \hfil     &&    $r_1=\dots =-r_N$     \hfil 
&&  $\{\Q_{i\al},\bQ_{N\al}\}_{i=1}^{N-1}$ \hfil   &&  $(N,B,-)$    \hfil    &\cr
height2pt&\omit&&\omit&&\omit&&\omit&& \omit &\cr
& \ cons. current  \ \hfil &&  $j+1$  \hfil     &&    $r_i=0$    \hfil 
&&  $\{\Q_{i2},\bQ_{N2},P_{22}\}_{i=1}^{N}$ \hfil   &&  $(N,{\rm cons.})$    \hfil    &\cr
height2pt&\omit&&\omit&&\omit&&\omit&& \omit &\cr
}
\hrule}
}

\newsec{Superconformal Characters for $SO(2N)$ $R$-symmetry}

A procedure for computing conformal characters for higher than two
dimensions and $\N=4$ superconformal characters for
four dimensions has been explained in detail elsewhere \refs{\mep, \char}.  
The procedure is also closely related to that in \fadhoy\ for constructing
 supermultiplets by
employing the Racah-Speiser algorithm. We proceed
by analogy with \refs{\mep, \char}.

Introducing variables $s,\,x,\,\y=(y_1,\dots,y_N)$, we may write the
character corresponding to the restricted Verma module, \restrictedverma\ for
  $V_{(\De;j;\rr)}\to \V^{\M}_{(\De;j;\rr)}$, as a formal trace, 
\eqn\vermachar{\eqalign{
C^{\M}_{(\De;j,\rr)}(s,x,\y)={}&{\widetilde \tr}_{\V^{\M}_{(\De;j;\rr)}}
\big(s^{2{D}}x^{2J_3} y_1{}^{H_1}\cdots y_N{}^{H_N}\big)\cr
={}&
s^{2\De}\,C_{2j}(x)\,C^{(N)}_{\rr}(\y)\cr
&\times 
\sum_{k_{\ga\de}\in \Bbb{N}} (s^2 x^2)^{k_{11}}s^{2 k_{12}}
(s^2 x^{-2})^{k_{22}}\cr
&\times\!\!\!\!\!\!\!\!\!\!
\sum_{\kappa_{n\al},{\bar \kappa}_{m\be}\in \{0,1\}}
\!\!\!\!\!\!\!\!\!\!
(s \,y_n \,x)^{\kappa_{n1}}
(s \,y_n \,x^{-1})^{\kappa_{n2}}(s\, y_m{}^{-1} x)^{{\bar \kappa}_{m1}}
(s\, y_m{}^{-1} x^{-1})^{{\bar \kappa}_{m2}}\, ,}
}
where the sum over $k_{\ga\de}$ gives the contributions of $P_{\ga\de}$ 
and that over $\kappa_{n\al}, \bar \kappa_{m\be}$ gives those of
$\Q_{n\al}, \bQ_{m}^{\be}$, and where,
\eqn\charsv{\eqalign{
C_j(x)&{}={x^{j+1}\over x-x^{-1}}\, , \cr 
C^{(N)}_\rr(\y)&{}={\ts\prod_{j=1}^N}y_j{}^{r_j+j-1}/\De(\y+\y^{-1})\, ,\qquad 
\De(\y)=\prod_{1\leq i<j\leq N}(y_i-y_j)\, , }
}
are the Verma module characters for $SU(2)$ and $SO(2N)$, giving contributions from the
$J_-$, $E^{-\pm}_{rs}$ generators, and the highest weight state, in \restrictedverma.

Once the correct generators are omitted from \restrictedverma,
so that various $\kappa_{n\al},{\bar \kappa}_{m\be},k_{\ga\de} $ are zero in \vermachar, the prescription 
for finding the characters of corresponding unitary
irreducible representations 
$\R^{\M}_{(\De;j;\rr)}$ is simply given by,{\foot{As discussed in  \refs{\mep, \char},
the action of the Weyl symmetriser on the Verma module character 
corresponds to quotienting out null states in the Verma module,
\restrictedverma\ for $V_{(\De;j;\rr)}\to \V^{\M}_{(\De;j;\rr)}$,  to obtain the irreducible module,
\restrictedverma\ for $V_{(\De;j;\rr)}\to \R^{\M}_{(\De;j;\rr)}$.}}
\eqn\irrepchar{
\chi^{\M}_{(\De;j;\rr)}(s,x,\y)={\tr}_{\R^{\M}_{(\De;j;\rr)}}
\big(s^{2D} \, x^{2 J_3} \, y_1{}^{H_1}\cdots y_N{}^{H_N}\big)
={\frak W}^{(N)}\,\, C^{\M}_{(\De;j;\rr)}(s,x,\y)\, ,
}
where $\frak W^{(N)}=\frak W^{\S_2} \frak W^{\S_N\ltimes (\S_2)^{N-1}}$
is the Weyl symmetriser for the maximal compact subgroup
of the superconformal group, $U(1)\times SU(2)\times SO(2N)$.  Here $\S_2$ and 
$\S_N\ltimes(\S_2)^{N-1}$
are the Weyl symmetry groups for $SU(2)$ and $SO(2N)$ and,
for some functions
$f(x)$, $f(\y)=f(y_1,\dots, y_N)$, the action of the relevant Weyl symmetrisers is given by,
\eqn\actionweyl{\eqalign{
\frak W^{\S_2}\, f(x)&{}=f(x)+f(x^{-1})\, ,\cr
\frak W^{\S_N\ltimes (\S_2)^{N-1}}\, \, f(\y)&{}=
{\ts \sum_{\vep_1,\dots,\vep_N=\pm 1\atop \prod_i\vep_i=1}}
{\ts \sum_{\si\in \S_N}}f(y_{\si(1)}{}^{\vep_1},\dots,y_{\si(N)}{}^{\vep_N})\, .}
}
It is important to realise that
the resulting characters may be expanded in terms of
$SU(2)\times SO(2N)$ characters using,
\eqn\chars{\eqalign{
\chi_j(x)&{}=\frak W^{\S_2} C_j(x)={x^{j+1}-x^{-j-1}\over x-x^{-1}}\,,\cr
\chi^{(N)}_\rr(\y)&{}=\frak W^{\S_N\ltimes (\S_2)^{N-1}}\,\, C^{(N)}_\rr(\y)\cr
&{}=
\big({\det}\big [y_i{}^{r_j+N-j}+y_i{}^{-r_j-N+j}\big ]
+ {\det}\big [y_i{}^{r_j+N-j}-y_i{}^{-r_j-N+j}\big ] \big)/ 2\, \De(\y+\y^{-1})\, ,}
}
the usual Weyl character formulae for $SU(2)$ and $SO(2N)$ finite dimensional,
irreducible representations.

Defining,
\eqn\defPQ{\eqalign{
P(s,x)&{}={1\over (1-s^2)(1-s^2 x^2)(1-s^2 x^{-2})}\, , \cr 
\Q_n(\y,x)&{}=\prod_{j=n+1}^N(1+y_j x)\, ,\qquad 
\bQ_n(\y,x)=\prod_{j=1}^n (1+y_j{}^{-1} x)\ ,}
}
then this prescription leads to the following character formulae for
 the unitary irreducible representations, using \vermachar\ with \irrepchar\
 and with the notation of Table 1,
\eqn\longsemi{\eqalign{
&\chi^{(N,i,n)}_{(\De;j;r_1,\dots,r_1,r_{n+1},\dots,r_N)}(s,x,\y)=\frak W^{(N)}\, 
C^{(N,i,n)}_{(\De;j;r_1,\dots,r_1,r_{n+1},\dots,r_N)}(s,x,\y)\cr
&=s^{2\De}P(s,x)\,{\frak W}^{(N)}\Big(C_{2j}(x)C^{(N)}_\rr(\y)\,
\R^{(N,i,n)}(s,x,\y)\,{\ts \prod_{\vep=\pm 1}}\bQ_N(s^{-1}\y,x^\vep) \Big)
\, ,\quad n<N\, ,\cr
&\chi^{(N,i,\pm)}_{(\De;j;r,\dots,r,\pm r)}(s,x,\y)=\frak W^{(N)}\, 
C^{(N,i,\pm)}_{(r+j+1;j;r,\dots,r,\pm r)}(s,x,\y)\cr
&=s^{2\De}P(s,x)\,{\frak W}^{(N)}\Big(C_{2j}(x)C^{(N)}_\rr(\y)\,
\R^{(N,i,\pm)}(s,x,\y)\,{\ts \prod_{\vep=\pm 1 }}\bQ_{N-1}(s^{-1}\y,x^{\vep})\Big)\, ,\cr}
}
where,
\eqn\longsemid{\eqalign{
\R^{(N,i,n)}(s,x,\y)&{}=\cases{\Q_0(s\y,x)\Q_n(s \y,x^{-1}) & for $i=A$,\cr
{\ts \prod_{\vep=\pm 1}}\Q_n(s\y,x^{\vep}) & for $i=B$,}\cr
\R^{(N,i,\pm)}(s,x,\y)&{}=\cases{\Q_0(s\y,x)(1+s y_N{}^{-1} x)(1+s y_N{}^{\mp 1} x^{-1})
& for $i=A$,\cr
{\ts \prod_{\vep=\pm 1}}(1+s y_N{}^{\mp 1} x^{\vep})& for $i=B$,}}
}
and appropriate $\De,j$ are as given in Table 1. (The conserved current multiplet
is discussed separately below.)

Using invariance of
$\prod_{\vep= \pm 1} \Q_0(s\y,x^\vep)\bQ_N(s^{-1}\y,x^\vep)$ under $\frak W^{(N)}$
then the long multiplet character is given by,
\eqn\long{\eqalign{
\chi^{(N,{\rm long})}_{(\De;j;\rr)}(s,x,\y)&{}=
\chi^{(N,A,0)}_{(\De;j;\rr)}(s,x,\y)\cr
&{}=s^{2\De}P(s,x)\chi_{2j}(x)\chi^{(N)}_\rr(\y)
{\ts \prod_{\vep=\pm 1}}
\Q_0(s \y,x^\vep)\bQ_N(s^{-1}\y,x^\vep)\, .}
}

This may be expanded in terms of $SU(2)\times SO(2N)$ characters using the
identities, 
\eqn\Pexpansion{
P(s,x)={1\over 1-s^4}\sum_{n=0}^\infty s^{2n}\,\chi_{2n}(x)\, ,
}
along with, for later use,
\eqn\Qexp{\eqalign{
&{\ts \prod_{j=1}^N}(1+t \,y_j)(1+t \,y_i{}^{-1})\cr
&={\ts \sum_{n=0}^{N-1}}
\big(t^n+t^{2N-n}\big)\chi^{(N)}_{(1^n,0^{N-n})}(\y)+t^N \,\chi^{(N)}_{(1^N)}(\y)
+t^N \,\chi^{(N)}_{(1^{N{-1}},-1)}(\y)\, .}
}

\noindent
{\bf Simplification of BPS Multiplet Characters} 

Half BPS characters may be simplified
by first writing, easily obtained from \longsemi,
\eqn\halfbpso{\eqalign{
&{\chi}^{(N,B,\pm)}_{(r;0;r,\dots,r,\pm r)}(s,x,\y)\cr
&=s^{2r}P(s,x)
\sum_{a_1,\dots,a_N=0}^2s^{a_1+\dots +a_N}
\chi_{j_{a_1}}(x)\cdots \chi_{j_{a_N}}(x)
\chi^{(N)}_{(r-a_1,r-a_2,\dots,\pm r\mp a_N)}(\y)\, ,}
}
where, $j_a=\half(1-(-1)^a)$ so that
\eqn\defja{
j_0=j_2=0\, , \qquad j_1=1\, . 
}
Some further manipulation shows that \halfbpso\ may be simlified further to,{\foot{
Another consistency check is for $N=4$,
or $SO(8)$ $R$-symmetry, whereby this formula leads directly to, 
$$
\eqalign{
&{\chi}^{(4,B,-)}_{(r;0;r,r,r,-r)}(s,x,\y)=s^{2r}P(s,x)\Big(\chi^{(4)}_{(r,r,r,-r)}(\y)+
s \chi_1(x)\chi^{(4)}_{(r,r,r,1{-r})}(\y)
+s^2\chi^{(4)}_{(r,r,r,2{-r})}(\y)\cr
&+s^2\chi_2(x)\chi^{(4)}_{(r,r,r{-1},1{-r})}(\y)
+s^3\chi_1(x)\chi^{(4)}_{(r,r,r{-1},2{-r})}(\y)+s^3\chi_3(x)\chi^{(4)}_{(r,r{-1},r{-1},1{-r})}(\y)\cr
&+s^4\chi^{(4)}_{(r,r,r{-2},2{-r})}(\y)
+s^4\chi_2(x)\chi^{(4)}_{(r,r{-1},r{-1},2{-r})}(\y)
+s^4\chi_4(x)\chi^{(4)}_{(r{-1},r{-1},r{-1},1{-r})}(\y)\cr
&+s^5\chi_1(x)\chi^{(4)}_{(r,r{-1},r{-2},2{-r})}(\y)
+s^5\chi_3(x)\chi^{(4)}_{(r{-1},r{-1},r{-1},2{-r})}(\y)
+s^6\chi^{(4)}_{(r,r{-2},r{-2},2{-r})}(\y)\cr
&+s^6\chi_2(x)\chi^{(4)}_{(r{-1},r{-1},r{-2},2{-r})}(\y)
+s^7\chi_1(x)\chi^{(4)}_{(r{-1},r{-2},r{-2},2{-r})}(\y) +s^8\chi^{(4)}_{(r{-2},r{-2},r{-2},2{-r})}(\y)\Big)
}
$$
which corresponds exactly to the graviton spectrum derived in \warner.
}}   
\eqn\halfbps{
\eqalign{
&{\chi}^{(N,B,\pm)}_{(r;0;r,\dots,r,\pm r)}(s,x,\y)\cr
&=s^{2r}P(s,x)
\sum_{0\leq a_1\leq \dots\leq  a_N\leq 2}s^{a_1+\dots +a_N}
\chi_{j_{a_1,\dots, a_N}}(x)\chi^{(N)}_{(r-a_1,r-a_2,\dots,\pm r\mp a_N)}(\y)\, ,}
}
where
\eqn\defjabc{
j_{a_1,\dots,a_N}=\half (N-(-1)^{a_1}-\dots-(-1)^{a_N})\, .
}

Further simplifications to the half BPS character \halfbps\ occur for $r<2$.  
For $r=1$ then,
\eqn\halfbpscons{\eqalign{
&{\chi}^{(N,B,\pm)}_{(1;0;1,\dots,{1},\pm {1})}(s,x,\y)\cr
&=\A_{(1,0)}(s,x)\,\,\chi^{(N)}_{(1^{N-1},\pm 1)}(\y)+\A_{({3\over 2},{1\over 2})}(s,x)\,\,
\chi^{(N)}_{(1^{N-1},0)}(\y)+\A_{(2,0)}(s,x)\,\,\chi^{(N)}_{(1^{N-1},\mp 1)}(\y)\cr
&\,\,\,\,\,+{\ts \sum_{n=0}^{N-2}}\D_{{1\over 2}(N-n)}(s,x)\,\,
\chi^{(N)}_{(1^n,0^{N-n})}(\y)\, ,}
}
where,
\eqn\longconstd{
\A_{(\De,j)}=s^{2\De}\chi_{2j}(x)P(s,x)\, ,\qquad
\D_{j}=s^{2j+2}\Big(\chi_{2j}(x)-s^2\chi_{2j-2}(x)\Big)P(s,x)\, ,
}
are the characters for unitary irreducible representations of the conformal group in three
dimensions, $SO(3,2)$, \dobsez\ - see also \mep\ - whereby $\A_{(\De,j)}$ corresponds to an 
unconstrained
spin $j$ field, conformal dimension $\De\geq j=1$, while $\D_j$ corresponds to
a conserved current with spin $j$, conformal dimension $j+1$, including
all their conformal descendants (or derivatives acting on fields).

Similarly, for $r=\half $ then,
\eqn\halfbpshalf{
{\chi}^{(N,B,\pm)}_{({1\over 2};0;{1\over 2},\dots,{1\over 2},\pm {1\over 2})}(s,x,\y)
=\D_{\rm Rac}(s,x)\chi^{(N)}_{({1\over 2},\dots,{1\over 2},\pm {1\over 2})}(\y)+
\D_{\rm Di}(s,x)\chi^{(N)}_{({1\over 2},\dots,{1\over 2},\mp {1\over 2})}(\y)
}
where,
\eqn\dirac{
\D_{\rm Rac}(s,x)={s+s^3\over (1-s^2 x^2)(1-s^2 x^{-2})}\, ,
\qquad \D_{\rm Di}(s,x)={s^2(x+x^{-1})\over (1-s^2 x^2)(1-s^2 x^{-2})} \, ,
}
are characters for the free field representations of $SO(3,2)$,
the so called `Di', respectively, `Rac', singleton 
representations \flato, corresponding to a free spin $\half $, respectively, scalar, 
field with conformal
dimension 1, respectively, $\half$, and all its descendants.

Finally, from \halfbps\ it may be shown that,
\eqn\halfpbsid{
{\chi}^{(N,B,\pm)}_{(0;0;0,\dots,0)}(s,x,\y)=1\, ,
}
the character for the identity representation.

For other BPS characters, from \longsemi, 
we may write, similarly to \halfbpso\ with \defja,
\eqn\otherbpso{\eqalign{
&{\chi}^{(N,B,n)}_{(r_1;0;r_1,r_1,\dots,r_1,r_{n+1},\dots,r_N)}(s,x,\y)\cr
&
=s^{2r_1}P(s,x)\!\!\!\!\!\!
\sum_{{a_1,\dots,a_N\atop {\bar a}_{n{+1}},\dots,{\bar a}_N}=0}^2
s^{a_1+\dots+a_N+\bar a_{n+1}+\dots+\bar a_N}
{\ts\prod_{i=1}^N}\chi_{j_{a_i}}(x) {\ts \prod_{i=n+1}^N}\chi_{j_{{\bar a}_{i}}}(x)\cr
&\qquad\qquad\qquad\qquad\qquad 
\times \chi^{(N)}_{(r_1-a_1,\dots,r_1-a_n,
r_{n{+1}}+{\bar a}_{n{+1}}-a_{n{+1}},\dots,r_N+{\bar a}_N-a_N)}(\y)\, ,}
}
and, similarly to \halfbps\ with \defjabc, using that $r_1=r_2=\dots=r_n$,
\eqn\otherbps{\eqalign{
&{\chi}^{(N,B,n)}_{(r_1;0;r_1,r_1,\dots,r_1,r_{n+1},\dots,r_N)}(s,x,\y)\cr
&=s^{2r_1}P(s,x)\cr
&\quad\!\!\!\! \times \sum_{0\leq a_1\leq\dots \leq a_{n}\leq 2}
\sum_{{a_{n{+1}},\dots,a_N\atop {\bar a}_{n{+1}},\dots,{\bar a}_N}=0}^2
s^{a_1+\dots+a_N+\bar a_{n+1}+\dots+\bar a_N}
\chi_{j_{a_1,a_2,\dots,a_n}}(x){\ts \prod_{i={n+1}}^N}\chi_{j_{a_{i}}}(x)
\chi_{j_{{\bar a}_{i}}}(x)\cr
&\qquad\qquad\qquad\qquad\qquad\qquad
\times\chi^{(N)}_{(r_1-a_1,\dots,r_1-a_n,
r_{n{+1}}+{\bar a}_{n{+1}}-a_{n{+1}},\dots,r_N+{\bar a}_N-a_N)}(\y)\, ,}
}
where,
\eqn\defjn{
j_{a_1,\dots,a_n}={\ts{1\over 2}}(n-(-1)^{a_1}-\dots-(-1)^{a_n})\, .
}
Without further restrictions on $r_i$, or $s,x,\y$,  there appear to be
no further simplifications to \otherbps. 

\noindent
{\bf The Conserved Current Multiplet Character}

The conserved current multiplet
(which has been excluded so far from the above discussion)
corresponding to the semi-shortening condition
\semishortconde, has character,
\eqn\conservchar{\eqalign{
&\!\!\!\chi^{(N,{\rm cons.})}_{(j+1;j;0,\dots, 0)}(s,x,\y)\cr
&\!\!\!=s^{2j+2}P(s,x){\frak W}^{(N)}\Big(C_{2j}(x)C^{(N)}_{0}(\y)(1-s^2 x^{-2})
\Q_0(s \y,x)\bQ_N(s^{-1}\y,x)\bQ_{N-1}(s^{-1}\y,x^{-1})\Big)\, ,}
}
which ensures that the contributions from the 
supercharges $\Q_{n2}, \,n=1,\dots,N$, $\bQ_{N2}$
along with $P_{22}$ are omitted from the Verma module character \vermachar.

To simplify, it is useful to observe that
$\Q_0(s \y,x)\bQ_N(s^{-1}\y,x)$ is invariant under action by 
${\frak W}^{\S_N\ltimes (\S_2)^{N-1}}$ and, further, that
${\frak W}^{\S_N\ltimes (\S_2)^{N-1}} C_0(\y)\bQ_{N-1}(\y,t)=1$
so that
\eqn\simplifdeto{\eqalign{
\chi^{(N,{\rm cons.})}_{(j+1;j;0,\dots, 0)}(s,x,\y)
={}&s^{2j+2}P(s,x){\frak W}^{\S_2}\Big(C_{2j}(x)(1-s^2 x^{-2})
\Q_0(s \y,x)\bQ_N(s^{-1}\y,x)\Big)\cr
={}&\sum_{n=0}^{N-1}\Big(\D_{j+{1\over 2} n}(s,x)
+\D_{j+N-{1\over 2}n}(s,x)\Big)\chi^{(N)}_{(1^n,0^{N-n})}(\y)\cr
&\quad\quad +
\D_{j+{1\over 2}N}(s,x)\Big(\chi^{(N)}_{(1^N)}(\y)+\chi^{(N)}_{(1^{N-1},-1)}(\y)\Big)\, ,}
}
using \Qexp, to expand in $x$, and subsequently
the expression for the conserved current character in \longconstd.

\noindent
{\bf Long Multiplet  Decompositions}

Using linearity of $\frak W^{(N)}$ we may easily obtain, 
for $j\geq{1\over 2}$, $r_1>r_2$,
\eqn\decomp{\eqalign{
\chi^{(N,{\rm long})}_{(r_1+j+1;j;r_1,r_2,\dots,r_N)}&(s,x,\y)\cr
&{}=\frak W^{(N)} (1+s\, y_1\,x^{-1})C^{(N,A,1)}_{(r_1+j+1;j;r_1,r_2,\dots,r_N)}(s,x,\y)\cr
&{}=\chi^{(N,A,1)}_{(r_1+j+1;j;r_1,r_2,\dots,r_N)}(s,x,\y)
+\chi^{(N,A,1)}_{(r_1+j+{3\over 2};j-{1\over 2};r_1+1,r_2,\dots,r_N)}(s,x,\y)\, ,}
}
which expresses the reducibility of a long multiplet with $\De=r_1{+j}{+1}$ into
a sum of semi-short multiplets.

{}For semi-short multiplet characters it may be shown that, for 
$r_{1}=r_2=\dots =r_n>r_{n+1}$, $r>0$,{\foot{This employs,
$$\eqalign{
&\!\!\!\!\!\!\!\!\!\!\!\!\!\!\!\!\!\!\!\!\!\!\!\!\!\!\!\!\!\!\!\!\!\!\!\!\!\!\!\!\!\!
\chi^{(N,A,1)}_{(r_1+j+1;j;r_1,\dots,r_1,r_{n+1},\dots,r_N)}(s,x,\y)\cr
&=\frak W^{(N)} {\ts{\prod_{i=2}^n}}(1+s\, y_i\,x^{-1})
C^{(N,A,n)}_{(r_1+j+1;j;r_1,\dots,r_1,r_{n+1},\dots,r_N)}(s,x,\y)\, ,}$$
along with the identity
$\chi^{(N)}_{\rr^w}(\y)=(-1)^{|w|}\chi^{(N)}_\rr(\y)$, 
$w\in \S_N\ltimes (\S_2)^{N-1}$,
where $\rr^w=w(\rr+\rho)-\rho$, with
$w\in \S_N\ltimes (\S_2)^{N-1}$, the Weyl group, and $\rho=(N{-1},N{-2},\dots, 0)$,
being the Weyl vector.  In particular, 
$\chi^{(N)}_{(r_1,\dots,r_1,r_1+1,\dots,r_N)}(\y)=0$ which implies 
$\chi^{(N,A,n)}_{(\De';j';r_1,\dots,r_1,r_1+1,\dots,r_N)}(s,x,\y)=0$.
The $(N,A,\pm)$ cases may be found similarly.}}
\eqn\strider{\eqalign{
\chi^{(N,A,1)}_{(r_1+j+1;j;r_1,\dots,r_1,r_{n+1},\dots,r_N)}(s,x,\y)
&{}=\chi^{(N,A,n)}_{(r_1+j+1;j;r_1,\dots,r_1,r_{n+1},\dots,r_N)}(s,x,\y)\, ,\cr
\chi^{(N,A,1)}_{(r+j+1,j,r,\dots, r, \pm r)}(s,x,\y)
&{}=\chi^{(N,A,\pm)}_{(r+j+1,j,r,\dots, r, \pm r)}(s,x,\y)\, .}
}
Similarly, for the conserved current multiplet character,
\eqn\stridero{
\chi^{(N,A,1)}_{(j+1;j;0,\dots,0)}(s,x,\y)
=\chi^{(N,{\rm cons.})}_{(j+1;j;0,\dots,0)}(s,x,\y)\, .
}
In order to avoid ambiguity, it is assumed in what follows that $r_n>r_{n+1}$,
for $(N,A,n)$
semishort multiplet characters, unless otherwise stated.

{}From \decomp\ with \strider\ and \stridero,  
then we have, for $j\geq{1\over 2}$, $r_1>r_{n+1}$, $n<N$,
\eqn\decompratchet{\eqalign{
&\chi^{(N,{\rm long})}_{(r_1+j+1;j;r_1,r_1,\dots,r_1,r_{n+1},\dots,r_N)}(s,x,\y)\cr
&{}=\chi^{(N,A,n)}_{(r_1+j+1;j;r_1,\dots,r_1,r_{n+1},\dots, r_N)}(s,x,\y)
+\chi^{(N,A,1)}_{(r_1+j+{3\over 2};j-{1\over 2};r_1+1,r_1,\dots,r_{1},r_{n+1},\dots,r_N)}(s,x,\y)\, ,}
}
along with, for $r>0$,
\eqn\decompress{\eqalign{
&\chi^{(N,{\rm long})}_{(r+j+1;j;r,\dots,r,\pm r)}(s,x,\y)\cr
&=\chi^{(N,A,\pm)}_{(r+j+1;j;r,\dots,r,\pm r)}(s,x,\y)
+\chi^{(N,A,1)}_{(r_1+j+{3\over 2};j-{1\over 2};r+1,r,\dots,r,\pm r)}(s,x,\y)\, ,}
}
and, 
\eqn\decompressop{
\chi^{(N,{\rm long})}_{(j+1;j;0,\dots,0)}(s,x,\y)
=\chi^{(N,{\rm cons.})}_{(j+1;j;0,\dots,0)}(s,x,\y)
+\chi^{(N,A,1)}_{(j+{3\over 2};j-{1\over 2};1,0,\dots,0)}(s,x,\y)\, .
}

It is also easily seen that,
\eqn\decompo{\eqalign{
\chi^{(N,A,1)}_{(r_1+{1\over 2};-{1\over 2};r_1,r_2,\dots,r_N)}(s,x,\y)&{}=\frak W^{(N)}
(1+s\, y_1\, x)C^{(N,B,1)}_{(r_1+{1\over 2};-{1\over 2};r_1,r_2,\dots,r_N)}(s,x,\y)\cr
&{}={\chi}^{(N,B,1)}_{(r_1+1;0;r_1+1,r_2,\dots,r_N)}(s,x,\y)\, ,}
}
using $\frak W^{\S_2}C_{-1}(x)=0$, $\frak W^{\S_2} x C_{-1}(x)
=\frak W^{\S_2}C_0(x)=1$,
so that, for $r_1>r_2$, 
\eqn\decompimp{\eqalign{
\chi^{(N,{\rm long})}_{(r_1+1;0;r_1,r_2,\dots,r_N)}(s,x,\y)
&=\frak W^{(N)} (1+s\, y_1\,x^{-1})C^{(N,A,1)}_{(r_1+1;0;r_1,r_2,\dots,r_N)}(s,x,\y)\cr
&=\chi^{(N,A,1)}_{(r_1+1;0;r_1,r_2,\dots,r_N)}(s,x,\y)
+\chi^{(N,B,1)}_{(r_1+2;0;r_1+2,r_2,\dots,r_N)}(s,x,\y)
\, ,}
}
which expresses the reducibility of a long multiplet with $\De=r_1{+1}$ into
a sum of a $(N,B,1)$ BPS and a semi-short multiplet.

Thus, from \decompimp\ with \strider\ and \stridero, we have, for $r_1=r_n>r_{n+1}$,
\eqn\decompump{\eqalign{
&\chi^{(N,{\rm long})}_{(r_1+1;0;r_1,\dots,r_1,r_{n+1},\dots, r_N)}(s,x,\y)\cr
&=\chi^{(N,A,n)}_{(r_1+1;0;r_1,\dots,r_1,r_{n+1},\dots,r_N)}(s,x,\y)
+\chi^{(N,B,1)}_{(r_1+2;0;r_1+2,r_1,\dots,r_1,r_{n+1},\dots, r_N)}(s,x,\y)
\, ,}
}
along with, for $r>0$,
\eqn\decomplump{\eqalign{
&\chi^{(N,{\rm long})}_{(r+1;0;r,\dots,r,\pm r)}(s,x,\y)\cr
&=\chi^{(N,A,\pm)}_{(r+1;0;r,\dots,r,\pm r)}(s,x,\y)
+\chi^{(N,B,1)}_{(r+2;0;r+2,r,\dots,r,\pm r)}(s,x,\y)
\, ,}
}
and,
\eqn\decomplumpop{
\chi^{(N,{\rm long})}_{(1;0;0,\dots,0)}(s,x,\y)
=\chi^{(N,{\rm cons.})}_{(1;0;0,\dots,0)}(s,x,\y)
+\chi^{(N,B,1)}_{(2;0;2,0,\dots,0)}(s,x,\y)
\, .}

\decomp, \decompratchet, \decompress, \decompressop,
\decompimp,
\decompump, \decomplump, \decomplumpop\  appear to exhaust all 
possibilities for long  multiplet decompositions
(and are also consistent with limits, as discussed in section 5)
and have important consequences for any superconformal field theory
in three dimensions.  In particular, they imply that 
all $(N,B,\pm)$ and $(N,B,n)$, $N>n>1$, short multiplet operators as well as certain 
$(N,B,1)$ short multiplet operators in 
 $\R_{(r_1+1,r_2,\dots,r_N)}$, $r_1\geq r_2$, $SO(2N)$ $R$-symmetry
representations must remain protected against
gaining anomalous dimensions.   The decomposition formulae \decomp, \decompimp\ have 
essentially appeared in \minwero\ where also 
comments about the protectedness
of certain operators in three dimensional superconformal field theories were made.
(Note that all the decomposition formulae here are consequences of the basic formula
\decomp.  This is not surprising as a similar thing happens for decomposition formulae of long multiplets for $\N=4$ superconformal
symmetry in four dimensions \mep.)

\newsec{Reduction to Subalgebra Characters}

Here are described certain limits that can be taken in the previous BPS and semi-short
multiplet characters that isolate contributions from fewer states in each multiplet
and hence lead to significant  simplifications.  
These limits are equivalent to reductions of the characters to those
for various subgroups of the superconformal group,
as explained in more detail in appendix A.

\noindent
{\bf BPS Limits}

\noindent
{\bf The $U(1)\otimes SO(2N{-2m})$ Sector}

By considering, 
\eqn\limitBPS{\eqalign{
&\!\!\!\!\chi^{\M}_{(\De;j; \rr)}(\de u^{1\over 2},
x, (\de^{-2}u)^{1\over m}\hat \y,\bar \y)\cr
&{}=\tr_{\R^{\M}_{(\De;j;\rr)}}(\de^{2\H_m}\,{u}^{\I_m} \,x^{2J_3} \,{\hat y}_1{}^{H_1}
\cdots {\hat y}_m{}^{H_m}\,{\bar y}_1{}^{H_{m+1}}\cdots{\bar y}_{N-m}{}^{H_N})\,,\cr
\hat\y&{}=({\de^{-2}u})^{-{1\over m}}(y_1,\dots,y_m)\, ,\quad {\ts \prod_{\ha =1}^m\hat y_\ha=1}\, ,
\qquad \bar \y=(y_{m+1},\dots,y_N)\, ,\cr
{\H_m}&{}=D-{\ts{1\over m}\sum_{\hat m=1}^m H_m}\, ,\qquad 
\I_m=D+{\ts{1\over m}\sum_{\hat m=1}^m H_m}\, ,}
}
in the limit $\de\to 0$, it is clear that only those states in 
$\R^{\M}_{(\De;j;\rr)}$ for which ${\H}_m$ 
has zero eigenvalue
contribute.  In particular, this applies to the highest weight state
in the $(N,B,m)$ BPS multiplet.

It can be shown that,
\eqn\limitBPSb{\eqalign{
\lim_{\de\to 0}
\chi^{\M}_{(\De;j; \rr)}(\de u^{1\over 2},
x, (\de^{-2}u)^{1\over m}\hat \y,\bar \y)=\chi^{U(1)\otimes SO(2N{-2m})}_{(R;\bar \rr)}(u,\bar \y)
=u^R\chi^{(N{-m})}_{\bar \rr}(\bar \y)\, ,}
}
in terms of $U(1)_{\I_m}\otimes SO(2N{-2m})$ characters, for appropriate $R$, $\bar \rr$,
see appendix A.

An identity which is useful for
simplifying the limits considered here is the following, for $\rr=(r_1,r_2,\dots,r_N)$,
\eqn\gyropter{\eqalign{
&\chi^{(N)}_\rr(\de^{-{2\over m}}\hat \u,\bar \y)\sim \de^{-{2\over m}(r_1+\dots+r_m)}
s_{\hat \rr}(\hat \u)\,\chi^{(N-m)}_{\bar \rr}(\bar \y)\, ,\cr
&\hat \rr=(r_1,\dots,r_m)\, ,\qquad \bar\rr=(r_{m+1},\dots,r_N)\, ,}
}
giving the leading behaviour as $\de\to 0$,
where $s_{\hat \rr}(\hat \u)$ is a Schur polynomial,{\foot{The identity \gyropter\
may be obtained by noting that, for small $\delta$,
$$\eqalign{
C^{(N)}_\rr(\de^{-{2\over m}}\hat \u,\bar \y)&\sim 
{\widetilde C}^{(m)}_{\hat \rr}(\de^{-{2\over m}}\hat \u)\,C^{(N-m)}_{\bar \rr}(\bar \y)\,,
\qquad  
{\widetilde C}^{(m)}_{\hat \rr}(\hat \u)={\ts \prod_{i=1}^m}\hat u_i{}^{\hat r_i+m-i}/\De(\hat \u)\, ,\cr
\frak W^{\S_N\ltimes (\S_2)^{N-1}} C^{(N)}_\rr(\de^{-{2\over m}}\hat \u,\bar \y)&\sim
\frak W^{\S_m}\,{ \frak W}^{\S_{(N-m)}\ltimes (\S_2)^{N-m-1}}
{\widetilde C}^{(m)}_{\hat \rr}(\de^{-{2\over m}}\hat \u)\,
C^{(N-m)}_{\bar \rr}(\bar \y)\, .}
$$
Here, ${\widetilde C}^{(m)}_{\hat \rr}(\hat \u)$ is equivalent to the $U(m)$ Verma module
character while $ \frak W^{\S_m}$ is the Weyl symmetriser for $U(m)$, acting
on $\hat \u$, so that for any $f(\hat \u)=f(\hat u_1,\dots, \hat u_m)$,
$\frak W^{\S_m}(\hat \u)=\sum_{\si\in \S_m}f(\hat u_{\si(1)},\dots,\hat u_{\si(m)})$. 
Hence, $\frak W^{\S_m}{\widetilde C}^{(m)}_{\hat \rr}(\hat \u)
=s_{\hat \rr}(\hat \u)$, the $U(m)$ Weyl character.
Here also, $\frak W^{\S_{N-m}\ltimes (\S_2)^{N-m-1}}$
 acts on $\bar \y$, so that
$\frak W^{\S_{N-m}\ltimes (\S_2)^{N-m-1}}
C^{(N-m)}_{\bar \rr}(\bar \y)=\chi^{(N-m)}_{\bar \rr}(\bar \y)$.}}
\eqn\schur{
s_{\hat \rr}(\hat \u)={\det}\big [\hat u_i{}^{r_j+n-j}\big ]/\De(\hat \u)\, ,
\qquad s_{(r,\dots,r)}(\hat \u)={\ts \prod_{i=1}^m}\hat u_i{}^r\, .
}

Using \gyropter, we may obtain from \longsemi\ that the limit taken in \limitBPS\
gives the following $U(1)\otimes SO(2N{-2m})$ characters, for $n\geq m$, see appendix A,
\eqn\limitbps{\eqalign{
\chi^{U(1)\otimes SO(2N{-2m})}_{(r_1;\bar \rr)}(u,\bar \y)&{}=
\lim_{\de\to 0} \chi^{(N,B,n)}_{(r_1;0; \rr)}(\de u^{1\over 2},
x, (\de^{-{2}}u)^{1\over m}\hat \y,\bar \y)=
u^{2r_1}\chi^{(N-m)}_{\bar \rr}(\bar \y)\, ,\cr
\chi^{U(1)\otimes SO(2N{-2m})}_{(r;r,\dots, r,\pm r)}(u,\bar \y)&{}
=\lim_{\de\to 0} \chi^{(N,B,\pm)}_{(r;0;r,\dots,r,\pm r )}(\de u^{1\over 2},
x, (\de^{-{2}}u)^{1\over m}\hat \y,\bar \y)=
u^{2r}\chi^{(N-m)}_{(r,\dots,r,\pm r)}(\bar \y)\, . }
}

For $m=1$, we have in addition to \limitbps\ that, 
\eqn\extralong{
\chi^{U(1)\otimes SO(2N{-2})}_{(r_1+2;\bar \rr)}(u,\bar \y)=
\lim_{\de\to 0} \chi^{(N,{\rm long})}_{(r_1+1;0; \rr)}(\de u^{1\over 2},
x, \de^{-{2}}u,\bar \y)\, ,
}
while other long multiplet characters, for $\De\geq r_1{+j}{+1}$, $j\neq 0$, along with
semi-short multiplet characters vanish, consistent with
\decompimp.   Since, in the limit taken in \limitBPSb,
long multiplet characters vanish for $m>1$, this provides further evidence,
apart from long multiplet decomposition formulae listed in the previous section,
that $(N,B,m)$, $m>1$ BPS operators must remain protected in any
three dimensional superconformal field theory. 

\noindent
{\bf The $U(1)$ Sectors}

Similarly, by considering,
\eqn\limithalfBPSo{\eqalign{
&\chi^{\M}_{(\De;j; \rr)}(\de u^{1\over 2},
x, (\de^{-{2}}
u)^{1\over N}{\hat y}_1\dots, (\de^{-{2}}
u)^{1\over N}{\hat y}_{N-1},
(\de^{-{2}}
u)^{\pm{1\over N}}{\hat y}_{N}{}^{\pm 1})\cr
&{}=\tr_{\R^{\M}_{(\De;j;\rr)}}(\de^{2{\H}_{\pm}}\,u^{{\I}_{\pm}} \,x^{2{ J_3}} 
\,{\hat y}_1{}^{H_1}
\cdots {\hat y}_N{}^{\pm H_N})\,,\cr
&{\H}_\pm={D}-{\ts{1\over N}}(H_1+\dots+H_{N-1}\pm H_N)\, ,\quad 
{\I}_\pm={D}+{\ts{1\over N}}(H_1+\dots+H_{N-1}\pm H_N)\, ,}
}
separately in the limit $\de\to 0$,
isolating states for which ${\H}_{\pm}$ has zero eigenvalues,
only the corresponding  $\half $ BPS character is non-zero giving,
\eqn\limithalfBPS{\eqalign{
\chi^{U(1)}_{(r)}(u)&{}=\lim_{\de\to 0} \chi^{(N,B,\pm)}_{(r;0; r,\dots,r,\pm r)}
(\de u^{1\over 2},
x, (\de^{-{2}}
u)^{1\over N}{\hat y}_1\dots, (\de^{-{2}}
u)^{1\over N}{\hat y}_{N-1},
(\de^{-{2}}
u)^{\pm{1\over N}}{\hat y}_{N}{}^{\pm 1})\cr
&{}=u^{2r}
\, .}
}

\noindent
{\bf Semi-short Limits}

\noindent
{\bf The $U(1)\otimes Osp(2N{-2m}|2)$ Sector}

Considering, for $\hat \y$, $\bar \y$ as in \limitBPS,
\eqn\limitSemishort{\eqalign{
&\chi^{\M}_{(\De;j; \rr)}(\de {\bar s}^{1\over 2}, \de^{-1}\bar s^{1\over 2}, (\de^{-2} u)^{1\over m}\hat \y, \bar \y)\cr
&
=\tr_{\R^{\M}_{(\De;j;\rr)}}(\de^{2{\J}_m}\, u^{\K_m}\,{\bar s}^{{{\bar D} } }
 \,{\hat y}_1{}^{H_1}
\cdots {\hat y}_m{}^{H_m}\,{\bar y}_1{}^{H_{m+1}}\cdots{\bar y}_{N-m}{}^{H_N})\,,\cr
&{\J}_m={D}-{J}_3 -{\ts{1\over m}}{\ts{\sum_{\ha=1}^m}}H_m\, ,
\quad \K_m={\ts{1\over m}}{\ts{\sum_{\ha=1}^m}}H_m\, ,\qquad 
{{\bar D}}={D}+{J}_3\, ,}
}
in the limit $\de\to 0$, again, it is clear that only those states in $\R^{\M}_{(\De;j;\rr)}$ 
for which the eigenvalue of ${\J}_m$ is zero contribute.

It can be shown that,
\eqn\limitSemishortob{\eqalign{
\lim_{\de\to 0}
\chi^{\M}_{(\De;j; \rr)}(\de {\bar s}^{1\over 2}, \de^{-1}\bar s^{1\over 2}, (\de^{-2} u)^{1\over m}\hat \y, \bar \y)
&{}=\chi^{(U(1)\otimes Osp(2N-2m|2),i)}_{(R;\bar \De;\bar \rr)}(u,\bar s,\bar \y)\cr
&{}=u^{R}\chi^{(Osp(2N{-2m}|2),i)}_{(\bar \De,\bar \rr)}(\bar s,\bar \y)\, ,}
}
in terms of $U(1)_{\K_m}\otimes Osp(2N{-2m}|2)$ characters, for appropriate $i$, $R$, $\bar \Delta$,
 see appendix A for notation. 

Similar to above, using \longsemi\ with \gyropter, and also, 
\eqn\chiropter{
\chi_{2j}(\de^{-1}\bb^{1\over 2} )\sim \de^{-2j}\,\bb^j\, ,
}
giving the leading behaviour as $\de\to 0$,
then we have, for semi-short and conserved multiplet cases, for $n\geq m$
and with $\bar \rr$ as in \gyropter,
\eqn\limitsemi{\eqalign{
&\chi^{(U(1)\otimes Osp(2N-2m|2),\rm long)}_{(r_1{+1};r_1{+2j}{+m}{+1};\bar \rr)}(u,\bar s,\bar \y)=
\lim_{\de\to 0} \chi^{(N,A,n)}_{(r_1+j+1;j;\rr)}(\de {\bar s}^{1\over 2}, \de^{-1}\bar s^{1\over 2}, 
(\de^{-2} u)^{1\over m}\hat \y, \bar \y)\cr
&\qquad\qquad\qquad\qquad\qquad\qquad\qquad
={u}^{r_1+1}\,{{\bar s}^{r_1+2j+m+1}\over 1-\bar s^2}\,\,
\chi^{(N-m)}_{\bar \rr}(\bar \y)\,
{\ts\prod_{\vep=\pm 1}}{\ts\prod_{i=1}^{N-m}}(1+\bar y_i{}^{\vep}\bar s)\,,\cr
&\chi^{(U(1)\otimes Osp(2N-2m|2),\rm long)}_{(r{+1};r{+2j}{+m}{+1};r,\dots,r,\pm r)}(u,\bar s,\bar \y)=
\lim_{\de\to 0} \chi^{(N,A,\pm)}_{(r+j+1;j;r,\dots,r,\pm r)}(\de {\bar s}^{1\over 2}, \de^{-1}\bar s^{1\over 2}, 
(\de^{-2} u)^{1\over m}\hat \y, \bar \y)\cr
&\qquad\qquad\qquad\qquad\qquad\qquad\qquad\!\!\!
={u}^{r+1}\,{{\bar s}^{r+2j+m+1}\over 1-\bar s^2}\,\,
\chi^{(N-m)}_{(r,\dots,r,\pm r)}(\bar \y)\,
{\ts\prod_{\vep=\pm 1}}{\ts\prod_{i=1}^{N-m}}(1+\bar y_i{}^{\vep}\bar s)\,,\cr
&\chi^{(U(1)\otimes Osp(2N-2m|2),\rm long)}_{(1;2j{+m}{+1};0,\dots,0)}(u,\bar s,\bar \y)=
\lim_{\de\to 0} \chi^{(N,{\rm cons.})}_{(j+1;j;0,\dots,0)}(\de {\bar s}^{1\over 2}, \de^{-1}\bar s^{1\over 2}, 
(\de^{-2} u)^{1\over m}\hat \y, \bar \y)\cr
&\qquad\qquad\qquad\qquad\qquad\qquad\qquad
={u}\,{{\bar s}^{2j+m+1}\over 1-\bar s^2}
{\ts\prod_{\vep=\pm 1}}{\ts\prod_{i=1}^{N-m}}(1+\bar y_i{}^{\vep}\bar s)\,.}
}
For $(N,B,n)$ BPS multiplet characters there are a number 
of cases to consider.
For $n<m$ we have,
for $r_n=\dots=r_1$ and $r_{n+1}=\dots=r_{m}=r_1{-1}\geq r_{m+1}\geq 
\dots \geq |r_N|$,
\eqn\limitsemioply{\eqalign{
&\chi^{(U(1)\otimes Osp(2N-2m|2),{\rm long})}_{(r_1;r_1{+m}{-n};\bar \rr)}
(u,\bar s,\bar \y)=
\lim_{\de\to 0} \chi^{(N,B,n)}_{(r_1;0;\rr)}
(\de {\bar s}^{1\over 2}, \de^{-1}\bar s^{1\over 2}, 
(\de^{-2} u)^{1\over m}\hat \y, \bar \y)\cr
&=u^{r_1}{{\bar s}^{r_1+m-n}\over 1-\bar s^2}\,\,\chi^{(N-m)}_{\bar \rr}(\bar \y)\,
{\ts \prod_{\vep=\pm 1}}{\ts \prod_{i=1}^{N-m}}(1+\bar y_i{}^{\vep}\,\bar s)\, ,
\,}
}
while for $n=m$, so that $r_1=\dots=r_n> r_{n+1}\geq \dots \geq |r_N|$,
\eqn\limitsemiopp{\eqalign{
\chi^{(U(1)\otimes Osp(2N-2m|2),{\rm long})}_{(r_1;r_1;\bar \rr)}(u,\bar s,\bar \y)
&{}=
\lim_{\de\to 0} \chi^{(N,B,m)}_{(r_1;0;\rr)}
(\de {\bar s}^{1\over 2}, \de^{-1}\bar s^{1\over 2}, 
(\de^{-2} u)^{1\over m}\hat \y, \bar \y)\cr
&{}=u^{r_1}{{\bar s}^{r_1}\over 1-\bar s^2}\,\chi^{(N-m)}_{\bar \rr}(\bar \y)\,
{\ts \prod_{\vep=\pm 1}}{\ts \prod_{i=1}^{N-m}}(1+\bar y_i{}^{\vep}\,\bar s)\, .}
}
For $n>m$ we have, using results from appendix A,
\eqn\limitsemio{\eqalign{
&\chi^{(U(1)\otimes Osp(2N-2m|2),n)}_{(r_1;r_1;\bar \rr)}(u,\bar s,\bar \y)=
\lim_{\de\to 0} \chi^{(N,B,n)}_{(r_1;0;\rr)}(\de {\bar s}^{1\over 2}, \de^{-1}\bar s^{1\over 2}, 
(\de^{-2} u)^{1\over m}\hat \y, \bar \y)\cr
&=u^{r_1}{{\bar s}^{r_1}\over 1-\bar s^2}\,\,
{\frak W}^{\S_{N-m}\ltimes (\S_2)^{N-m-1}}
\Big(C^{(N-m)}_{\bar \rr}(\bar \y)\,{\ts \prod_{i=n-m+1}^{N-m}}(1+\bar y_i\,\bar s)\,
{\ts\prod_{i=1}^{N-m}}(1+\bar y_i{}^{-1}\bar s)\Big)\, ,}
}
so that
\eqn\limitsemioppop{\eqalign{
&\chi^{(U(1)\otimes Osp(2N-2m|2),n)}_{(r_1;r_1;\bar \rr)}(u,\bar s,\bar \y)
\cr
&=u^{r_1}{{\bar s}^{r_1}\over 1-\bar s^2}\sum_{0\leq a_1\leq \dots \leq a_{n-m}\leq 1}
\sum_{{a_{n-m+1},\dots,a_{N-m}\atop
\bar a_{n-m+1},\dots,\bar a_{N-m}}=0}^1s^{a_1+\dots + a_{N-m}+\bar a_{n-m+1}+\dots +\bar a_{N-m}}\cr
&\qquad \qquad\qquad\qquad
\times \chi^{(N-m)}_{(r_1-a_1,\dots,r_1-a_{n-m},r_{n+1}+\bar a_{n-m+1}-a_{n-m+1},\dots,r_N+\bar a_{N-m}
-a_{N-m})}(\bar \y)
\,.}
}
Finally, for the $\half$ BPS cases,
\eqn\limitsemihalf{\eqalign{
&\chi^{(U(1)\otimes Osp(2N-2m|2),\pm)}_{(r;r;r,\dots, r,\pm r)}(u,\bar s,\bar \y)
=\lim_{\de\to 0} \chi^{(N,B,\pm)}_{(r;0;r,\dots,r,\pm r)}
(\de {\bar s}^{1\over 2}, \de^{-1}\bar s^{1\over 2}, 
(\de^{-2} u)^{1\over m}\hat \y, \bar \y)\cr
&=u^r{{\bar s}^{r}\over 1-\bar s^2}\
\sum_{0\leq a_1\leq\dots\leq a_{N-m}\leq 1}\bar s^{a_1+\dots+a_{N-m}}
\chi^{(N-m)}_{(r-a_1,\dots,r-a_{N-m-1},\pm r\mp a_{N-m})}(\bar \y)\, .}
}
In particular, this implies,
\eqn\limitsemihalfo{\eqalign{
\chi^{(U(1)\otimes Osp(2N-2m|2),\pm)}_{({1\over 2};{1\over 2};{1\over 2},\dots, 
{1\over 2},\pm {1\over 2})}(u,\bar s,\bar \y)={}&\lim_{\de\to 0} 
\chi^{(N,B,\pm)}_{({1\over 2};0;{1\over 2},\dots,{1\over 2},\pm {1\over 2})}(\de {\bar s}^{1\over 2}, 
\de^{-1}\bar s^{1\over 2}, 
(\de^{-2} u)^{1\over m}\hat \y, \bar \y)\cr
={}&u^{1\over 2}{{\bar s}^{{1\over 2}}\over 1-\bar s^2}
\Big(\chi^{(N-m)}_{({1\over 2},\dots,{1\over 2},\pm {1\over 2})}(\bar \y)+\bar s\,
\chi^{(N-m)}_{({1\over 2},\dots,{1\over 2},\mp {1\over 2})}(\bar \y)\Big)\, ,}
}
agreeing with \halfbpshalf, in this limit, using \gyropter\
with \schur\ for $r=\half$.

For $m>1$, we have for long multiplet characters,
\eqn\longretp{\eqalign{
&\chi^{(U(1)\otimes 
Osp(2N-2m|2),{\rm long})}_{(r_1{+1};r_1{+2j}{+m}{+1};r_{m+1},\dots, r_N)}
(u,\bar s,\bar \y)
\cr
&=
\lim_{\de\to 0} \chi^{(N,{\rm long})}_{(r_1+j+1;j;r_1,\dots, r_1,r_{m+1},\dots, r_N)}
(\de {\bar s}^{1\over 2}, \de^{-1}\bar s^{1\over 2}, 
(\de^{-2} u)^{1\over m}\hat \y, \bar \y)\, ,}
}
which is in accord with \decomp\ with \strider.
For $m=1$, we have that
\eqn\longretp{
(1+u)\chi^{(U(1)\otimes Osp(2N-2|2),{\rm long})}_{(r_1{+1};r_1{+2j}{+2};r_{2},\dots, r_N)}
(u,\bar s,\bar \y)
=
\lim_{\de\to 0} \chi^{(N,{\rm long})}_{(r_1+j+1;j;r_1,\dots, r_N)}
(\de {\bar s}^{1\over 2}, \de^{-1}\bar s^{1\over 2}, 
\de^{-2} u, \bar \y)\, ,
}
compatible with \decomp.

\noindent
{\bf The $U(1)\otimes SU(1,1)$ Sectors}

Similarly, by considering,
\eqn\limithalfBPSoppp{\eqalign{
&\chi^{\M}_{(\De;j; \rr)}(\de {\bar s}^{1\over 2},
\de^{-1}{\bar s}^{1\over 2}, (\de^{-{2}}
u)^{1\over N}{\hat y}_1\dots, (\de^{-{2}}
u)^{1\over N}{\hat y}_{N-1},
(\de^{-{2}}
u)^{\pm{1\over N}}{\hat y}_{N}{}^{\pm 1})\cr
&{}=\tr_{\R^{\M}_{(\De;j;\rr)}}(\de^{2{\J}_{\pm}}\,u^{{\K}_{\pm}} \,{\bar s}^{{\bar D}} 
\,{\hat y}_1{}^{H_1}
\cdots {\hat y}_N{}^{\pm H_N})\,,\cr
&{\J}_\pm={D}-J_3-{\ts{1\over N}}(H_1+\dots+H_{N-1}\pm H_N)\, ,\quad 
{\K}_\pm={\ts{1\over N}}(H_1+\dots+H_{N-1}\pm H_N)\, ,}
}
separately in the limit $\de\to 0$, we have for the corresponding $(N,A,\pm)$
multiplet character,
\eqn\firstsemipm{\eqalign{
&
\chi^{U(1)\otimes SU(1,1)}_{(r{+1};r{+2j}{+1}{+N})}(u,\bar s)\cr
&=
\lim_{\de\to 0} \chi^{(N,A,\pm)}_{(r+j+1;j;r,\dots,r,\pm r)}
(\de {\bar s}^{1\over 2},
\de^{-1}{\bar s}^{1\over 2}, (\de^{-{2}}
u)^{1\over N}{\hat y}_1\dots, (\de^{-{2}}
u)^{1\over N}{\hat y}_{N-1},
(\de^{-{2}}
u)^{\pm{1\over N}}{\hat y}_{N}{}^{\pm 1})
\cr
&
={u}^{r+1}\,{{\bar s}^{r+2j+N+1}\over 1-\bar s^2}\,,}
}
and, for the conserved current character, in either limit in \limithalfBPSoppp,
\eqn\firstsemipmpo{\eqalign{
&
\chi^{U(1)\otimes SU(1,1)}_{({1};{2j}{+1}{+N})}(u,\bar s)\cr
&=
\lim_{\de\to 0} \chi^{(N,{\rm cons.})}_{(j+1;j;0,\dots,0)}(\de {\bar s}^{1\over 2},
\de^{-1}{\bar s}^{1\over 2}, (\de^{-{2}}
u)^{1\over N}{\hat y}_1\dots, (\de^{-{2}}
u)^{1\over N}{\hat y}_{N-1},
(\de^{-{2}}
u)^{\pm{1\over N}}{\hat y}_{N}{}^{\pm 1})
\cr
&
={u}\,{{\bar s}^{2j+N+1}\over 1-\bar s^2}\,.}
}
For $(N,B,n)$, $n\leq N{-1}$, BPS characters
with  $r_n=\dots=r_1$ and $r_{n+1}=\dots=r_{N}=\pm r_1{\mp 1}$
and taking corresponding limit in \limithalfBPSoppp\ then,
\eqn\firstsemipmpo{\eqalign{
&\chi^{U(1)\otimes SU(1,1)}_{({r_1};{r_1}{+N}{-n})}(u,\bar s)\cr
&=
\lim_{\de\to 0} \chi^{(N,B,n)}_{(r_1;0;\rr)}(\de {\bar s}^{1\over 2},
\de^{-1}{\bar s}^{1\over 2}, (\de^{-{2}}
u)^{1\over N}{\hat y}_1\dots, (\de^{-{2}}
u)^{1\over N}{\hat y}_{N-1},
(\de^{-{2}}
u)^{\pm{1\over N}}{\hat y}_{N}{}^{\pm 1})
\cr
&
={u^{r_1}}\,{{\bar s}^{r_1+N-n}\over 1-\bar s^2}\,.}
}
For the $(N,B,\pm)$ half BPS cases and in the corresponding limit
in \limithalfBPSoppp\ we have,
\eqn\firstsemipmpopo{\eqalign{
&\chi^{U(1)\otimes SU(1,1)}_{({r};{r})}(u,\bar s)\cr
&=
\lim_{\de\to 0} \chi^{(N,B,n)}_{(r;0;r,\dots, r,\pm r)}(\de {\bar s}^{1\over 2},
\de^{-1}{\bar s}^{1\over 2}, (\de^{-{2}}
u)^{1\over N}{\hat y}_1\dots, (\de^{-{2}}
u)^{1\over N}{\hat y}_{N-1},
(\de^{-{2}}
u)^{\pm{1\over N}}{\hat y}_{N}{}^{\pm 1})
\cr
&
={u^{r}}\,{{\bar s}^{r}\over 1-\bar s^2}\,,}
}
along with,
\eqn\firstsemipmpopo{\eqalign{
&\chi^{U(1)\otimes SU(1,1)}_{({{1\over 2}};{{3\over 2}})}(u,\bar s)\cr
&=
\lim_{\de\to 0} \chi^{(N,B,n)}_{({1\over 2};0;{1\over 2},\dots, {1\over 2},\mp {1\over 2})}
(\de {\bar s}^{1\over 2},
\de^{-1}{\bar s}^{1\over 2}, (\de^{-{2}}
u)^{1\over N}{\hat y}_1\dots, (\de^{-{2}}
u)^{1\over N}{\hat y}_{N-1},
(\de^{-{2}}
u)^{\pm{1\over N}}{\hat y}_{N}{}^{\pm 1})
\cr
&
={u^{1\over 2}}\,{{\bar s}^{3\over 2}\over 1-\bar s^2}\,.}
}
All other characters, apart from
the long multiplet one corresponding to \decomplump, 
in the limit \limithalfBPSoppp\
vanish.

\noindent
{\bf The Superconformal Index}

The (single particle) superconformal indices \refs{\minwero,\mald}
may be computed by taking the limit $u\to -1$ in the
$U(1)\otimes Osp(2N{-2}|2)$ characters above, {\it i.e.}
for $m=1$ in the semi-short limits above.{\foot{
It is easily checked that for $N=3$, $m=1$, then \limitsemihalfo\
for $(u,\bar s,y_2,y_3)\to (-1,-x,y_1,y_2)$ matches with the index for fundamental fields
computed in section 2 of \minw.}}    

In particular, from \longretp,
this limit ensures that $Osp(2N|4)$ long multiplet characters vanish,
and hence do not  contribute in the decomposition of partition functions,
in the same limit,
in terms of $Osp(2N|4)$ characters.  Thus, partition functions
evaluated in this limit receive contributions from protected operators
only.  It should be noted however that the magnitude of the numbers obtained for
counting operators by expansion in terms of characters, in this limit, provide only
a lower bound on the numbers of actual protected operators 
due to the $(-1)^F$ sign in the index, for fermion number $F$.

\newsec{Partition Functions for $\N=6$ Superconformal Chern-Simons Theory}

In \abjm, a class of $\N=6$ superconformal Chern Simons theories,
with gauge group $U(n)\times U(n)$, was proposed.  For
levels $\pm k$ in the Chern Simons terms, these theories admit a
dual description in terms of M theory compactified on 
AdS${}_4\times S^7/\Bbb{Z}_k$.

Here, the free field partition function of the theory, where the effective 
't Hooft coupling $n/k\to 0$, so that $k\to \infty$, in the large $n<k$ limit,
is computed using appropriate superconformal characters.

The supergravity
partition function, obtained using the duality proposed in \abjm\
in the $n/k\to \infty$ limit, for $n>k \to \infty$, is also computed using
appropriate superconformal characters.

\noindent
{\bf Free Field Theory}

For  $\N=6$ $U(n)\times U(n)$ superconformal Chern-Simons theory
 the dynamical field content consists of scalars $\phi_1$, $\phi_2$
 and spin half fermions $\psi_1$, $\psi_2$ belonging to the 
 `Rac', respectively  `Di', representations
of $SO(3,2)$, mentioned after \dirac.  The fields are listed in Table 2
showing  also their $SO(6)$ $R$-symmetry eigenvalues and
gauge group representations.

\medskip
\vbox{
\hskip1.5cm   Table 2
\nobreak

\hskip1.5cm
\vbox{\tabskip=0pt \offinterlineskip
\hrule
\halign{&\vrule# &\strut \ \hfil#\  \cr
height2pt&\omit&&\omit&&\omit&&\omit&& \omit &\cr
&\   field  \ \hfil   && \  $ \De $ \ \hfil && \ $ SO(3,2) $ rep. \  \hfil && \ $SO(6)$ rep. \ \hfil  
&& \ $U(n)\times U(n)$ rep. \ \hfil &\cr
height2pt&\omit&&\omit&&\omit&&\omit&& \omit &\cr
\noalign{\hrule}
height2pt&\omit&&\omit&&\omit&&\omit&& \omit &\cr
& \ $\phi_1$  \ \hfil &&  $\half$  \hfil    &&  `Rac' \hfil  &&   $(\half,\half,\half)$      \hfil            
 &&  $(n,{\overline{n}})$    \hfil    &\cr
height3pt&\omit&&\omit&&\omit&&\omit&& \omit &\cr
& \ $\psi_1$  \ \hfil &&  $1$  \hfil     &&  `Di' \hfil &&   $(\half,\half,-\half)$    \hfil 
  &&  $(n,{\overline{n}})$    \hfil    &\cr
height3pt&\omit&&\omit&&\omit&&\omit&& \omit &\cr
& \ $\phi_2$ \ \hfil &&  $\half$  \hfil     &&    `Rac' \hfil   && $(\half,\half,\half)$     \hfil 
&&   $({\overline{n}},n)$    \hfil    &\cr
height3pt&\omit&&\omit&&\omit&&\omit&& \omit &\cr
& \ $\psi_2$  \ \hfil &&  $1$  \hfil     &&   `Di' \hfil   &&   $(\half,\half,-\half)$     \hfil 
&&  $({\overline{n}}, n)$    \hfil    &\cr
height2pt&\omit&&\omit&&\omit&&\omit&& \omit &\cr
}
\hrule}
}

Here the $SO(6)$ orthonormal basis
labels $(r,q,p)$ are related to $SU(4)$ Dynkin labels $[a,b,c]$ by,
 \eqn\jumbawaba{
 (r,q,p)\to [q{+p},r{-q},q{-p}]\, ,
 }
so that $(\half,\half,\half)\to [1,0,0]$, $(\half,\half,-\half)\to
[0,0,1]$.

As may be evident from \halfbpshalf, $\phi_1$ and $\psi_1$, respectively 
$\phi_2$ and $\psi_2$, belong to the half BPS multiplet $(3,B,+)$,
respectively $(3,B,-)$.  (In both cases, the scalar is
the primary field of the multiplet.)

For free field theory, the single particle partition function
is given by, 
\eqn\singpartfun{
Y_{\rm free}(s,x,\y,\u,\v)=\tr\Big(s^{2D} x^{2J_3} y_1{}^{H_1} y_2{}^{H_2} y_3{}^{H_3}
u_1{}^{L_1}\cdots u_n{}^{L_n} v_1{}^{M_1}\cdots v_n{}^{M_n}\Big)\, ,
}
where the trace is over states corresponding to
$\phi_1,\phi_2$, $\psi_1,\psi_2$ and all their superconformal
descendants, of the form \restrictedverma, and  $s,x,\y,\u,\v$ are `fugacities' 
with
 $L_1,\dots,L_n$, $M_1,\dots,M_n$ being usual $U(n)\times U(n)$
Cartan subalgebra generators.

Thus we may write, in terms of characters, 
\eqn\singleparticle{
Y_{\rm free}(s,x,\y,\u,\v)=
f_+(s,x,\y)
p_1(\u)p_1(\v^{-1})+
f_-(s,x,\y)
p_1(\u^{-1})p_1(\v)\, ,
}
defining, for subsequent use,
\eqn\defpn{
p_j(\u)=\sum_{i=1}^n u_i{}^j\, ,\qquad p_j(\u^{-1})=p_{-j}(\u)\, ,
}
so that $p_1(\u)p_1(\v^{-1}),\, p_1(\u^{-1})p_1(\v)$ correspond to the characters 
for the $(n,\overline n)$, respectively, $(\overline n,n)$ representations
of $U(n)\times U(n)$,
and where 
\eqn\deffpm{\eqalign{
f_\pm(s,x,\y)={}&
{\chi}^{(3,B,\pm)}_{({1\over 2};0;{1\over 2},{1\over 2},\pm {1\over 2})}(s,x,\y)\cr
={}&
\big(y_1\,y_2\,y_3\big)^{\mp {1\over 2}}\left({\ts \sum_{j=1}^3 y_j{}^{\pm 1}}
+(y_1\,y_2\,y_3)^{\pm 1}\right)\D_{\rm Rac}(s,x)\cr
&+
\big(y_1\,y_2\,y_3\big)^{\pm {1\over 2}}\left({\ts \sum_{j=1}^3 y_j{}^{\mp 1}}+(y_1\,y_2\,y_3)^{\mp 1}\right)
\D_{\rm Di}(s,x)\, ,}
}
being half BPS characters \halfbpshalf\ for $SO(6)$ $R$-symmetry.
 
The multi-particle partition function, receiving contributions 
from only gauge invariant operators, is given by the usual integral
over the gauge group, namely, 
\eqn\multip{
Z^{(n)}_{\rm free}(s,x,\y)=\int_{U(n)}\!\!\!\!\!\!\d \mu(\u)\int_{U(n)}\!\!\!\!\!\d \mu(\v)
\exp\bigg(\sum_{j=1}^\infty{1\over j}Y_{\rm free}(s^j,(-1)^{j+1}x^j,\y^j,\u^j,\v^j)\bigg)\, ,
}
where the signs on $x$ take account
of particle statistics.

The integral may be evaluated  in the large $n$ limit
by using a method described  in \counting\ (see also
\indexfh\ for other applications).
We may first expand,
\eqn\multpo{\eqalign{
&Z^{(n)}_{\rm free}(s,x,\y)\cr
&=\sum_{\blambda,\brho}{1\over z_\blambda z_\brho}f_{+\,\blambda}(s,x,\y) 
f_{-\,\brho}(s,x,\y)
\int_{U(n)}\!\!\!\!\!\!\d \mu(\u)\int_{U(n)}\!\!\!\!\!\d \mu(\v) p_\blambda(\u)p_{\brho}(\u^{-1})
p_{\brho}(\v)p_\blambda(\v^{-1})\, ,}
}
in terms of partitions,
\eqn\Dparts{\eqalign{
\blambda=(\lambda_1,\dots,\lambda_j,\dots)\, ,&\qquad {\ts \sum_{j\geq 1}j\lambda_j}=|\blambda|\in
\Bbb{N}\, ,\cr
\brho=(\rho_1,\dots,\rho_j,\dots)\, ,&\qquad {\ts \sum_{j\geq 1}j\rho_j}=|\brho|\in
\Bbb{N}\, ,}
}
where, for $\bsi=(\si_1,\dots, \si_j,\dots)$,
\eqn\defs{
z_\bsi=\prod_{j\geq 1}{ \si_j! j^{\si_j}}\, ,\quad
f_{\pm\, \bsi}(s,x,\y)=\prod_{j\geq 1} f_{\pm}(s^j,(-1)^{j+1}x^j,\y^j){}^{\si_j}\, ,\quad
p_\bsi(\x)=\prod_{j\geq 1} p_j(\x){}^{\si_j}\, .
}
Using the orthogonality relation for power symmetric polynomials,
\eqn\ortho{
\int_{U(n)}\d \mu(\u)p_\blambda(\u)p_{\brho}(\u^{-1})=z_\blambda 
\de_{\blambda\,\brho}\, ,
\qquad |\blambda|,|\brho|\leq n\, ,
}
then we trivially obtain in the large $n$ limit,
\eqn\largeNo{\eqalign{
Z^{(n)}_{\rm free}(s,x,\y)\sim
 Z_{\rm free}(s,x,y)&{}=\sum_{\blambda}f_{+\,\blambda}(s,x,\y)f_{-\,\blambda}(s,x,\y)
\cr
&=\prod_{j\geq 1}
\sum_{\lambda_j\geq 0}\big(f_+(s^j,(-1)^{j+1}x^j,\y^j)f_-(s^j,(-1)^{j+1}x^j,\y^j)\big)^{\lambda_j}\cr
&=\prod_{j\geq 1}{1\over1- f_+(s^j,(-1)^{j+1}x^j,\y^j)f_-(s^j,(-1)^{j+1}x^j,\y^j)}\, .}
}
This result was also derived in
 \minw\ by using
saddle point methods, in the context of showing superconformal 
index matching. 

By decomposing the
product of half BPS characters \deffpm, 
\eqn\decompl{
f_+(s,x,\y)f_-(s,x,\y)=\chi^{(3,B,2)}_{(1;0;1,1,0)}(s,x,\y)
+\sum_{j=0}^\infty\chi^{(3,\rm cons.)}_{(j+1;j;0,0,0)}(s,x,\y)\, ,
}
it is interesting to note that \largeNo\ may be obtained by considering a
field theory with fields in the $(3,B,2)$ representation and 
conserved currents transforming in the adjoint representation
of $U(m)$ for large $m$.  For this theory, using \decompl,
\eqn\yupoinjnnj{
Y(s,x,\y,\z)=f_+(s,x,\y)f_-(s,x,\y)p_1(\z)p_{1}(\z^{-1})\, ,
}
so that the corresponding multi-particle partition function
is given by,
\eqn\buhvibibgi{
Z(s,x,\y)=\int_{U(m)}\!\!\!\!\!\d \mu(\z)
\exp\bigg(\sum_{j=1}^\infty{1\over j}Y(s^j,(-1)^{j+1}x^j,\y^j,\u^j,\z^j)\bigg)\, .
}
In the large $m$ limit, it is easily seen that $Z(s,x,\y)$ matches with \largeNo.
It may be interesting to understand this equivalence from a gauge theory
or string theory perspective.

\noindent
{\bf Supergravity Limit}

In the strong coupling limit, large $n/k$, $n>k\to \infty$, 
as explained in \abjm, using results of \pope, the single particle
states (gravitons) effectively belong to scalar superconformal representations
with conformal dimensions $r$ and in the $\R_{(r,r,0)}$ $SO(6)$
representations, for $r\in \Bbb{N}$, $r>0$, so that, in terms of notation
here, they belong to $(3,B,2)$ BPS multiplets. 

Taking account of superconformal 
descendants,  we may thus write for the single particle
partition function,
\eqn\singlpart{
Y_{\rm sugra}(s,x,\y)=\sum_{r=1}^\infty \chi^{(3,B,2)}_{(r;0;r,r,0)}(s,x,\y)\, ,
}
where using \longsemi,  for $\frak W^{\S_3\ltimes (\S_2)^2}$ 
acting on $\y=(y_1,y_2,y_3)$,
\eqn\singpartsu{
\chi^{(3,B,2)}_{(r;0;r,r,0)}(s,x,\y)=s^{2r}P(s,x)\frak W^{\S_3\ltimes (\S_2)^2}\bigg(
C^{(3)}_{(r,r,0)}(\y)
\prod_{\vep=\pm1 }(1+s y_3 x^{\vep})\prod_{j=1}^3(1+s y_i{}^{-1}x^\vep)\bigg)\, .
}
Thus, using the definition of the $SO(6)$ Verma module character \charsv,
for $N=3$, and summing over $r$ in \singlpart\ with \singpartsu, we may
write,
\eqn\uypoiy{\eqalign{
&Y_{\rm sugra}(s,x,\y)\cr
&=s^2 P(s,x)\frak W^{\S_3\ltimes (\S_2)^2}\bigg( {y_1{}^3 y_2{}^2
\prod_{\vep=\pm1 }(1+s y_3 x^{\vep})\prod_{i=1}^3(1+s y_i{}^{-1}x^\vep)
\over (1-s^2 y_1y_2)\prod_{1\leq j<k\leq 3}(y_j{}^{-1}-y_k{}^{-1})(1-y_j y_k)}\bigg)\, .}
}
Simplifying the $\S_3$ part of the action of $\frak W^{\S_3\ltimes (\S_2)^2}$
in the latter we end up with the more succinct formula, 
\eqn\bjvg{\eqalign{
&Y_{\rm sugra}(s,x,\y)\cr
&=f(s,x,y_1,y_2,y_3)+f(s,x,{\ts {1\over y_1}},{\ts{1\over y_2}},y_3)+
f(s,x,{\ts{1\over y_1}},y_2,{\ts{1\over y_3}})+
f(s,x,y_1,{\ts{1\over y_2}},{\ts{1\over y_3}})-1\, ,}
}
where
\eqn\mkjoioi{\eqalign{
f(s,x,y_1,y_2,y_3)=P(s,x){\prod_{\vep=\pm 1}(1+s^3\, y_1 \,y_2\, y_3 \, x^\vep)
\prod_{i=1}^3(1+s \, y_i x^{\vep})
\over \prod_{1\leq j<k\leq 3}^3(1-s^2 y_j y_k)(1-y_j{}^{-1}y_k{}^{-1})}\, .
}
}
We then have the multi-particle (free graviton gas) partition function given by
the usual plethystic exponential, also taking into account particle statistics,
\eqn\multigrav{
Z_{\rm sugra}(s,x,\y)=\exp\bigg(\sum_{j=1}^\infty{1\over j}
Y_{\rm sugra}(s^j,(-1)^{j+1}x^j,
y_1{}^j,y_2{}^j,y_3{}^j)\bigg)
\, .
}

\newsec{Counting Multi-Trace Gauge Invariant Operators}

The limits in $Osp(2N|4)$
characters discussed in section 5, giving reductions to 
subalgebra characters, are equivalent to decoupling
limits, that isolate sectors of operators in  partition
functions.  By taking such limits, for $N=3$,
 in \largeNo, \multigrav, and decomposing in terms of
characters in the same limits we are able to count corresponding 
multi-trace gauge invariant operators
in the free field, supergravity  limits. 
For free field theory, the counting numbers obtained provide an upper bound
on the numbers of protected operators in the interacting theory,
while for the supergravity limit they are expected to count
actually protected ones.   

{}For $f_{\pm}(s,x,\y)\to f_{\pm}^G(u,\bar \y)$,
$Y_{\rm sugra}(s,x,\y)\to   Y_{\rm sugra}^G(u,\bar \y)$,
in the BPS limits, where $G$ is the corresponding subgroup,
and $f_{\pm}(s,x,\y)\to f_{\pm}^H(u,\bar s,\bar \y)$,
$Y_{\rm sugra}(s,x,\y)\to   Y_{\rm sugra}^H(u,\bar s,\bar \y)$,
in the semi-short limits, where $H$ is the corresponding subgroup,
the limits generically give,
\eqn\singpartbpssemicases{\eqalign{
f_{\pm}^G(u,\bar \y)=
\chi^{(3,B,\pm)}_{({1\over 2};0;{1\over 2},{1\over 2},\pm {1\over 2})}(u,\bar \y)\, ,&
\qquad
Y_{\rm sugra}^G(u,\bar \y)=
{\ts \sum_{r=1}^\infty}
\chi^{(3,B,2)}_{(r;0;r,r,0)}(u,\bar \y)\, ,\cr
f_{\pm}^{H}(u,\bar s,\bar \y)=
\chi^{(3,B,\pm)}_{({1\over 2};0;{1\over 2},{1\over 2},\pm {1\over 2})}(u,\bar s,\bar \y)\, ,&
\qquad Y_{\rm sugra}^H(u,\bar s,\bar \y)={\ts \sum_{r=1}^\infty}
\chi^{(3,B,2)}_{(r;0;r,r,0)}(u,\bar s,\bar \y)\, ,}
} 
where, for the superconformal characters,
\eqn\defcharlimitsft{
\chi^{\M}_{(\De;j;\rr)}(s,x,\y)\to \chi^{\M}_{(\De;j;\rr)}(u,\bar \y)\, ,\qquad
\chi^{\M}_{(\De;j;\rr)}(s,x,\y)\to \chi^{\M}_{(\De;j;\rr)}(u,\bar s, \bar \y)\, ,
}
in the same limits.
An important point for the corresponding 
multi-particle partition functions is properly taking into account particle statistics
which differs in both cases.
For $Z_{\rm free}(s,x,\y)\to Z^G_{\rm free}(u,\bar \y)$,
$Z_{\rm sugra}(s,x,\y)
\to Z^G_{\rm sugra}(u,\bar \y)$,
in the BPS limits and 
$Z_{\rm free}(s,x,\y)\to Z^H_{\rm free}(u,\bar s,\bar \y)$,
$Z_{\rm sugra}(s,x,\y)
\to Z^H_{\rm sugra}(u,\bar s, \bar \y)$, 
in the semishort limits,
consistency requires,{\foot{No signs are necessary for the BPS limits as the $x$ dependence
drops out. For the traces in \limitSemishort, \limithalfBPSoppp,
as the $\J_m$, $\J_\pm$ eigenvalues are
zero for states contributing to
 subalgebra characters, in the limit as $\de\to 0$,
then $u\to u \al$, $\bar s\to \bar s/\al$ introduces a factor of $\al^{-2J_3}$
in the trace.  For $\al=-1$ this is equivalent to a sign change
in the original variable $x$ before the $\de\to 0$ limit is taken.}}
\eqn\multipartbpscases{\eqalign{
Z^G_{\rm free}(u,\bar \y)
&{}={\ \prod_{j=1}^\infty}{1\over 
1-f_+^G(u^j,\bar \y^j)
f_-^G(u^j,\bar \y^j)}\,,\cr 
Z^G_{\rm sugra}(u,\bar \y)&{}=\exp\bigg({\sum_{j=1}^\infty}
{{1\over j}}Y^G_{\rm sugra}(u^j,\bar \y^j)\bigg)\, ,\cr
Z^H_{\rm free}(u,\bar s,\bar \y)
&{}={\prod_{j=1}^\infty}{1\over 1-f_+^H(u^j\al_j ,\bar s^j/\al_j, \bar \y^j)
f^H_-(u^j\al_j,\bar s^j/\al_j, \bar \y^j)}\bigg|_{\al_j=(-1)^{j+1}}\,, \cr
Z^H_{\rm sugra}(u,\bar s,\bar  \y)&{}=\exp\bigg({\sum_{j=1}^\infty}
{{1\over j}}Y^H_{\rm sugra}(u^j\al_j,\bar s^j/\al_j,\bar \y^j)
\Big|_{\al_j=(-1)^{j+1}}\bigg)\, .}
}

\noindent
{\bf BPS Cases}

\noindent
{\bf The $U(1)$ Sectors}

Corresponding to \limithalfBPS\  we
consider the $(3,B,+)$ limit, whereby
\eqn\bpsul{
\chi^{(3,B,+)}_{(r;0;r,r,\pm r)}(u)=\chi^{U(1)}_{(r)}(u)= u^{2r},
} 
so that 
$f^{U(1)}_+(u)=u$, $f^{U(1)}_-(u)=0$.
It is also easily seen that $Y^{U(1)}_{\rm sugra}(u)=0$
in the limit \limithalfBPS.
Thus, from  \multipartbpscases,
\eqn\partyo{
Z_{\rm free}^{U(1)}(u)=Z_{\rm sugra}^{U(1)}(u)=1\, ,
}
so that, clearly, there are no $(3,B,+)$ BPS gauge invariant operators
in the free field or supergravity spectrum,
apart from the identity operator.
The same result applies to $(3,B,-)$ BPS operators.

\noindent
{\bf The $U(1)\otimes U(1)$ Sector}

Corresponding to \limitbps\ for $N=3$, $m=2$, we have that, 
\eqn\bpslo{\eqalign{
\chi^{(3,B,2)}_{(r;0;r,r,q)}(u,y)&{}=\chi^{U(1)\otimes U(1)}_{(r;q)}(u,y)=u^{2r}y^{q}\, ,
\cr
\chi^{(3,B,\pm)}_{(r;0;r,r,\pm r)}(u,y)&{}=\chi^{U(1)\otimes U(1)}_{(r;\pm r)}(u,y)=u^{2r}y^{\pm r}\, ,
}}
so that $f^{U(1)\otimes U(1)}_\pm(u,y)= u y^{\pm {1\over 2}}$.
We also have that,
\eqn\ysugrasec{
Y^{U(1)\otimes U(1)}_{\rm sugra}(u,y)=\sum_{r=1}^\infty u^{2r}={u^2\over 1-u^2}\, .
}
Thus, from \multipartbpscases,
\eqn\largeNlo{
Z_{\rm free}^{U(1)\otimes U(1)}(u,y)=
Z_{\rm sugra}^{U(1)\otimes U(1)}(u,y)=\prod_{j\geq 1}{1\over 1-u^{2j}}\,.
}
Thus, expanding over the characters in \bpslo,
\eqn\numbero{
Z_{\rm free}^{U(1)\otimes U(1)}(u,y)=Z_{\rm sugra}^{U(1)\otimes U(1)}(u,y)
=1+\sum_{r=1}^\infty 
N^{(3,B,2)}_{r}\chi^{(3,B,2)}_{(r;0;r,r,0)}(u,y)\, ,
}
where
\eqn\fcaseo{
N^{(3,B,2)}_{r}=p(r)=1,2,3,5, \dots,\quad r=1,2,3,4,\dots\, ,
}
where $p(r)$ is the usual partition number for $r$. 
The agreement \largeNlo\ is expected as $(3,B,2)$ operators
are protected due to long multiplet decomposition rules discussed in 
section 4.

\noindent
{\bf The $U(1)\otimes SO(4)$ Sector}

Due to \decompimp, we expect disagreement between counting of
$(3,B,1)$ operators in the free field and supergravity limits and thus
we split the discussion here.
Corresponding to \limitbps\ for $N=3$, $m=1$, we have that, 
\eqn\bpslo{\eqalign{
\chi^{(3,B,1)}_{(r;0;r,q,p)}(u,u_+,u_-)
&{}=\chi^{U(1)\otimes SO(4)}_{(r;q,p)}(u,u_+,u_-)=u^{2r}
\chi_{q+p}(u_+)\chi_{q-p}(u_-)\, ,\cr
\chi^{(3,B,2)}_{(r;0;r,r,p)}(u,u_+,u_-)&{}
=\chi^{U(1)\otimes SO(4)}_{(r;r,p)}(u,u_+,u_-)=u^{2r}
\chi_{r+p}(u_+)\chi_{r-p}(u_-)\, ,\cr
\chi^{(3,B,\pm )}_{(r;0;r,r,\pm r)}(u,u_+,u_-)&{}
=\chi^{U(1)\otimes SO(4)}_{(r;r,\pm r)}(u,u_+,u_-)
=u^{2r}\chi_{2r}(u_{\pm})\, ,}
}
in terms of $SU(2)$ characters in \chars, using $SO(4)\simeq SU(2)\otimes SU(2)$,
so that, for $SO(4)$ characters,
\eqn\sofourchs{
\chi^{(2)}_{(q,p)}({\bar y}_1,{\bar y}_2)= \chi_{q+p}(u_+)\chi_{q-p}(u_-)\, ,\qquad
{\bar y}_1=u_+u_-\, ,\quad {\bar y}_2=u_+/u_-\, .
}

We have $f^{U(1)\otimes SO(4)}_{\pm}(u)= u\chi_1(u_\pm)$, where
$\chi_1(u_\pm)=u_\pm+u_\pm{}^{-1}$, so that,
 from \multipartbpscases,
\eqn\largeNlo{
Z_{\rm free}^{U(1)\otimes SO(4)}(u,u_+,u_-)=
\prod_{j=1}^{\infty}{1\over 1-u^{2j}\chi_1(u_+{}^j)\chi_1(u_-{}^j)}\, .
}
The numbers of multi-trace $(3,B,1)$ BPS operators 
are then determined
by expansions over the characters \bpslo, using \fcaseo,
\eqn\chexp{\eqalign{
&Z_{\rm free}^{U(1)\otimes SO(4)}(u,u_+,u_-)\cr
&{}=1+\sum_{r=1}^\infty p(r)
\chi^{(3,B,2)}_{(r;0;r,r,0)}(u,u_+,u_-)  +
\sum^\infty_{r,q, |p|\geq 0}
N^{(3,B,1)}_{{\rm free},(r,q,p)}\,\chi^{(3,B,1)}_{(r;0;r,q,p)}(u,u_+,u_-)\, ,}
}
for which formulae are obtained in appendix B.  
We may determine for the first few cases, for $r=0,1,2,\dots,$
\eqn\fcaset{
N^{(3,B,1)}_{{\rm free},(r+1 , r, \pm 1)}
=\sum_{j=0}^{r}p(j)-p(r{+1})=0,1,2,5,8\dots \, ,
}
and 
\eqn\fcaseth{
N^{(3,B,1)}_{{\rm free},(r+2,r, 0)}=
2\sum_{j=1}^{r}\sum_{k=0}^{j}p(k)-\sum_{j=0}^{r+1}p(j)+p(r{+2})
=2,5,12,23,44,\dots \, ,
}
along with,
\eqn\fcasethth{
N^{(3,B,1)}_{{\rm free},(r+4,r+2, \pm 2)}=
2\sum_{j=1}^{r}\sum_{k=0}^{[{j\over 2}]}p({\ts{1\over 2}}(1{-(-1)}^j)+2k)+
\sum_{j=0}^{r+2}p(j)-p(r{+3})=3,6,15,26,49,\dots \, .
}

Note that generally, from results in appendix
B, $N^{(3,B,1)}_{{\rm free},(r,q,p)}$ is a 
potentially non-zero
integer only for $N^{(3,B,1)}_{{\rm free},(r, r{-s}{-t},s{-t})}$, $r\in \Bbb{N}$,
$s,t=0,\dots, [r/2]$, $s\, t\neq 0$.

(From \jumbawaba, $(r,r{-s}{-t},t{-s})\to [r{-2s},s{+t},r{-2 t}]$, in terms 
of $SU(4)$ Dynkin labels.)

For the supergravity limit we may determine, 
\eqn\sugrajni{
Y^{U(1)\otimes SO(4)}_{\rm sugra}(u,u_+,u_-)=
\sum_{r=1}^\infty \chi^{(3,B,2)}_{(r;0;r,r,0)}(u,u_+,u_-)
={1-u^4\over \prod_{\vep,\eta=\pm 1}(1-u^2 u_+{}^{\vep}
u_-{}^{\eta})}-1\, ,
}
so that, from \multipartbpscases,
\eqn\multigrav{
Z^{U(1)\otimes SO(4)}_{\rm sugra}(u,u_+,u_-)=\prod_{j=1}^{\infty}
{1\over \prod_{k,l=0}^j(1-u^{2j}u_+{}^{2k-j}
u_-{}^{2l-j})}\, .
}
Again we may expand, 
\eqn\chexpo{\eqalign{
& Z_{\rm sugra}^{U(1)\otimes SO(4)}(u,u_+,u_-)\cr
&{}=1+\sum_{r=1}^\infty p(r)
\chi^{(3,B,2)}_{(r;0;r,r,0)}(u,u_+,u_-)  +
\sum^\infty_{r,q, |p|\geq 0}
N^{(3,B,1)}_{{\rm sugra},(r,q,p)}\,\chi^{(3,B,1)}_{(r;0;r,q,p)}(u,u_+,u_-)\, ,}
}
to find for the first few cases, using results from appendix B, for
$r=0,1,2,\dots,$ 
\eqn\uquarto{
N^{(3,B,1)}_{{\rm sugra},(r+1 ,r,\pm 1)}=N^{(3,B,1)}_{{\rm free},(r+1 ,r,\pm 1)}
}
while,
\eqn\fcasethuip{
N^{(3,B,1)}_{{\rm sugra},(r+2, r, 0)}=
\sum_{j=0}^{r}\sum_{k=0}^{j}p(k)-\sum_{j=0}^{r+1}p(j)+p(r{+2})=1,2,5,9,18,\dots\, ,
}
along with,
\eqn\fcaseththo{
N^{(3,B,1)}_{{\rm sugra},(r+4,r+2, \pm 2)}=
\sum_{j=1}^{r}\sum_{k=0}^{[{j\over 2}]}p({\ts{1\over 2}}(1{-(-1)}^j)+2k)+
\sum_{j=0}^{r+2}p(j)-p(r{+3})=2,4,10,17,32,\dots \, .
}

As emphasised, the matching \uquarto\ is expected from \decompimp\
along with the free field restrictions implied for general $r,q,p$ in 
$N^{(3,B,1)}_{{\rm{free}},(r,q,p)}$ mentioned
above. 

 The first few numbers of operators may be easily obtained
by performing series expansions to low orders in $u$ and using the
orthogonality relation for $SU(2)$ characters in appendix B
and are listed in the following table.

\medskip
\vbox{
\hskip0.5cm Table 3
\nobreak

\hskip0.5cm
\vbox{\tabskip=0pt \offinterlineskip
\hrule
\halign{&\vrule# &\strut \ \hfil#\  \cr
height2pt&\multispan5 &\cr
&\multispan5 \hfil BPS primary operators \hfil&\cr
height2pt&\multispan5 &\cr
&\multispan5\hrulefill& \cr
height2pt&\omit&&\omit&&\omit&\cr
&\ $\De$ \ \hfil   &&\  Supergravity limit \hfil &&
Remaining operators \ \ \hfil &\cr
height2pt&\omit&&\omit&&\omit&\cr
\noalign{\hrule}
height2pt&\omit&&\omit&&\omit&\cr
& \ 1  \ \hfil && \ $\R_{(1,1,0)}$ \hfil &
& \  \ \hfil &\cr
height1pt&\omit&&\omit&&\omit&\cr
& \ 2  \ \hfil&& \ $2\R_{(2,2,0)}$,
$\R_{(2,0,0)}$ \hfil &
& \ $\R_{(2,0,0)}$ \ \hfil &\cr
height1pt&\omit&&\omit&&\omit&\cr
& \ 3  \ \hfil&& \ $3\R_{(3,3,0)}$,
$\R_{(3,2,\pm 1)}$,  $2\R_{(3,1,0)}$ \hfil &
& \ $3\R_{(3,1,0)}$ \ \hfil &\cr
height1pt&\omit&&\omit&&\omit&\cr
& \ 4  \ \hfil&& \ $5\,\R_{(4,4,0)}$,
$2\,\R_{(4,3,\pm 1)}$, $2\,\R_{(4,2,\pm 2)}$,
$5\,\R_{(4,2,0)}$\hfil  &
& \ $\R_{(4,2,\pm 2)}$,
$7\R_{(4,2,0)}$\ \hfil &\cr
&&& \ $\R_{(4,1,\pm 1)}$,
$3\R_{(4,0,0)}$ \ \hfil  &
& \   $3\R_{(4,1,\pm 1)}$,
$6\R_{(4,0,0)}$ \ \hfil &\cr
height2pt&\omit&&\omit&&\omit&\cr
}
\hrule}

{\eightpoint
{\parindent 1.5cm{\narrower
\noindent
BPS primary operators with 
conformal dimensions $\De$ belonging to $SO(6)$ 
representations $\R_{(r,q,p)}$, 
as obtained from expansion of partition functions. 
(For the free field case, the extra operators appear
in the rightmost column.)

}}}}

Note that more generally, as may be easily argued from results in 
appendix B, $N^{(3,B,1)}_{{\rm sugra},(r, q, p)}$ is potentially non-vanishing
only for $N^{(3,B,1)}_{{\rm sugra},(r,r{-s}{-t},s{-t})}$,
$r\in \Bbb{N}$, $s,t=0,\dots,[r/2]$, 
$s\,t\neq 0$, consistent with the free field theory result.

\noindent
{\bf Semi-short Cases}

\noindent
{\bf The $U(1)\otimes SU(1,1)$ Sectors}

In these sectors, where the relevant limits in characters are given by
\limithalfBPSoppp, for $N=3$, fermion contributions become
important in multi-particle partition functions.

The surviving characters, for the $(N,A,+)$ limit, are given by,
\eqn\survivesemi{\eqalign{
\chi^{(3,B,-)}_{({1\over 2};0;{1\over 2},{1\over 2},-{1\over 2})}
(u,\bar s)={u^{1\over 2}\bar s^{3\over 2}\over 1-\bar s^2}\, ,&
\qquad
\chi^{(3,B,+)}_{(r;0;r,r,r)}
(u,\bar s)={u^{r}\bar s^{r}\over 1-\bar s^2}\, ,\cr
\chi^{(3,B,2)}_{(r;0;r,r,r-1)}
(u,\bar s)={u^{r}\bar s^{r+1}\over 1-\bar s^2}\, ,&\qquad
\chi^{(3,B,1)}_{(r;0;r,r-1,r-1)}
(u,\bar s)={u^{r}\bar s^{r+2}\over 1-\bar s^2}\, ,\cr
\chi^{(3,{\rm cons.})}_{(j+1;j;0,0,0)}
(u,\bar s)={u\,\bar s^{2j+4}\over 1-\bar s^2}\, ,&\qquad
\chi^{(3,A,+)}_{(r+j+1;j;r,r,r)}
(u,\bar s)={u^{r+1}\bar s^{r+2j+4}\over 1-\bar s^2}\, .}
}

Thus, due to $f^{U(1)\otimes SU(1,1)}_+(u,\bar s)=(u \bar s)^{1\over 2}/(1-\bb^2)$,
$f^{U(1)\otimes SU(1,1)}_-(u,\bar s)=u^{1\over 2} \bar s^{3\over 2}/(1-\bb^2)$, 
from \multipartbpscases,
\eqn\largeNloppppp{
Z_{\rm free}^{U(1)\otimes SU(1,1)}(u,\bar s)=\prod_{j=1}^{\infty}
{1\over 1+(-1)^j u^j\,\bb^{2j}(1-\bb^{2j})^{-2}}\, .
}
Expanding in $u$ we find,
\eqn\expansionul{\eqalign{
Z_{\rm free}^{U(1)\otimes SU(1,1)}(u,\bar s)={}& 1+{u \, s^2\over (1-s^2)^2}+
{4 u^2\,s^6\over(1-s^2)^2(1-s^4)^2}+O(u^3,s^6)\cr
={}&   1+\chi^{(3,B,2)}_{(1;0;1,1,0)}
(u,\bar s)+\sum_{j=0}^\infty\chi^{(3,{\rm cons.})}_{(j+1;j;0,0,0)}
(u,\bar s)\cr
{}&  +\sum_{r=1}^\infty\sum_{j={r\over 2}-1\atop
j\geq 0}^\infty N_{{\rm free},(r,j)}^{(3,A,+)}\, \chi^{(3,A,+)}_{(r+j+1;j;r,r,r)}
(u,\bar s)\, .}
}

It appears non-trivial to determine $N_{{\rm free},(r,j)}^{(3,A,+)}$
generally, however we may easily determine,
\eqn\reqwith{\eqalign{
N_{{\rm free},(1, j)}^{(3,A,+)}&{}=2 [\half j+1][\half j+2]=4,4,12,12,\dots,
\qquad j=\half,{\ts{3\over 2}},{\ts{5\over 2}},{\ts{7\over 2}}\, ,\dots,\cr
N^{(3,A,+)}_{{\rm free},(r, {r\over 2}-1)}&{}=1\, ,\quad 
r=2,3,4,\dots,
\qquad  N^{(3,A,+)}_{{\rm free},(r,{r\over 2}-1+n)}>1\, ,\quad n=1,2,3,\dots\, .}
}

In this limit, the supergravity single particle partition function reduces to,
\eqn\trippety{
Y_{\rm sugra}^{U(1)\otimes SU(1,1)}(u,\bar s)=\chi^{(3,B,2)}_{(1;0;1,1,0)}(u,\bar s)
={u\,\bb^2\over 1-\bb^2}\, ,
}
so that, from \multipartbpscases,
\eqn\hippety{
Z_{\rm sugra}^{U(1)\otimes SU(1,1)}(u,\bar s)=\prod_{j=1}^{\infty}(1+u \,\bb^{2j})\, ,
}
which is a reflection of $Y_{\rm sugra}^{U(1)\otimes SU(1,1)}(u,\bar s)$
receiving purely fermionic operator contributions (it is odd under $(u,\bar s)\to -(u,\bar s)$).

Using the identities,
\eqn\usefulpentag{
\prod_{j=1}^\infty(1+z q^j)=\sum_{n=0}^\infty {q^{{1\over 2}n(n+1)}z^n\over 
\prod_{j=1}^n(1-q^j)}\, ,\qquad \prod_{j=2}^n{1\over 1-q^j}
=\sum_{m=0}^\infty P_n(m)q^m\, ,
}
where $P_n(m)$ is the number of partitions of $m$ into no more that $n$ parts,
each part $\geq 2$, we may write,
\eqn\exparnsion{
Z_{\rm sugra}^{U(1)\otimes SU(1,1)}(u,\bar s)=1+\chi^{(3,B,2)}_{(1;0;1,1,0)}(u,\bar s)
+\sum_{r=1}^\infty \sum_{j={1\over 2}r^2+r-1}^\infty 
N^{(3,A,+)}_{{\rm sugra},(r, j)}\, \chi^{(3,A,+)}_{(r+j+1;j;r,r,r)}
(u,\bar s)\, ,
}
where,
\eqn\determinecoeffs{
N^{(3,A,+)}_{{\rm sugra},(r, j)}=P_{r+1}(j+1-\half r^2-r)\, .
}
Using the foregoing we may find relatively simple formulae for at least the
first few cases,
\eqn\somevalues{\eqalign{
N^{(3,A,+)}_{{\rm sugra},(1, j)}&{}=\cases{1\, , & $j-{\ts{1\over 2}}
= 0\,\,  {\rm mod} \,\, 2$,
\cr 0\, , & 
$j-{\ts{3\over 2}}= 0\,\,  {\rm mod} \,\, 2$, }
\quad j={\ts{1\over 2}},{\ts{3\over 2}}, \dots, \cr
N^{(3,A,+)}_{{\rm sugra},(2, j)}&{}=
1,0,1,1,\dots=\cases{[{\ts{1\over 6}} j-\half]\, , & 
$j=4\,\,{\rm mod}\,\, 6$,
\cr [{\ts{1\over 6}} j+\half]\, , & 
otherwise,}\quad j=3,4,5,6,\dots, \cr
N^{(3,A,+)}_{{\rm sugra},(r, {1\over 2}r^2+r-1)}&{}=1\, .}
}

Assuming that the partition functions in the  supergravity limit
receive contributions from only protected operators then a
number of observations from \expansionul, \reqwith, \exparnsion\ and 
\somevalues\ are
evident.  

First, there are no $\chi^{(3,B,1)}_{(r;0;r,r-1,r-1)}
(u,\bar s)$ contributions and one contribution
from $\chi^{(3,B,2)}_{(r;0;r,r,r-1)}(u,\bar s)$, for $r=1$, both consistent
with previous analysis for BPS cases.
Second, the conserved current contributions in \expansionul\
disappear from the spectrum in
the strong coupling limit, as happens for $\N=4$ super Yang Mills.
Third, there is one scalar $(3,A,+)$ semishort primary
operator in the $SO(6)$ representation $\R_{(2,2,2)}$,
and its conformal descendants,
contributing in the free field theory limit, due to
$N^{(3,A,+)}_{{\rm free},(2, 0)}=1$ in 
\expansionul, and no such contribution in the supergravity limit,
from \exparnsion.  In the interacting theory, these operators, 
from the long multiplet
decomposition formulae \decomplump,  must then pair up with a
$(3,B,1)$ BPS primary operator
in the $SO(6)$ representation $\R_{(4,2,2)}$ ,
and its descendants.
This is consistent with Table 3 as there is precisely one such
remaining $(3,B,1)$ primary operator.  (Similarly, the scalar conserved current,
and
its descendants,
contributing to \expansionul\ pairs with the single 
$(3,B,1)$ BPS primary operator in the  $SO(6)$ representation
$\R_{(2,0,0)}$, and
its descendants, 
listed in Table 3.)  

Note that the analysis for $(3,A,-)$ semishort operators
is very similar and gives the same result for counting
of BPS and conserved current multiplets along with,
\eqn\numbersd{
N^{(3,A,-)}_{{\rm free},(r, j)}=N^{(3,A,+)}_{{\rm free},(r, j)}\, ,
\qquad N^{(3,A,-)}_{{\rm sugra},(r, j)}=N^{(3,A,+)}_{{\rm sugra},(r, j)}\, .
}
In particular, again there is one 
scalar $(3,A,-)$ primary semishort operator
contributing in the free field theory limit, absent from the supergravity
spectrum,  that pairs up with  a
$(3,B,1)$ BPS primary operator in the $\R_{(4,2,-2)}$ $SO(6)$ representation,
consistent with Table 3.

\noindent
{\bf The $U(1)\otimes Osp(2|2)$ Sector}

In the limit \limitSemishort, for $N=3$, $m=2$, the
non-vanishing characters are, defining $Q(\bb ,y)=(1+\bb y)(1+\bb y^{-1})/(1-\bb^2)$,
\eqn\jcharsd{\eqalign{
\chi^{(3,B,\pm)}_{(r;0;r,r,\pm r)}
(u,\bar s,y)={u^{r}\bar s^{r}\over 1-\bb^2}y^{\pm r}(1+\bar s\,y^{\mp 1})
\, ,&\qquad
 \chi^{(3,B,2)}_{(r;0;r,r,q)}
(u,\bar s,y)={(u\,\bb)^r}\, y^q\, Q(\bb,y) \, ,\cr
\chi^{(3,B,1)}_{(r;0;r,r-1,q)}
(u,\bar s,y)=u^r{\bb^{r+1}}\, y^q\, Q(\bb,y)
\, , &\qquad
\chi^{(3,{\rm cons.})}_{(j+1;j;0,0,0)}
(u,\bar s,y)=u \,{\bb^{2j+3}} Q(\bb,y)
\, ,\cr
\chi^{(3,A,\pm)}_{(r+j+1;j;r,r,\pm r)}
(u,\bar s,y)&{}=u^{r+1}{\bb^{r+2j+3}}\, y^{\pm r}\, Q(\bb,y)
\, ,\cr
\chi^{(3,A,2)}_{(r+j+1;j;r,r,q)}
(u,\bar s,y)&{}=u^{r+1}{\bb^{r+2j+3}}\, y^{q}\, Q(\bb,y)\, .}
}

Using that $f_{\pm}^{U(1)\otimes Osp(2|2)}(u,\bar s,y)=(u\, \bb)^{1\over 2}
(y^{\pm {1\over 2}}+\bb\, y^{\mp {1\over 2}})/(1-\bar s^2)$
we have,
\eqn\freefieldobama{
Z_{\rm free}^{U(1)\otimes Osp(2|2)}(u,\bar s,y)=\prod_{j=1}^{\infty}
{1\over 1-u^j\,\bb^{j}(1-(-1)^j\bb^j y^j)(1-(-1)^j\bb^j y^{-j}) (1-\bb^{2j})^{-2}}\, .
}

Using the previous results for counting of BPS operators, 
we may determine,
\eqn\frippetybit{
W_{{\rm free}}^{(3,B,2)}(u,\bb,y)=\sum_{r=1}^\infty 
p(r)\chi^{(3,B,2)}_{(r;0;r,r,0)}(u,\bar s,y)=
\bigg(\prod_{k=1}^\infty {1\over 1-(u\,\bb)^j}-1\bigg)Q(\bb,y)\, ,}
for contributions from $(3,B,2)$ multiplets, using \fcaseo,
and
\eqn\gippetybit{\eqalign{
&W_{{\rm free}}^{(3,B,1)}(u,\bb,y)\cr
&{}=\sum_{r=1}^\infty\bigg(
N^{(3,B,1)}_{{\rm free},(r,r{-1},1)}\chi^{(3,B,1)}_{(r;0;r,r-1,1)}(u,\bar s,y)
+N^{(3,B,1)}_{{\rm free},(r,r{-1},-1)}\chi^{(3,B,1)}_{(r;0;r,r-1,-1)}(u,\bar s,y)\bigg)\cr
&{}=\bb (y+y^{-1})
\bigg({2u\bb-1\over 1-u\bb}\prod_{k=1}^\infty {1\over 1-(u\,\bb)^j}+1\bigg)Q(\bb,y)\, ,}
}
for contributions of $(3,B,1)$ multiplets, using
the form of the corresponding characters in \jcharsd, along with \fcaset.
Similarly, using the free field results for the $U(1)\otimes SU(1,1)$ sector in 
\expansionul, we may determine,
\eqn\bippetybit{
W_{{\rm free}}^{(3,{\rm cons.})}(u,\bb,y)
=\sum_{j=0}^\infty\chi^{(3,{\rm cons.})}_{(j+1;j;0,0,0)}(u,\bar s,y)=
{u\,\bb^3\over 1-\bb^2}Q(\bb,y)\, ,
}
for conserved current multiplet contributions and,
\eqn\aippetybit{\eqalign{
&W_{{\rm free}}^{(3,A,\pm)}(u,\bb,y)\cr
&=\sum_{r=1}^\infty\sum_{j={r\over 2}-1\atop
j\geq 0}^\infty N_{{\rm free},(r,j)}^{(3,A,\pm)}\, \chi^{(3,A,\pm)}_{(r+j+1;j;r,r,\pm r)}
(u,\bar s,y)\cr
&=y^{\mp 1}\bb^{-1}(1+\bb\, y)(1+\bb\,y^{-1})
\bigg(\prod_{k=1}^\infty {1\over 1+(-1)^j(u y^{\pm 1} \bb^2)^j(1{-\bb^{2j}})^{-2}}
-{u\,y^{\pm 1}\bb\over (1-\bb^2)^2}-1\bigg)\, ,}
}  
for $(3,A,\pm)$ semishort multiplets, evident by using \expansionul,
\numbersd, and the form of the corresponding characters in \jcharsd.
We may then determine numbers $N_{{\rm free},(r,j,q)}^{(3,A,2)}$
 of $(3,A,2)$ semishort operators
from,
\eqn\garrulous{\eqalign{
Z_{\rm free}^{U(1)\otimes Osp(2|2)}(u,\bar s,y)
={}&1+W_{{\rm free}}^{(3,B,2)}(u,\bb,y)
+W_{{\rm free}}^{(3,B,1)}(u,\bb,y)
+W_{{\rm free}}^{(3,{\rm{cons.}})}(u,\bb,y)\cr
&
+W_{{\rm free}}^{(3,A,+)}(u,\bb,y)
+W_{{\rm free}}^{(3,A,-)}(u,\bb,y)\cr
&+\sum_{r\geq 1,2j\geq 0
\atop 0\leq |q|<r }N_{{\rm free},(r,j,q)}^{(3,A,2)}\,\chi^{(3,A,2)}_{(r+j+1;j;r,r,q)}
(u,\bar s,y)\, .}
}
Finding general formulae for  $N_{{\rm free},(r,j,q)}^{(3,A,2)}$
is nontrivial, however series expansion of \garrulous, using
Mathematica, suggests the following results for particular cases,
for $r=0,1,2,\dots$,
\eqn\particularint{
N_{{\rm free},(r+1,0,0)}^{(3,A,2)}=2\sum_{j=0}^{r}\sum_{k=0}^jp(k)+
\sum_{j=0}^{r+1}p(j)=4,10,21,40,71,\dots\, ,
}
along with,
\eqn\particularinto{
N_{{\rm free},(r+3,0,\pm 2)}^{(3,A,2)}=2\sum_{j=0}^{r+1}
\sum_{k=0}^{[{j\over 2}]}p({\ts{1\over 2}}(1{-(-1)}^j)+2k)
-\sum_{j=0}^{r+3}p(j)+p(r{+4})=2,5,10,19,33,\dots\, .
}

For the $U(1)\otimes Osp(2|2)$
sector, the supergravity single particle partition function reduces to,
\eqn\trippety{
Y_{\rm sugra}^{U(1)\otimes Osp(2|2)}(u,\bar s,y)=\sum_{r=1}^\infty
\chi^{(3,B,2)}_{(r;0;r,r,0)}(u,\bar s,y)={u\, \bb
(1+\bb\, y)(1+\bb\,y^{-1})\over (1-\bb^2)(1-u\, \bb)}
\, ,
}
so that, from \multipartbpscases,
\eqn\hippety{
Z_{\rm sugra}^{U(1)\otimes Osp(2|2)}(u,\bar s,y)
=\prod_{j,k=1}^\infty {(1+u^{j} \bb^{j+2k-1}y)(1+u^{j} \bb^{j+2k-1}y^{-1})
\over (1-u^{j} \bb^{j+2k-2})(1-u^{j} \bb^{j+2k})}\, .
}

This time, we may determine,
\eqn\vippetybit{\eqalign{
& W_{{\rm sugra}}^{(3,B,2)}(u,\bb,y)
=W_{{\rm free}}^{(3,B,2)}(u,\bb,y)\, ,\cr
& W_{{\rm sugra}}^{(3,B,1)}(u,\bb,y)
=W_{{\rm free}}^{(3,B,1)}(u,\bb,y)\, ,\qquad  
W_{{\rm sugra}}^{(3,{\rm cons.})}(u,\bb,y)
=0\, ,}
}
due to the relevant sector of BPS operators remaining
protected and the conserved current multiplet operators
disappearing from the spectrum, as shown previously.
Similarly, using the supergravity limit results for the $U(1)\otimes SU(1,1)$ sector in 
\exparnsion, we may determine,
\eqn\nippetybit{\eqalign{
&W_{{\rm free}}^{(3,A,\pm)}(u,\bb,y)\cr
&=\sum_{r=1}^\infty\sum_{j={r\over 2}-1\atop
j\geq 0}^\infty N_{{\rm sugra},(r,j)}^{(3,A,\pm)}\, \chi^{(3,A,\pm)}_{(r+j+1;j;r,r,\pm r)}
(u,\bar s,y)\cr
&=y^{\mp 1}\bb^{-1}(1+\bb\, y)(1+\bb\,y^{-1})
\bigg(\prod_{k=1}^\infty (1+u\,y^{\pm 1}\,\bb^{2j})
-{u\,y^{\pm 1}\bb\over (1-\bb^2)^2}-1\bigg)\, ,}
}
for $(3,A,\pm)$ semishort multiplets, evident by using also
\numbersd, and the form of the corresponding characters in \jcharsd.
We may then determine numbers $N_{{\rm sugra},(r,j,q)}^{(3,A,2)}$
 of $(3,A,2)$ semishort operators
from,
\eqn\garrulous{\eqalign{
Z_{\rm sugra}^{U(1)\otimes Osp(2|2)}(u,\bar s,y)
={}&1+W_{{\rm sugra}}^{(3,B,2)}(u,\bb,y)
+W_{{\rm sugra}}^{(3,B,1)}(u,\bb,y)\cr
&
+W_{{\rm sugra}}^{(3,A,+)}(u,\bb,y)
+W_{{\rm sugra}}^{(3,A,-)}(u,\bb,y)\cr
&+\sum_{r\geq 1,2j\geq 0
\atop 0\leq |q|<r }N_{{\rm sugra},(r,j,q)}^{(3,A,2)}\,\chi^{(3,A,2)}_{(r+j+1;j;r,r,q)}
(u,\bar s,y)\, .}
}
Again, general formulae for  $N_{{\rm sugra},(r,j,q)}^{(3,A,2)}$
seem nontrivial to obtain, however using
Mathematica suggests, for particular cases, for $r=0,1,2,\dots$,
\eqn\iuytropi{
N_{{\rm sugra},(1,j,0)}^{(3,A,2)}
=1\, ,\qquad j=0,1,2,\dots\, ,
}
and
\eqn\particularinty{
N_{{\rm sugra},(r+1,0,0)}^{(3,A,2)}=\sum_{j=0}^{r}\sum_{k=0}^jp(k)
=1,3,7,14,26,\dots\, ,
}
along with,
\eqn\particularintyo{
N_{{\rm sugra},(r+3,0,\pm 2)}^{(3,A,2)}=\sum_{j=0}^{r+1}
\sum_{k=0}^{[{j\over 2}]}p({\ts{1\over 2}}(1{-(-1)}^j)+2k)
-\sum_{j=0}^{r+3}p(j)+p(r{+4})=0,0,1,2,5,\dots\, .
}

A consistency check is provided by \decompump\ which implies that
the unprotected scalar $(3,A,2)$
semishort primary operators in $SO(6)$ representation $\R_{(r,r,0)}$, 
respectively $\R_{(r,r,\pm 2)}$,
and their conformal descendants, counted
above, should pair with unprotected $(3,B,1)$ BPS primary operators
in the $SO(6)$ representation $\R_{(r+2,2,0)}$, 
respectively $\R_{(r+2,r,\pm 2)}$, and
their descendants, counted previously.
Using \fcaseth, \fcasethuip, \particularint\ and \particularinty,
along with \fcasethth, \fcaseththo,  \particularint\ and \particularintyo,
we find,
\eqn\agreementonecase{\eqalign{
N^{(3,B,1)}_{{\rm free},(r+2, r, 0)}-
N^{(3,B,1)}_{{\rm sugra},(r+2, r, 0)}&{}=
N_{{\rm free},(r,0,0)}^{(3,A,2)}-
N_{{\rm sugra},(r,0,0)}^{(3,A,2)}=\sum_{j=0}^{r}\sum_{k=0}^j p(k)\, ,\cr
N^{(3,B,1)}_{{\rm free},(r+2, r, \pm 2)}-
N^{(3,B,1)}_{{\rm sugra},(r+2, r, \pm 2)}&{}=
N_{{\rm free},(r,0,\pm 2)}^{(3,A,2)}-
N_{{\rm sugra},(r,0,\pm 2)}^{(3,A,2)}\cr
&{}=\sum_{j=0}^{r-2}
\sum_{k=0}^{[{j\over 2}]}p({\ts{1\over 2}}(1{-(-1)}^j)+2k)\, ,}
}
expressing perfect agreement with this expectation.

\noindent
{\bf The $U(1)\otimes Osp(4|2)$ Sector}

This sector is nontrivial to analyse in similar terms as above, 
due to necessary and nontrivial expansions over $SO(4)$ characters,
as done for the $U(1)\otimes SO(4)$ sector above, so here are simply
given formulae for \singpartbpssemicases.
Using \limitsemio,  for $N=3$, $m=1$, with \sofourchs, we have,
\eqn\charform{\eqalign{
&\chi^{(3,B,2)}_{(r;0;r,r,0)}(u,\bb,u_+,u_-)=
\chi^{(U(1)\otimes Osp(2|2),2)}_{(r;r;r,0)}(u,\bb,u_+,u_-)\cr
&={(u\,\bb)^r\over 1-\bb^2}\sum_{\vep,\eta=\pm 1}{(u_+){}^{\vep r}(u_-){}^{\eta r}
(1+\bb\,u_+{}^{\vep}u_-{}^{-\eta})(1+\bb \,u_+{}^{-\vep}u_-{}^{\eta})
(1+\bb \,u_+{}^{-\vep}u_-{}^{-\eta})\over (1-u_+{}^{-2\vep})(1-u_-{}^{-2\eta})}\, ,}
}
so that, with \limitsemihalfo,
\eqn\yupioloi{\eqalign{
&f^{U(1)\otimes Osp(4|2)}_{\pm}(u,\bb,u_+,u_-)
={(u\,\bb)^{1\over 2}\over 1-\bb^2}\Big(\chi_1(u_\pm)+\bb
\chi_1(u_\mp)\Big)\, ,\cr
& Y^{U(1)\otimes Osp(4|2)}_{\rm sugra}(u,\bb,u_+,u_-)\cr
&{}={1\over 1-\bb^2}\sum_{\vep=\pm 1}{(1+u \,\bb^2u_+{}^{2\vep})
(1+\bb (u_+\,u_-)^{-\vep})(1+\bb(u_+/u_-)^{-\vep})\over
(1-u_+{}^{-2\vep})
(1-u\,\bb (u_+\,u_-)^{\vep})(1-u\,\bb(u_+/u_-)^{\vep})}-1
\, .}
}
\yupioloi\
is consistent with \bjvg\ in the limit \limitSemishort\ and the formula
in the second line can also be
shown to be symmetric under exchange of $u_+$, $u_-$,
as is necessary.

\newsec{Conclusion}

While much progress has been made recently in terms of determining 
the spectra of superconformal field theories, there are many open questions,
for instance, for the new superconformal Chern Simons theories 
or $\N=4$ super Yang Mills.

Focussing on the former, while the spectra of the $\N=6$ superconformal
Chern Simons theory
at zero (effective) 't Hooft coupling and in the large $n,k$ limits has been partially
addressed here by use of character methods, it may be interesting to
investigate operator counting for finite $n,k$.  

For large $n,k$,
the results here provide extra confirmation of expectations that
the primary operators dual to Kaluza Klein modes,
and multi-traces of these operators, in $[r,0,r]$ $SU(4)$ representations,
conformal dimension $r$, are protected, as argued in another way
in \abjm.  Also, the counting here implies that these
 are the only 
gauge invariant multi-trace primary operators in the $(3,B,2)$ superconformal
representation.  

For the $(3,B,1)$ representations, the only gauge invariant
primary operators belong to $[r{-2s},s{+t},r{-2t}]$, $s,t=0,\dots, [r/2]$, $s\,t\neq 0$,
$SU(4)$ representations, for which
generating functions for counting are given in
appendix B.  Furthermore, in accord with long multiplet
decomposition rules, counting of  $(s,t)=(1,0)$ and $(0,1)$ cases
shows matching between the free field and supergravity limits,
providing a consistency check of the character procedure used here and
further evidence, perhaps, in favour of the duality proposed in \abjm. 

Turning to semishort cases, conserved current multiplet
operators disappear from the spectrum in the supergravity limit
while the primary operators for $(3,A,\pm)$
semishort operators, belonging to $[2r,0,0]$, $[0,0,2r]$ $SU(4)$ representations,
have spins $j\geq \half r-1$ in the free field limit, while for the supergravity
limit,  $j\geq \half r^2+r-1$, a reflection of the very simple partition function
obtained in \hippety.

While only partial counting is obtained here for
semishort operators, the numbers of
primary operators obtained are consistent with the
following formula, implied by  \decompump, \decomplump, \decomplumpop,
\eqn\followingformula{
N^{\M}_{{\rm prot},(r,q)}=N^{\M}_{{\rm free},(r,q)}
-N^{(3,B,1)}_{{\rm free},(r+2,r,q)}+N^{(3,B,1)}_{{\rm prot},(r+2,r,q)}\, ,
}
where
$N^{\M}_{{\rm prot},(r,q)}$, $N^{\M}_{{\rm free},(r,q)}$
denote numbers of corresponding protected/free {\it scalar}
semi-short primary operators, in appropriate 
$\R_{(r,r,q)}$
$SO(6)$ representations (so that 
for $\M=(3,{\rm cons.})$, then $r,q=0$, for $\M=(3,A,\pm)$ then $r=\pm q\neq 0$
and for $\M=(3,A,2)$ then $r>|q|>0$)
while   $N^{(3,B,1)}_{{\rm prot},(r+2,r,q)}$,
$N^{(3,B,1)}_{{\rm free},(r+2,r,q)}$ denote numbers of 
protected/free primary operators in $(3,B,1)$ BPS multiplets,
in $\R_{(r+2,r,q)}$
$SO(6)$ representations.

Note also that, for free field theory in the large $n$ limit,
 $r,q$ are further restricted to $r=2$, $q=\pm 2$ for the $\M=(3,A,\pm)$
cases and $r\in \Bbb{N}$, $q=0,\pm 2$, $r>|q|$, for the $\M=(3,A,2)$ cases.

\followingformula\ implies that
it is possible to compute the 
partition function corresponding to such protected operators
using only free field theory and
the knowledge of which $(3,B,1)$ BPS operators remain protected.

In particular this applies to the 
$\N=6$ Chern Simons theory, having finite $n$ and large levels $k$,
in which case allowable $r,q$ in \followingformula, deriving from free field theory,
 may change.
The generating functions,
\eqn\doublecontour{\eqalign{
F_q(z)={}&\sum_{r\geq 0} N^{\M}_{{\rm free},(r,q)}z^r\, ,\cr
G_q(z)={}&\sum_{r\geq 0} \Big(N^{(3,B,1)}_{{\rm free},(r+2,r,q)}-
N^{(3,B,1)}_{{\rm prot},(r+2,r,q)}\Big)z^r\, ,}
}
may then be determined using the free field partition functions
$Z^{(n),U(1)\otimes Osp(2|2)}_{\rm free}(u,\bb,y)$, 
$Z^{(n),U(1)\otimes SO(4)}_{\rm free}(u,u_+,u_-)$, denoting \multip\
evaluated in the relevant $H\subset Osp(6|4)$ sector,
and $Z^{(n),U(1)\otimes SO(4)}_{\rm prot}(u,u_+,u_-)$, 
the partition function for protected
$(3,B,1)$ operators evaluated in the $U(1)\otimes SO(4)$ sector.{\foot{
It is evident from the form of the characters in \jcharsd, that,
$$
F_q(z)={1\over (2\pi i)^2} \oint\oint{1-x^2\over x^3 y^{q}(x+y)(1+x y)}
\bigg(Z^{(n),U(1)\otimes Osp(2|2)}_{\rm free}(z/x,x,y)-1\bigg)\d x\, \d y\, ,
$$
in terms of a double contour integral, where the contribution from
the identity operator is subtracted.  (Depending on free field constraints
for finite $n$ it may also be necessary to subract contributions from
other short multiplet representations as well.)  
$G_q(z)$ may, in 
principle, be computed in terms of a double contour integral similar
to that in appendix B, where $f(u,x,y)=
Z^{(n),U(1)\otimes SO(4)}_{\rm free}(u^{1\over 2},x,y)
-Z^{U(1)\otimes SO(4)}_{\rm prot}(u^{1\over 2},x,y)$ in (B.1).}}
(Presumably the latter partition function is equivalent to the chiral
ring partition function, for finite $n$.)

These generating functions may then be used to write,
using also \longsemi, \strider, \stridero\ for $N=3$,
\eqn\cinqport{\eqalign{
Z_q(s,x,y_1,y_2,y_3)={}&\sum_{r\geq 0}N^{\M}_{{\rm prot},(r,q)}
\chi^{(3,A,1)}_{(r+1;0;r,r,q)}(s,x,y_1,y_2,y_3)\cr
={}&{\frak W}^{(3)}\Big(H_q(s^2 y_1 y_2)\,y_3{}^q\,
C^{(3,A,1)}_{(1;0;0,0,0)}(s,x,y_1,y_2,y_3)\Big)\, ,\cr
H_q(z)={}& F_q(z)-G_q(z)\, ,}
}
for the partition function restricted to relevant protected 
scalar primary operators, and their superconformal
descendants.

Such partition functions should be consistent with the large $n$
result where, using that there are no relevant scalar $(3,{\rm cons.})$,
$(3,A,\pm)$  primary operators in the supergravity spectrum and
that $N^{(3,A,2)}_{{\rm prot},(r,q)}=N^{(3,A,2)}_{{\rm sugra},(r,0,q)}$
with  \particularinty, \particularintyo, then 
\eqn\numiopl{\eqalign{
H_0(z)={}& \sum_{r=1}^\infty
{N^{(3,A,2)}_{{\rm sugra},(r,0,0)}}\,z^r=
{z\over (1-z)^2}\prod_{k=1}^\infty{1\over 1-z^k}\, ,\cr
H_{\pm 2}(z)={}& \sum_{r=3}^\infty N^{(3,A,2)}_{{\rm sugra},(r,0,\pm 2)}\,z^r=
\bigg({z^2\over (1-z)(1-z^2)}-{1\over 1-z}+{1\over z}\bigg)
\prod_{k=1}^\infty{1\over 1-z^k}-{1\over z}\, ,}
}
with other $H_q(z)$ vanishing.

Turning to $\N=4$ super Yang Mills, with gauge group $U(N)$, there are still many
outstanding questions regarding operator spectra that perhaps may
be more easily answered for $\N=6$ superconformal Chern Simons
theories, due to the latter having fewer decoupled sectors, as demonstrated here.  
One concerns
finite $N$ counting for protected operators.  (Addressing this question may help
answer the difficult black hole entropy/microstate counting
problem.)

Recently, the latter issue received some attention in \grant, where, 
among other things,  the problem of 
counting certain gauge
invariant operators, 
consisting of products of
single trace chiral ring operators acted on by derivatives,  
was considered. 
(These operators should remain protected and
certainly enumerating them is interesting and worthwhile.)

The simpler problem of counting chiral ring operators 
has been addressed from the perspective of computing Hilbert series 
for the ring of symmetric polynomials in \bofeng\ 
(equivalently, the latter issue has been 
investigated from a perhaps 
more formal  perspective in \domokos)  whereby the non-trivial part of 
the computation is in taking account of syzygies (or trace identities in terms of matrices).  
The same difficulty applies to counting operators of the sort, referred to above,
considered in \grant\ and perhaps could be circumvented by a judicious choice
of basis for the operators.

In appendix C,
a more combinatorial approach is described using a natural basis for
multi-trace operators involving commuting matrices, the simplest sort of
chiral ring that is relevant for $\N=4$ super Yang Mills.
The technique involves
 employing the Polya enumeration theorem, applied to counting graph
colourings where the relevant graphs, in this case, have a particular wreath
product group automorphism symmetry.  (This is similar to the approach
used for counting single trace operators, in the large $N$ limit, for $\N=4$
SYM, \refs{\sund, \poly}, where the relevant graphs there were necklaces with
cyclic group automorphism symmetry.)  The result obtained by this approach 
(equivalent to the Hilbert series mentioned above)
is naturally expressed in terms of cycle index polynomials for the symmetric 
permutation group.  (For alternative polynomial expansions of
chiral ring partition functions, useful in the context
of asymptotics, see, for example, \luc.) 

It may be worth trying to find other combinatorial methods
of counting chiral ring operators more generally, perhaps
using known extensions of the Polya enumeration theorem,
taking into account symmetry in colours \harary, for example.
In any case, Hilbert series appear to have very interesting
connections with counting graph colourings that perhaps
may lead to even other extensions of the Polya 
enumeration theorem.

\noindent
{\bf{Acknowledgements}}

I am very grateful to Nicolas Boulanger, 
James Lucietti, Thomas Quella, Oliver Rosten, Per Sundell and particularly to Hugh
Osborn for useful
comments and discussions.  This research is supported by the 
Netherlands Foundation for Fundamental Research on
Matter (FOM). 
\vfil
\eject

\appendix{A}{$Osp(2N|4)$ Subalgebras} 

Corresponding to the shortening conditions
 \shortcond,  \shortcondo,  \semishortcond, 
\semishortcondo, with notation as in Table 1, there are subalgebras which
are now discussed.  The characters for these subalgebras lead to the
limits considered in section 5.

Corresponding to \shortcond\ for $(N,B,m)$ short multiplets we have,
\eqn\casso{
Osp(2N|4)\supset \big ( SU(2|m)\otimes SO(2N{-2m})\big ) \ltimes 
U(1)_{\I_m}\, .
}

The generators of $SU(2|m)$ are $M_\al{}^{\be}$, $\Q_{\ha\al}$,
${\bar \S}_{\hb}{}^{\be}$, $\ha,\hb=1,\dots,m$, of section 2, along with
$\H_m$
as in \limitBPS,
and
$T_{\ha\hb}$, generators of $SU(m)\subset SO(2m)$, with, in terms of the generators
in \crl,
\eqn\defT{
T_{\ha\ha}=H_{\ha}-{\ts {1\over m}}\sum_{\hb=1}^mH_{\hb}\, ,\qquad
T_{\ha\hb}=\cases{{\ts{i\over 2}}E^{+-}_{\ha\hb}\, ,\quad & $\ha<\hb$ \cr 
-{\ts{i\over 2}}E^{-+}_{\hb\ha}\, , \quad& $\hb<\ha$}\, ,
}
so that $\sum_{\ha=1}^mT_{\ha\ha}=0$ and, for $\hc,\hd=1,\dots, m$,
\eqn\relatioc{
[T_{\ha\hb},T_{\smash{\hc\hd}}]=\de_{\smash{\hb\hc}}\,T_{\smash{\ha\hd}}-\de_{\smash{\ha\hd}}\,
T_{\smash{\hb\hc}}\, .
}
$SU(2|m)$ has usual algebra with, in particular,
\eqn\commo{
\{\Q_{\ha\al},{\bar \S}_{\hb}{}^\be\}=2i (M_\al{}^\be\, \de_{\ha\hb}- \de_\al{}^\be\, 
T_{\ha\hb}
+\de_\al{}^\be\,\de_{\ha\hb}\,\H_m)\, ,
}
and
\eqn\central{
[\H_m,\Q_{\ha\al}]=({\ts{{1\over 2}-{1\over m}}})\Q_{\ha\al}\, ,
\qquad [\H_m,{\bar \S}_{\hb}{}^\be]=-({\ts{{1\over 2}-{1\over m}}}){\bar \S}_{\hb}{}^\be\, .
}
For $m=2$, $\H_m$ is evidently  then a central extension.

The $SO(2N{-2m})$ subgroup in \casso\ is generated by 
${R}_{\ba\,\bb}$ $\ba,\bb,\bar t,\bar u 
=2m{+1},\dots,
2N$, 
\eqn\redsym{
[R_{\ba \bb},R_{\bar t \bar u}]=i(\de_{\bar r\bar t}\,R_{\bar s\bar u}-\de_{\bar s\bar t}
\,R_{\bar r\bar u}-\de_{\bar r\bar u}\,R_{\bar s\bar t}+\de_{\bar s\bar u}\,R_{\bar r\bar t})\, ,
}
while
$\I_m$
in \limitBPS\ is
an external automorphism with,
\eqn\externo{
[\I_m,\Q_{\ha\al}]=({\ts{{1\over 2}+{1\over m}}}) \Q_{\ha\al}\, ,\qquad 
[\I_m,{\bar \S}_{\hb}{}^\be]=-({\ts{{1\over 2}+{1\over m}}}) {\bar \S}_{\hb}{}^\be\, .
}

Corresponding to \shortcond\ for $n=N$ for $(N,B,+)$ multiplets and,
separately, 
\shortcondo\ for $(N,B,-)$ multiplets we have,
\eqn\casso{
Osp(2N|4)\supset SU(2|N) \ltimes 
U(1)_{\I_\pm}\, .
}
The generators of $SU(2|N)$ are $M_\al{}^{\be}$, $\Q_{\ha\al}$,
${\bar \S}_{\hb}{}^{\be}$, for $\ha,\hb=1,\dots,N{-1},+$, for the $(N,B,+)$ case,
and $\ha,\hb=1,\dots,N{-1},-$, for the $(N,B,-)$ case, 
where we define 
\eqn\reput{
\Q_{+\al}=\Q_{N\al}\, ,\qquad \Q_{-\al}={\bar \Q}_{N\al}\, ,
\qquad {\bar \S}_{+}{}^{\al}={\bar \S}_{N}{}^{\al}\, ,
\qquad {\bar \S}_-{}^{\al}=\S_N{}^{\al}\, ,
}
 along with
$\H_\pm$
as in \limithalfBPSo,
and
$T_{\ha\hb}$,
generators of $SU(N)$ given by, for $\ha<N$,
\eqn\defTryu{\eqalign{
T_{\ha\ha}={}&H_{\ha}-{\ts {1\over N}}(H_1+\dots+H_{N{-1}}\pm H_N) \, ,\cr
T_{\pm\pm}={}&\pm H_N-{\ts {1\over N}}(H_1+\dots+H_{N{-1}}\pm H_N)\, ,\cr
T_{\ha \pm}={}&{\ts{i\over 2}}E^{+\pm}_{\ha N}\, ,\quad T_{\pm \ha}=-{\ts{i\over 2}}E^{-\pm}_{\ha N}\, ,\quad 
T_{\ha\hb}=\cases{{\ts{i\over 2}}E^{+-}_{\ha\hb}\, ,\quad & $\ha<\hb<N$ \cr 
-{\ts{i\over 2}}E^{-+}_{\hb\ha}\, , \quad& $\hb<\ha<N$}\, ,}
}
satisfying the same commutation relation \relatioc\ for 
$\ha,\hb,\hc,\hd=1,\dots,N{-1},\pm$.   

The $\Q_{\ha\al}$,
${\bar \S}_{\hb}{}^{\be}$ generators satisfy \commo\ for $\H_m$
replaced by $\H_{\pm}$, as appropriate, and $\ha,\hb=1,\dots,N{-1},\pm$,
and with,
\eqn\centralextensio{
[\H_\pm,\Q_{\ha\al}]=({\ts{{1\over 2}-{1\over N}}})\Q_{\ha\al}\, ,
\qquad 
[\H_\pm,{\bar \S}_{\hb}{}^\be]=-({\ts{{1\over 2}-{1\over N}}}){\bar \S}_{\hb}{}^\be\, ,
}
so that for $R$-symmetry group $SO(4)$, {\it i.e.} $ N=2$, 
$\H_\pm$ is a central extension.

$\I_{\pm}$ in \limithalfBPSo\ acts and as external automorphism with,
\eqn\externam{
[\I_{\pm},\Q_{\ha\al}]=({\ts{{1\over 2}+{1\over N}}}) \Q_{\ha\al}\, ,\qquad 
[\I_{\pm},{\bar \S}_{\hb}{}^\be]=-({\ts{{1\over 2}+{1\over N}}}) {\bar \S}_{\hb}{}^\be\, .
}

The expression \limitBPSb\ may be understood as follows.  By taking the limit $\de\to 0$,
only $(N,B,n)$, $n\geq m$, and $(N,B,\pm)$ BPS multiplets have states, including
the highest weight state, with zero $\H_m$ eigenvalues.  Characters for the 
$( SU(2|m)\otimes SO(2N{-2m})\big ) \ltimes 
U(1)_{\I_m}$ subalgebra, when restricted to these representations, reduce to
$U(1)_{\I_m}\otimes SO(2N{-2m})$ characters, as these representations are trivial
representations of the $SU(2|m)$ subalgebra, evident from section 3.

The limit in \limithalfBPS\ may be understood similarly to
other BPS cases whereby the characters reduce in the half BPS cases
to $U(1)_{\I_\pm}$ characters.

Corresponding to \semishortcond\ for $(N,A,m)$ semi-short multiplets we have,
\eqn\casso{
Osp(2N|4)\supset \big ( SU(1|m)\otimes Osp(2N{-2m}|2)\big ) \ltimes 
U(1)_{\K_m}\, .
}

The generators of $SU(1|m)$ are $\Q_{\ha 2},\bS_{\hb}{}^2,T_{\ha\hb}$,
$\ha,\hb=1,\dots,m$, along with 
$\J_m$
as in \limitSemishort\
where $SU(1|m)$ has usual algebra with in particular,
\eqn\genrat{
\{\Q_{\ha 2},\bS_{\hb}{}^2\}=2i(-T_{\ha\hb}+\de_{\ha\hb}\,\J_m)\, ,
}
with,
\eqn\centex{
[\J_m,\Q_{\ha 2}]=(1-{\ts{{1\over m}}})\Q_{\ha 2}\, ,
\qquad [\J_m,\bS_{\hb}{}^2]=-(1-{\ts{{1\over m}}})\bS_{\hb}{}^2\, .
}
For $m=1$, $\J_m$ is then a central extension. 

The generators of $Osp(2N{-2m}|2)$ are $Sp(2,{\Bbb{R}})\simeq SU(1,1)$ 
generators $P_{11}$, $K^{11}$
of section 2, along with,
$
 \bar D=D+J_3
$
in \limitSemishort\
satisfying,
\eqn\algcomp{
[\bar D,P_{11}]=2P_{11}\, ,\quad [\bar D,K^{11}]=-2 K^{11}\, ,
\quad [K^{11},P_{11}]=4 \bar D\, ,
}
along with $Q_{\ba 1}$, $S_{\bb}{}^1$, $\ba,\bb=2m{+1},\dots,
2N$,
of section 2, and $SO(2N-2m)$ generators 
$R_{\ba\bb}$ as before, satisfying,
\eqn\mixed{\eqalign{
[\bar D,Q_{\ba 1}]=Q_{\ba 1}&\, ,\qquad
[\bar D, S_{\bb}{}^{ 1}]=-S_{\bb}{}^1\, , \cr
[K^{11},Q_{\ba 1}]=2i S_{\ba}{}^1&\, ,\qquad 
[P_{11},S_{\bb}{}^1]=-2iQ_{\bb1}\, ,}
} 
and
\eqn\othercrs{
\{Q_{\ba1},Q_{\bb1}\}=2\de_{\ba\bb}P_{11}\, ,\qquad 
\{S_{\ba}{}^{1},S_{\bb}{}^{1}\}=2\de_{\ba\bb}K^{11}\, ,\qquad 
\{Q_{\ba1},S_{\bb}{}^{1}\}=2i\,\de_{\ba\bb}\bar D+2R_{\ba\bb}\, .
}

In \casso,
$\K_m$ as in \limitSemishort,
is an outer automorphism with,
\eqn\outerautosem{
[\K_m,\Q_{\ha 2}]={\ts{1\over m}}\Q_{\ha 2}\, ,\quad 
 [\K_m,\bS_{\hb}{}^{ 2}]=-{\ts{1\over m}}\bS_{\hb}{}^{ 2}\, ,\quad
 [\K_m,Q_{\ba 1}]=[\K_m,S_{\bb}{}^1]=0\, .
 }

Corresponding to \semishortcond\ for $n=N$ for $(N,A,+)$ multiplets, and 
separately, \semishortcondo\ for $(N,A,-)$ semi-short multiplets we have,
\eqn\casso{
Osp(2N|4)\supset \big ( SU(1|N)\otimes SU(1,1)\big ) \ltimes 
U(1)_{\K_\pm}\, .
}
The generators of $SU(1|N)$ are $\Q_{\ha 2},\bS_{\hb}{}^2,T_{\ha\hb}$,
for $\ha,\hb=1,\dots,N{-1},+$, for the $(N,A,+)$ case, and
$\ha,\hb=1,\dots,N{-1},-$, for the $(N,A,-)$ case, with the definitions \reput\
for $\al=2$, with  $T_{\ha\hb}$ as in
\defTryu\ and
$\J_\pm$
as in \limithalfBPSoppp. 

The supercharges satisfy \genrat\ 
with $\J_m$ replaced by $\J_{\pm}$, as appropriate,
and $\ha,\hb=1,\dots,N{-1},\pm$,
with,
\eqn\centex{
[\J_\pm,\Q_{\ha 2}]=(1-{\ts{{1\over N}}})\Q_{\ha 2}\, ,
\qquad [\J_\pm,\bS_{\hb}{}^2]=-(1-{\ts{{1\over N}}})\bS_{\hb}{}^2\, ,
}
so that, for $R$ symmetry group $SO(2)$, {\it i.e.} $N=1$, $\J_\pm$ is
a central
extension.
The $SU(1,1)$ generators are $P_{11}$, $K^{11}$ and $\bar D$, as above,
satisfying \algcomp.
Also,  $\K_{\pm}$ as given in \limithalfBPSoppp\ is an outer automorphism with,
\eqn\outerautosemopi{
[\K_\pm,\Q_{\ha 2}]={\ts{1\over N}}\Q_{\ha 2}\, ,\quad 
 [\K_\pm,\bS_{\hb}{}^{ 2}]=-{\ts{1\over N}}\bS_{\hb}{}^{ 2}\, .
 }

\limitSemishortob\ can be understood similarly to BPS limits.
 Again by taking the limit $\de\to 0$, only those states in
 relevant semishort multiplets and various BPS
 multiplets with zero $\J_m$ eigenvalue
 contribute to corresponding characters.  Note, however, that for semi-short
 multiplets, the states do not include highest weight states, but some
 superconformal descendants.  Such characters, when restricted to
 the $\big ( SU(1|m)\otimes Osp(2N{-2m}|2)\big ) \ltimes 
U(1)_{\K_m}$ subalgebra, reduce to $U(1)_{\K_m}\otimes Osp(2N{-2m}|2)$
characters as these representations are trivial representations of the $SU(1|m)$
subalgebra.

To see the equivalence in
terms of $Osp(2N{-2m}|2)$, characters, we may, as in section 2, make a 
change of basis
for $Q_{\ba 1},S_{\bb}{}^1$, $R_{\ba\bb}$ to the orthonormal basis, as in
\orthoalg\ and \supercharges, to $\{\Q_{\bar m 1}$, $\bQ_{\bar m 1}$, $\S_{\bar n}{}^1$, 
$\bS_{\bar n}{}^1$,
$H_{\bar m}$, $E^{\pm\pm}_{\bar m\bar n}$,  $E^{\pm\mp}_{\bar m\bar n}$,
$\bar m,\bar n=m{+1},\dots, N\}$,
in an
obvious way.  
Denoting highest weight states by $|\bar \De;\bar \rr\rangle^{\rm h.w.}$, where
\eqn\extrahighy{
(K^{11}, S_\bb{}^1)|\bar \De;\bar \rr\rangle^{\rm h.w.}=0\, ,\qquad 
(\bar D,H_{\bar m})|\bar \De;\bar \rr \r^{\rm h.w.}=(\bar \Delta, r_{\bar m})
|\bar \De;\bar \rr\rangle^{\rm h.w.}\, ,
}
unitarity requires $\bar \Delta\geq r_{m+1}$.  Compatible 
shortening conditions are given by, for  $\bar n=m{+1},\dots, {N}$,
\eqn\extrashortt{\eqalign{
\Q_{\bar n 1}|\bar \De;\bar \rr\rangle^{\rm h.w.}=0
\qquad \Rightarrow& \qquad \bar \Delta =r_{m{+1}}
=\dots =r_{\bar n}\, ,\cr
\bQ_{N 1} |\bar \De;\bar \rr\rangle^{\rm h.w.}=0\qquad 
\Rightarrow& \qquad \bar \Delta=r_{m{+1}}=\dots=r_{N{-1}}=-r_{N}\, .}
}

We may follow a similar procedure as for $Osp(2N|4)$ characters in section 4
to find $Osp(2N|2)$ characters for representations  $\R_{(\bar \Delta,\bar \rr)}$.
The corresponding characters for irreducible representations are 
given by,{\foot{Here the maximal
compact subgroup is $U(1)\otimes U(1)\otimes SO(2N{-2m})$ so that the relevant 
Weyl symmetriser, acting on 
Verma module
characters, is $\frak W^{\S_{N-m}\ltimes (\S_2)^{N-m-1}}$.}}
\eqn\drapes{\eqalign{
&{}\chi^{(Osp(2N{-2m}|2),i)}_{(\bar \De,\bar \rr)}(\bar s,\bar \y)
=\tr_{\R_{(\bar \Delta,\bar \rr)}}(\bar s^{{{\bar D} }}\bar y_1{}^{H_{m+1}}\cdots
\bar y_{N-m}{}^{H_N})\cr
&{}={\bar s^{\bar \Delta}\over 1-\bar s^2}\frak W^{\S_{N-m}\ltimes (\S_2)^{N-m-1}}
\Big(C^{(N-m)}_{\bar \rr}(\bar \y)\R^{(i)}(\bar s,\bar \y)
{\ts \prod}_{j=1}^{N-m-1}(1+\bar s\,\bar y_{j}{}^{-1})\Big)\, ,}
}
where, corresponding to the action of supercharges on the highest weight state for long
and short representations $\R_{(\bar \Delta,\bar \rr)}$, 
\eqn\defnewR{\eqalign{
\R^{(i)}(\bar s,\bar \y)=\cases{(1+\bar s \,(\bar y_{N-m}){}^{-1}){\ts \prod_{j=1}^{N-m}}(1+\bar s\, \bar y_{j})\, ,& 
for $i=\rm long$\cr
(1+\bar s \,(\bar y_{N-m}){}^{-1}){\ts \prod_{j=\bar n+1-m}^{N-m}}(1+\bar s \,\bar y_{j})\, ,& 
for $i=\bar n$\cr
(1+\bar s\, (\bar y_{N-m}){}^{\mp1})\, ,& for $i=\pm$,}}
}
where for $i={\rm long}$ then  $\bar \De\geq r_{m+1}$, for long multiplets,
while for $i=\bar n$ then $\bar \De=r_{m+1}=\dots=r_{\bar n}> r_{\bar n+1}$
and for $i=\pm$ then $\bar \De=r_{m+1}=\dots=r_{N-1}=\pm  r_n$.

\appendix{B}{Character Expansions}

Defining,
\eqn\deff{
f(u,x,y)=\sum_{r,s,t\geq 0}N_{r s t}u^r \chi_s(x)\chi_t(y)\, ,
}
and using usual the orthogonality
relation for $SU(2)$ characters in \chars,
\eqn\sutorthyuop{
-{1\over 4 \pi i}
\oint {\d z\over z}(z-z^{-1})^2\chi_s(z)\chi_t(z)= \de_{st}\, ,
}
which may be equivalently expressed by, due to $\chi_j(z)=\chi_j(z^{-1})$,
\eqn\sutorth{
-{1\over 2 \pi i}
\oint \chi_s(z)\chi_t(z)(z-z^{-1})\d z= \de_{st}\, ,
} 
we have that,
\eqn\orthapp{
N_{rst}={1\over (2 \pi i)^3}\oint \oint \oint f(u,x,y)u^{-r-1}\chi_s(x)\chi_t(y)(x-x^{-1})
(y-y^{-1})\,\d u \,\d x \,\d y\,,
}
where each contour is the relevant unit circle.
Using $\chi_s(z)=-\chi_{-s-2}(z)$, we have,
\eqn\sym{
N_{r s t}=-N_{r\,\,{-s-2}\,\,t}=-N_{rs\,\,{-t-2}}=N_{r\,\,{-s-2}\,\,{-t-2}}\, .
}
It is convenient below to construct generating functions,
\eqn\dreftoo{
F_{r s}(z)=\sum_{r=0}^\infty N_{r\,r{-2s}\,r{-2t}} z^r\, ,
}
so that using \orthapp, summing over $r$ and performing the subsequent contour integral
over $u$ for $|z|<|x y|$, 
\eqn\dreft{\eqalign{
F_{rs}(z)={}& {1\over (2 \pi i)^2}\oint \oint \bigg(x^{1-2 s} y^{1-2t}f(z x y,x,y)-x^{1-2s}y^{-1+2t} f(z x/y, x, y)\cr
&\qquad \qquad\,\, -x^{-1+2s}y^{1-2t}f(z y/x, x, y)+x^{-1+2s}y^{-1+2t} f(z/(x y),x, y)\bigg)\d x\,\d y\, .}
} 
Note that for application to the main text,
in terms of the notation used below,
\eqn\yopihnm{
N_{{\rm{free}},(r,r{-s-t},t-s)}=N_{{\rm F}, r\, r{-}2s\, r{-}2t}\, ,\qquad
N_{{\rm{sugra}},(r,r{-s-t},t-s)}=N_{{\rm S}, r\, r{-}2s\, r{-}2t}\, .
}

\noindent
{\bf The Free Field Case} 

{}For the free field case, due to \largeNlo\ with \chexp, we consider,
\eqn\expansionjio{
f_{\rm F}(u,x,y)=\prod_{j\geq 1}{1\over 1-u^{j}\chi_1(x{}^j)\chi_1(y^j)}
=\sum_{r,s,t\geq 0}N_{{\rm F},r s t}u^r\chi_s(x)\chi_t(y)\, ,
}
which may be expanded as,
\eqn\expansionji{
f_{\rm F}(u,x,y)=\sum_{\blambda}
u^{|\blambda|}p_{\blambda}(x,x{}^{-1})p_{\blambda}(y,y{}^{-1})\,,
}
in terms of power symmetric polynomials $p_\blambda(\z)$, \defs\ with \defpn.
Using an expansion formula for power symmetric polynomials
in terms of Schur polynomials, we may expand the latter in terms of 
$SU(2)$ characters via,
\eqn\schurexp{
p_\blambda(z,z^{-1})=
\sum_{m,n\geq 0\atop m+2n=|\blambda|}\omega_\blambda{}^{(m,n)}
\chi_m(z)\, ,
}
where $\omega_\blambda{}^{(m,n)}$ are symmetric group
characters.
Introducing the notation,
$\left({\blambda\atop\brho}\right)
=\prod_{j\geq 1}\left({\lambda_j\atop \rho_j}\right)$
then it is easily determined from \schurexp\ that (other formulae
for these characters may be found in in \wyb\ but the following is the most
useful for purposes here, and is perhaps simpler), 
\eqn\formch{
\omega_\blambda^{(|\blambda|,0)}=1\, ,\qquad
\omega_\blambda{}^{(|\blambda|-2n,n)}=
\sum_{\brho\atop |\brho|=n}\left({\blambda\atop\brho}\right)
-\sum_{\brho\atop |\brho|=n-1}\left({\blambda\atop\brho}\right)\, ,
\quad n=1,\dots, [|\blambda|/2],
}
with $\omega^{(n,m)}_{\blambda}$ being otherwise zero.
Thus, using \sutorth, \orthapp\ for \expansionji\ with \schurexp, 
\eqn\genres{
N_{{\rm F}, r\, r{-}2s\, r{-}2t}=
\sum_{\blambda\atop|\blambda|=r}\omega_\blambda{}^{(r-2s,s)}
\omega_\blambda{}^{(r-2t,t)}\, ,\quad s,t=0,\dots,[r/2]\, .
}
Thus, $N_{{\rm F}, r\, r{-}2s\, r{-}2t}$ is a potentially non-zero, non-negative integer 
only for $r,s=0,\dots,[r/2]$.

A useful identity, which may be easily generalised, 
is the following, namely, for
\eqn\defNr{
n_{rij}=\sum_{\blambda\atop |\blambda|=r}
\left(\lambda_j\atop i\right)\, ,
}
then,{\foot{This may be seen in a simple way by first introducing,
$h(x,z)={1\over 1-x z^j}\prod_{k=1\atop k\neq j}^\infty{1\over 1-z^k}\, ,
$
so that, in a series expansion,
$
h(x,z)=\sum_{\blambda} x^{\lambda_j}z^{|\blambda|}\,.
$
We then have
$\lim_{x\to 1}{1\over i!}{\d^i\over \d x^i}h(x,z)= g_{ij}(z)$
using 
${1\over i!}{\d^i\over \d x^i}{1\over 1-x z^j}={z^{i j}
\over (1-x z^j)^{i+1}}$.}
}
\eqn\cardiagram{
g_{i j}(z)=\sum_{r=0}^\infty n_{rij}\, z^r=
{z^{ij}\over (1-z^j)^i}\prod_{k=1}^{\infty}{1\over 1-z^k}\, .
}

The simplest case is for $s=t=0$, whereby for \dreftoo\ we find, using \formch\ along with \cardiagram,
\eqn\generato{
F_{{\rm F},00}(z)=g_{00}(z)=\prod_{k=1}^\infty {1\over 1-z^k}\, ,
}
and thus have, giving \fcaseo,
\eqn\formcho{
N_{{\rm F},r r r}=n_{r00}=p(r)\, .
}

Similarly, for \dreftoo\ using \formch\ along with \cardiagram,
\eqn\generatt{
F_{{\rm F}, 10}(z)=F_{{\rm F}, 01}(z)
=g_{11}(z)-g_{00}(z)={2z-1\over 1-z}\prod_{k=1}^\infty {1\over 1-z^k}\,,
}
and so, agreeing with \fcaset,
\eqn\formcht{
N_{{\rm F},r\,r{-2}\,r}=N_{{\rm F}, r\,r\,r{-2}}=n_{r11}-p(r)=\sum_{j=0}^{r-1}p(j)-p(r)\, .
}

Similarly, for \dreftoo\ using \formch\ along with \cardiagram,
\eqn\generatt{
F_{{\rm F}, 11}(z)=2g_{21}(z)-g_{11}(z)+g_{00}(z)={4z^2-3z+1\over (1-z)^2}\prod_{k=1}^\infty {1\over 1-z^k}\,,
}
and so, giving \fcaseth,
\eqn\formcht{
N_{{\rm F}, r\,r{-2}\,r{-2}}=2n_{r21}-n_{r11}+p(r)\, .
}

Notice, that by a very similar argument, it may be shown that
from \formch\ with \genres\ for $t=0$,
\eqn\quarterbpspartoiu{\eqalign{
&F_{{\rm F},s\,0}(z)=F_{{\rm F},0\,s}(z)\cr
&=\sum_{\rho_1,\dots,\rho_s\geq 0}
\Big(z^s\de_{\rho_1+\dots+s\rho_s,s}-z^{s-1}\de_{\rho_1+\dots+s\rho_s,s-1}\Big)
{1\over \prod_{j=1}^s(1-z^j)^{\rho_j}}\prod_{k=1}^\infty{1\over 1-z^k}\, ,}
}
enabling determination of $N_{{\rm F},r\,r-2s\,r-2t}$ for $s\,t=0$.

\noindent
{\bf The Supergravity Case}

In this case we consider, due to \multigrav\ with \chexpo,
\eqn\deff{
f_{{\rm S}}(u,x,y)=\prod_{j\geq 1}{1\over \prod_{k,l=0}^j(1-u^j x^{2k-j}y^{2l-j})}
=\sum_{r,s,t\geq 0}N_{{\rm S},r s t}u^r \chi_s(x)\chi_t(y)\, ,
}
and we use \dreft, \dreftoo, to determine generating functions for low $s,t$.

For application to \dreftoo, we have the leading terms,
\eqn\krampmystyle{\eqalign{
f_{{\rm S}}(z x y,x, y)={}&\prod_{k=1}^\infty {1\over 1-z^k}+\dots \, ,\cr
f_{{\rm S}}(zx/y,x,y)={}&\prod_{k=1}^\infty {1\over (1-z^k)(1-z^k y^{-2})}+\dots \, ,\cr
f_{{\rm S}}(zy/x,x,y)={}&\prod_{k=1}^\infty {1\over (1-z^k)(1-z^k x^{-2})}+\dots \, ,\cr
f_{\rm S}(z/(x y),x y)={}&
\prod_{k=1}^\infty {1\over (1-z^k)(1-z^k x^{-2})(1-z^k y^{-2})(1-z^k x^{-2}y^{-2})}+\dots \, ,
}
}
where $\dots$ denotes terms that in a series expansion
in $x,y$ involve powers $x^{2a},y^{2b}$, $a,b\in \Bbb{Z}$, $a, b\neq 0,-1$,
that do not contribute below.  

We have, from \krampmystyle,
\eqn\fourparto{\eqalign{
{1\over (2 \pi i)^2}\oint \oint x y f_{\rm S}(z x y,x,y)\d x\,\d y&{}=0\, ,\cr
{1\over (2 \pi i)^2}\oint \oint {x\over y} f_{\rm S}(z x/y, x, y)\d x\, \d y&{}=0\, ,\cr
{1\over (2 \pi i)^2}\oint \oint {y\over x} f_{\rm S}(z y/x, x, y)\d x\,\d y&{}=0\, ,\cr
{1\over (2 \pi i)^2}\oint \oint {1\over x y} f_{\rm S}(z/(x y),x, y)\d x\,\d y
&{}= \prod_{k=1}^\infty {1\over 1-z^k}\, ,}
}
so that for \dreftoo\ we have, from \dreft,
\eqn\defGoo{
F_{{\rm S}, 00}(z)=g_{00}(z)\, .
}
agreeing with \generato.
Hence, $N_{{\rm S}, r r r}=N_{{\rm F}, r rr }$ in \formcho, 
leading to the first equation in
\uquarto.

Similarly, using \krampmystyle\ and 
\eqn\identyuio{
\prod_{k=1}^\infty {1\over 1-t z^k}=1+{z t\over 1-z}+O(t^2,z^2)\, ,
}
we have,
\eqn\fourpartt{\eqalign{
{1\over (2 \pi i)^2}\oint \oint {x\over y} f_{\rm S}(z x y,x,y)\d x\,\d y{}&=0\, ,\cr
{1\over (2 \pi i)^2}\oint \oint {y x} f_{\rm S}(z x/y, x, y)\d x\, \d y&{}=0\, ,\cr
{1\over (2 \pi i)^2}\oint \oint {1\over x y} f_{\rm S}(z y/x, x, y)\d x\,\d y
&{}= \prod_{k=1}^\infty {1\over 1-z^k}\, ,\cr
{1\over (2 \pi i)^2}\oint \oint {y\over x } f_{\rm S}(z/(x y),x, y)\d x\,\d y&{}=  
{z\over 1-z}\prod_{k=1}^\infty {1\over 1-z^k}\, ,}
}
so that for \dreftoo, from \dreft,
\eqn\defGoop{
F_{{\rm S}, 01}(z)=g_{11}(z)-g_{00}(z)\, ,
}
agreeing with \generatt, so that $N_{{\rm S}, r \,r\, r{-2}}=N_{{\rm F}, r\, r\, r {-2}}$ 
in \formcht.  It is easy to show that $F_{{\rm S},10}(z)=F_{{\rm S},01}(z)$
so that $N_{{\rm S}, r \,r{-2}\, r}=N_{{\rm S}, r \,r\,r{-2}}$.  Thus we have \uquarto.

Similarly, using \krampmystyle\ with \identyuio,
\eqn\fourpartt{\eqalign{
{1\over (2 \pi i)^2}\oint \oint {1\over x y} f_{\rm S}(z x y,x,y)\d x\,\d y&{}
= \prod_{k=1}^\infty {1\over 1-z^k}\, ,\cr
{1\over (2 \pi i)^2}\oint \oint {y\over x} f_{\rm S}(z x/y, x, y)\d x\, \d y&{}
=
{z\over 1-z}\prod_{k=1}^\infty {1\over 1-z^k}\, ,\cr
{1\over (2 \pi i)^2}\oint \oint {x\over y} f_{\rm S}(z y/x, x, y)\d x\,\d y&{}
= 
{z\over 1-z}\prod_{k=1}^\infty {1\over 1-z^k}\, ,\cr
{1\over (2 \pi i)^2}\oint \oint {x y } f_{\rm S}(z/(x y),x, y)\d x\,\d y&{}
= \bigg( {z\over 1-z}+{z^2\over (1-z)^2}\bigg)
\prod_{k=1}^\infty {1\over 1-z^k}\, ,}
}
so that for \dreftoo, from \dreft,
\eqn\defGoop{
F_{{\rm S},11}(z)=g_{21}(z)-g_{11}(z)+g_{00}(z)
={3z^2-3z+1\over (1-z)^2}\prod_{k=1}^\infty {1\over 1-z^k}\,,
}
giving, from \cardiagram,
\eqn\reftyop{
N_{{\rm S},r\, r{-2}\,r{-2}}=n_{r 21}-n_{r 11}+p(r)\, ,
}
leading to \fcasethuip.

\appendix{C}{Counting Multi-traces of Commuting Matrices}

In \domokos\ it was shown that a basis for multiple traces of
$J$ commuting $N\times N$ matrices $X_k$, $k=1,\dots,J$,
is provided by,
\eqn\basisdiff{
\tr\, {\U}_1 \, \, \tr \, {\U}_2\cdots \tr \, {\U}_N\, , \qquad 
{\U}_i=\prod_{j=1}^{n}Y_{ij}\, ,\qquad Y_{i j}\in \{X_k,\,k=0,\dots, J\}\, ,
}
for $n\to \infty$, where $X_0=\Bbb{I}$, the identity matrix,
which, of course, also commutes with $X_k$, $k=1,\dots, J$,
so that initially we treat it on an equal footing with the other matrices. 
\basisdiff\ is linearly independent up to the action of the automorphism
symmetry group described below.

Each $\U_i$ has an automorphism symmetry $\S_n$
since ${ \U}_i=\prod_{j=1}^n Y_{ij}=({\U}_i)^{\si}=
\prod_{j=1}^n Y_{i \si(j)}$  for every $\si\in \S_n$ 
(since $X_k$ commute).  Hence each $\tr\,\U_i$ corresponds to a graph
with $\S_n$ symmetry $K_n$ where $Y_{ij}$ to corresponds to
the $j^{\rm th}$ vertex.  Here, $K_n$ is taken to be the complete graph
on $n$ vertices.{\foot{This is the well known graph where any two vertices are joined
by an edge.  $K_n$ is used here for
illustrative purposes - we could in fact consider any graph with $\S_n$ automorphism
symmetry, for instance the graph complement of $K_n$, composed of just
$n$ vertices, no edges.}}
Particular $\tr\,\U_i$ corresponds to a colouring of the vertices of
$K_n$, in colours $c_k$, $k=0,\dots, J$,
with the exact value of $Y_{ij}\in \{X_k,\,k=0,\dots, J\}$ 
corresponding to the colour of
the $j^{\rm th}$ vertex.

$\tr\, { \U}_1 \, \, \tr \, {\U}_2\cdots 
\tr \, { \U}_N$ corresponds to $N$ copies of $K_n$, denoted $K_n{}^N$.
This graph has a corresponding automorphism group, in graph theory
known as the wreath product group, in this case given by
the semi-direct product $(\S_n)^N\rtimes \S_N$.
For this group, for $\si=(\si_1,\dots, \si_N)\in (\S_n)^N$,  $\tau\in \S_N$, and 
defining $\si^\tau=(\si_{\tau^{-1}(1)},\dots,\si_{\tau^{-1}(N)})$, then for
 $(\si,\tau)$, $(\si',\tau')\in (\S_n)^N\rtimes \S_N$,
group multiplication is defined by
$(\si,\tau)(\si',\tau')=
(\si\si'{}^{\tau^{-1}},\tau\tau')$, so that, with ${1}_{(\S_n)^N\ltimes \S_N}
=
({1}_{\S_n},\dots,{1}_{\S_n},{1}_{\S_N})$, then
the inverse $(\si,\tau)^{-1}=((\si^{-1})^{\tau},\tau^{-1})$. 

The action of $(\S_n)^N\rtimes \S_N$ on \basisdiff\  is given by 
$(\tr\, { \U}_1 \, \, \tr \, {\U}_2\cdots 
\tr \, { \U}_N)^{(\si,\tau)}
=
\tr({ \U}_{\tau(1)})^{\si_1} \, \, \tr ( {\U}_{\tau(2)})^{\si_2}\cdots 
\tr ( { \U}_{\tau(N)})^{\si_N}$, $(\U_i)^{\si_k}=\prod_{j=1}^n Y_{i\si_k(j)}$,
with elements of the basis related by such a group transformation
being identical. To achieve  a linearly independent basis it is necessary to
divide out by this automorphism symmetry.

Generically, depending on the graph $G$,
with $n$ vertices, having automorphism group symmetry $\Gamma$,
the generating function
for the number $N_{n_0,\dots,n_J}$ of colourings of the vertices of $G$, 
in $n_k$ colours 
$c_k$, $k=0,\dots, J$, $\sum_{k=0}^Jn_k=n$, may be written as,
\eqn\generating{
C_{\Gamma}(x_0,\dots,x_J)=\sum_{n_0,\dots,n_J\geq 0\atop
n_0+\dots +n_J=n}N_{n_0,\dots,n_J}x_0{}^{n_0}\cdots x_J{}^{n_J}\, .
}

We may use the Polya enumeration theorem to determine the generating
function for the number of colourings of $K_n{}^N$, 
equivalent to the number of independent multiple trace operators, 
\basisdiff, subject
to the automorphism symmetry $(\S_n)^N\rtimes \S_N$, 
in terms of cycle indices for the
symmetric permutation group.

The cycle index for the symmetric permutation group $\S_n$ is
given by, with the notation of section 6,
\eqn\cycleindexsym{
Z_{\S_n}(s_1,\dots,s_n)=\sum_{\blambda\atop |\blambda|=n}
{\ts \prod_{j=1}^n}
{1\over j^{\lambda_j}\lambda_j !}s_j{}^{\lambda_j}
={1\over 2\pi i}\oint {\d z\over z^{n+1}}
{\exp\bigg(\sum_{j\geq 1}{1\over j}z^j s_j\bigg)}\, ,}
and the generating function for the number of colourings of $K_n$
is, by the Polya
enumeration theorem,
\eqn\identyu{\eqalign{
C_{\S_n}(x_0,x_1,\dots, x_J)&{}=Z_{\S_n}(s_1,\dots, s_n)\, , 
\qquad s_i=p_i(x_0,\dots,x_J)\, ,\cr 
&{}={1\over 2 \pi i}\oint{\d z\over z^{n+1}}{1\over (1- z x_0)\dots (1-z x_J)}\, ,}
}
where $p_i(\x)=\sum_{k=0}^J x_{k}{}^i$, a power symmetric polynomial.
Similarly, for $K_n{}^N$, with automorphism group $(\S_n)^N\rtimes \S_N$, 
the generating function for the number of colourings is given by,
using the Polya enumeration theorem applied to wreath product groups
\harary,
\eqn\petwreath{
C_{(\S_n)^N\rtimes 
\S_N}(x_0,x_1,\dots,x_J)=Z_{\S_N}(\widetilde{s}_1,\dots,{\widetilde{s}}_N)\,,
}
where
\eqn\juiop{
 {\widetilde{s}}_i=Z_{\S_n}(s_{1i},\dots,s_{ni})=C_{\S_n}(x_0{}^i,\dots,x_J{}^i)\, ,
\qquad s_{ji}=p_j(x_0{}^i,x_1{}^i,\dots,x_J{}^{i})\,.
}

In order that contributions to $\lim_{n\to \infty}C_{\S_n}(x_0,x_1,\dots,x_J)$
from finite numbers of $X_k$, $k=1,\dots,J$
in $\tr\,U_i$  not vanish then we must also take $x_0\to1$. 
Following from \identyu,
\eqn\takelimt{
\lim_{n\to \infty}C_{\S_n}(1,x_1,\dots,x_J)
=\lim_{s\to 1}(1-s)\sum_{n=0}^{\infty}s^n C_{\S_n}(1,x_1,\dots,x_J)
={1\over (1-x_1)\cdots (1-x_J)}\,
.
}
Thus the generating function for the number of distinct basis elements of the form
\basisdiff\ as $n\to \infty$ is,
\eqn\genbasismulti{\eqalign{
\lim_{n\to \infty}
C_{(\S_n)^N\rtimes \S_N}(1,x_1,\dots,x_J)={}&
Z_{\S_N}(\widetilde{s}_1,\dots,{\widetilde{s}}_N)\, , \qquad \widetilde{s}_i
={1\over (1-x_1{}^i)\cdots (1-x_J{}^i)}\, ,\cr
={}& {1\over 2\pi i}\oint {\d z\over z^{N+1}}\exp\bigg(\sum_{n\geq 1}{1\over n}z^n
\widetilde{s}_n\bigg)\cr
={}&{1\over 2\pi i}\oint {\d z\over z^{N+1}}\prod_{j_1,\dots,j_J=0}^\infty
{1\over (1-z x_1{}^{j_1}\cdots x_J{}^{j_J})}\, ,
}
}
which matches exactly with the chiral ring partition function obtained using the
plethystic approach \bofeng.

\listrefs

\bye